\documentclass[12pt,a4paper,oneside]{report}

\usepackage[T1]{fontenc}
\usepackage[utf8]{inputenc}
\usepackage{lmodern}
\usepackage{microtype}
\usepackage{psl-cover/psl-cover}

\usepackage[a4paper,margin=1in]{geometry}
\usepackage{setspace}
\usepackage{amsmath,amssymb,amsthm,mathtools}
\usepackage{bm}
\usepackage{physics}
\usepackage{siunitx}
\usepackage{braket}

\usepackage{graphicx}
\graphicspath{{figures/}}
\usepackage{caption}
\usepackage{subcaption}
\usepackage{booktabs}
\usepackage{enumitem} %
\usepackage{comment} %

\usepackage{color}
\definecolor{Blue}{rgb}{0.00, 0.00, 1.00}
\definecolor{Red}{rgb}{1.00, 0.00, 0.00}
\definecolor{Green}{rgb}{0.00, 0.70, 0.00}
\newcommand{\red}{\color{Red}}
\newcommand{\blue}{\color{Blue}}
\newcommand{\green}{\color{Green}}

\usepackage[hidelinks]{hyperref}
\usepackage[nameinlink,noabbrev]{cleveref}

\usepackage[
  backend=bibtex,
  style=numeric-comp,
  sorting=none
]{biblatex}
\newcommand{\pubref}[1]{\hyperref[pub:#1]{#1}}
\newcommand{\Pone}{\pubref{P1}}
\newcommand{\Ptwo}{\pubref{P2}}
\newcommand{\Pthree}{\pubref{P3}}
\newcommand{\Pfour}{\pubref{P4}}

\providecommand{\D}{\mathrm{d}}
\providecommand{\I}{\mathrm{i}}

\theoremstyle{plain}

\theoremstyle{definition}

\pslassetspath{psl-cover}

\title{Fluctuations and multifractality in stochastic models of interface growth and population dynamics}
\author{Maximilien Bernard}

\institute{ENS - PSL}
\doctoralschool{Physique en Île-de-France}{564}
\specialty{Physique Théorique}
\date{26 Juin 2026}

\frabstract{
Cette thèse étudie comment le désordre et les fluctuations façonnent les phénomènes de croissance à grande échelle dans trois contextes liés : les interfaces fluctuantes, les chaînes de ressorts aléatoires et la croissance multiplicative aléatoire.

La première partie concerne la \textbf{croissance d'interfaces}. Dans ce contexte, une interface initialement plate devient rugueuse et développe des corrélations sur des échelles de longueur croissantes. Ce comportement est caractérisé par des exposants d'échelle, qui décrivent comment les fluctuations de hauteur croissent dans l'espace et le temps. Ces exposants peuvent être mesurés soit par des observables globales, sondant l'ensemble de l'interface, soit par des observables locales restreintes à une fenêtre finie.  Cependant, dans de nombreux systèmes expérimentaux et numériques, ces deux descriptions ne concordent pas, un phénomène connu sous le nom de lois d'échelle anormales. Pour clarifier son origine, nous étudions deux modèles dans lesquels ce décalage a des causes différentes. Le premier est une ligne élastique hétérogène, où nous montrons que l'anomalie apparente est un pur effet statistique. Le second est l'équation des milieux poreux stochastique, un modèle fortement non linéaire dans lequel l'anomalie est réelle : les interfaces affichent des lois locales et globales distinctes que nous déterminons.

La deuxième partie porte sur la \textbf{localisation d'Anderson}, qui apparaît ici à travers les propriétés spectrales des lignes élastiques hétérogènes. Dans une chaîne homogène, les modes propres de vibration sont étendus sur tout le système. En présence de désordre gelé, ils peuvent au contraire se localiser spatialement, avec des amplitudes concentrées près d'un centre de localisation et décroissant à mesure que l'on s'en éloigne.  Nous étudions des chaînes présentant à la fois des masses et des constantes de ressort aléatoires, en particulier dans le régime de désordre fort, où les développements perturbatifs à faible désordre faible échouent. Nous développons une nouvelle méthode combinatoire pour explorer ce régime.

La dernière partie concerne la \textbf{croissance multiplicative aléatoire} avec redistribution. Dans de tels modèles, les variations de richesse ou de population par exemple, sont proportionnelles à la quantité déjà présente, de sorte que de faibles différences initiales se trouvent amplifiées au cours du temps. Ce processus génère naturellement des distributions larges et fournit un mécanisme simple pour l'émergence d'inégalités et de phénomènes de concentration, avec des applications à la dynamique des populations, à la distribution des richesses, à la croissance des villes et à l'écologie. Nous étudions en particulier des modèles dans lesquels chaque site possède son propre taux de croissance gelé, représentant un avantage ou un désavantage persistant, tout en étant soumis à des fluctuations transitoires. La redistribution entre en compétition avec ces mécanismes en homogénéisant le système. Cette compétition conduit à des diagrammes de phase riches, présentant des phases localisées et délocalisées, ainsi qu'une nouvelle phase intermédiaire, que nous qualifions de partiellement localisée, et qui résulte de l'interaction entre l'hétérogénéité gelée et le bruit temporel.
}

\enabstract{
This thesis studies how disorder and fluctuations shape large-scale growth phenomena in three related settings: fluctuating interfaces, random spring chains, and random multiplicative growth.

The first part concerns \textbf{interface growth}. In standard kinetic roughening, an initially flat interface becomes rough and develops correlations over increasing length scales. This behavior is characterized by scaling exponents, which describe how height fluctuations grow in space and time. These exponents can be measured either through global observables, probing the whole interface, or through local observables restricted to a finite window. In many experimental and numerical systems, however, these two descriptions do not agree, a phenomenon known as anomalous scaling. To clarify its origin, we study two models in which this mismatch has different causes. The first is a heterogeneous elastic line, where we show that the apparent anomaly is a purely statistical effect. The second is the stochastic porous medium equation, a strongly nonlinear model in which the anomaly is genuine: the interfaces display distinct local and global laws that we determine.

The second part concerns \textbf{Anderson localization}, which appears here through the spectral properties of the heterogeneous elastic lines. In a homogeneous chain, vibrational eigenmodes are extended over the whole system. In the presence of quenched disorder, they may instead become spatially localized, with amplitudes concentrated near a localization center and decaying away from it. We study chains with both random masses and random spring constants, with particular emphasis on the strong-disorder regime, where standard weak-disorder expansions break down. We develop a new combinatorial method to probe this regime.

The last part concerns \textbf{random multiplicative growth} with redistribution. In such models, changes in wealth, population, or mass, for instance, are proportional to the amount already present, so that small differences are amplified over time. This naturally generates broad distributions and provides a simple mechanism for the emergence of inequality and concentration, with applications to population dynamics, wealth distribution, city growth, and ecology. We in particular study models in which each site has its own quenched growth rate, representing a persistent advantage or disadvantage, and is also subject to transient fluctuations. Redistribution competes with these mechanisms by homogenizing the system. This competition leads to rich phase diagrams, with localized and delocalized phases, as well as a new intermediate phase, which we call partially localized, and which arises from the interplay between quenched heterogeneity and temporal noise.
}
\frkeywords{croissance d'interface; lois d'échelle anormales ;  localisation d'Anderson ; croissance multiplicative aléatoire ;  localisation}
\enkeywords{Interface grwoth;  anomalous scaling;  Anderson localization; multiplicative growth; localization}

\jurymember{1}{C\'ecile, Monthus}{Institut de Physique Th\'eorique, CEA}{Pr\'esidente}
\jurymember{2}{Stefano, Zapperi}{Universit\`a degli Studi di Milano}{Rapporteur}
\jurymember{3}{Romualdo, Pastor-Satorras}{Universitat Polit\`ecnica de Catalunya}{Rapporteur}
\jurymember{4}{Aleksandra, Pektovic}{Universit\'e de Toulouse}{Examinatrice}
\jurymember{5}{Yan, Fyodorov}{King's College London}{Examinateur}
\jurymember{6}{Pierre, Le Doussal}{Ecole Normale Supérieure}{Directeur de th\`ese}
\jurymember{7}{Alberto, Rosso}{LPTMS}{Directeur de th\`ese}
\jurymember{8}{Jean-Philippe, Bouchaud}{CFM}{Invit\'e}

\begin{document}

\pagenumbering{roman}
\maketitle{}
\chapter*{Acknowledgements}

I would first like to thank my PhD supervisors, Pierre Le Doussal and Alberto Rosso. It has been a real pleasure to work with them, and I have learned a great deal from both of them over the course of this thesis. I particularly appreciated the freedom they gave me, always finding the right balance between guiding me and letting me explore on my own. I would also like to thank Andrei Fedorenko, with whom I very much enjoyed working on the stochastic porous medium equation, Christophe Texier, who introduced me to the world of Anderson localization and generously spent a lot of time explaining it to me, and Jean-Philippe Bouchaud, who introduced me to the very rich set of problems studied in the second part of this thesis. I also thank the members of my \emph{comité de suivi}, Camille Scalliet and Frederic van Wijland, for following my work and checking on me each year.

I spent these years between two laboratories, LPENS and LPTMS. I thank the administrative and technical staff of both labs for all their help over the years. I also had the pleasure to teach under the supervision of Guilhem Semerjian and Giulio Biroli, whom I would like to thank. More generally, I would like to thank all the PhD students and postdocs of LPENS and LPTMS whom I have had the pleasure to meet over the years. They bring life to the labs and make the PhD experience much more enjoyable. They are too many to name individually, but I am sure they will recognize themselves.  At LPENS, Paolo deserves a special mention for organizing the coffee meetings and helping to hold together the daily life of the lab. At LPTMS, I would like to thank Andrey and Florent for organizing the journal club. I would particularly like to thank my office mates, Benoît and Yueheng, whose company I greatly appreciated throughout these years.

Outside the lab, often at the climbing gym, I would like to thank the friends with whom I shared many great moments, in particular Sydney, Perceval, and Ali. Finally, I would of course like to thank my family, my brothers, my parents, and my grandparents, for their help and support throughout the years.
\clearpage

\chapter*{Publications}
\addcontentsline{toc}{chapter}{Publications}

The results presented in this thesis are based in part on the following peer-reviewed publications:
In the manuscript, these publications are referred to as \Pone--\Pfour.

\begin{enumerate}[label=P\arabic*,ref=P\arabic*]
\item \label{pub:P1} Maximilien Bernard, Pierre Le Doussal, Alberto Rosso, and Christophe Texier.
\textit{Anomalous scaling of heterogeneous elastic lines: A picture from sample-to-sample fluctuations}.
Physical Review E \textbf{110}, 014104 (2024).
\href{https://doi.org/10.1103/PhysRevE.110.014104}{doi: 10.1103/PhysRevE.110.014104}.

\item \label{pub:P2} Maximilien Bernard, Andrei A. Fedorenko, Pierre Le Doussal, and Alberto Rosso.
\textit{Stochastic porous-medium equation in one dimension}.
Physical Review E \textbf{112}, L043501 (2025).
\href{https://doi.org/10.1103/yn11-gdk9}{doi: 10.1103/yn11-gdk9}.

\item \label{pub:P3} Maximilien Bernard and Christophe Texier.
\textit{Disordered harmonic chains with random masses and springs: A combinatorial approach}.
Physical Review E \textbf{113}, 014143 (2026).
\href{https://doi.org/10.1103/bc9p-fhyz}{doi: 10.1103/bc9p-fhyz}.

\item \label{pub:P4} Maximilien Bernard, Jean-Philippe Bouchaud, and Pierre Le Doussal.
\textit{A mean-field theory for heterogeneous random growth with redistribution}.
Physical Review E \textbf{113}, L032101 (2026).
\href{https://doi.org/10.1103/nchz-flkj}{doi: 10.1103/nchz-flkj}.
\end{enumerate}

Some results presented in Chapters~7 and~8 are not yet published and will be presented in forthcoming publications:

\begin{enumerate}
\item Maximilien Bernard, Jean-Philippe Bouchaud, and Pierre Le Doussal.
\textit{Heterogeneous random growth with mean-field interactions}.
In preparation.

\item  Maximilien Bernard, Giuseppe Del Vecchio Del Vecchio, Jean-Philippe Bouchaud, and Pierre Le Doussal.
\textit{Mean-field growth with quenched heterogeneity and stochastic resetting}.
In preparation.
\end{enumerate}

\clearpage

\tableofcontents
\clearpage

 \clearpage
 \pagenumbering{arabic}

 \chapter*{General Introduction}
\addcontentsline{toc}{part}{General Introduction}

This thesis studies three topics in the statistical physics of disordered systems: interface growth, Anderson localization, and random multiplicative growth. Although these problems belong to different physical settings, they share the same basic question: how does disorder affect the large-scale behavior? As will become clear throughout the manuscript, they are also related at a mathematical and conceptual level.

The first part of the thesis, Part~I, is devoted to \textit{Interface growth}. A starting point is Mandelbrot's discussion of the \textit{coastline paradox} \cite{mandelbrot1967coast}. Suppose one wishes to measure the length of a coastline. Using a coarse ruler, one takes into account only the largest bends of the land, effectively straightening every bay or cove smaller than the ruler itself. If the measurement is repeated with a finer ruler, these smaller irregularities become visible, and the measured length increases. With an even finer ruler, even smaller structures are taken in to account, and the length increases once more.

Hence, the relevant information is not the length itself, but rather the way the measured length varies with the resolution. This dependence defines a roughness exponent.

Interface growth provides a dynamical version of this problem. Such interfaces arise in many contexts, including thin-film deposition, erosion, bacterial colonies, and magnetic domain walls. In these systems, the interface is a nonequilibrium boundary whose geometry results from the competition between roughening mechanisms, such as thermal noise or quenched disorder, and smoothing mechanisms, such as elasticity or surface tension.

The evolution of these interfaces is commonly described within the {Family-Vicsek scaling} \cite{family1985scaling} framework. Starting from a flat initial condition, height fluctuations at distant points are initially uncorrelated. As time progresses, smoothing mechanisms transmit information along the interface, so that correlations gradually build up over larger and larger distances. This defines a time-dependent correlation length, which grows until it reaches the system size. The roughening process is therefore characterized not only by a roughness exponent, describing how the amplitude of fluctuations scales with system size, but also by a dynamical exponent, governing the spread of correlations in time. Determining these scaling exponents has been one of the central goals of nonequilibrium statistical physics over the past decades.

A remarkable feature of kinetic roughening is the universality of the resulting scaling laws. Very different microscopic growth mechanisms may display the same large-scale behavior, provided they share a few basic ingredients such as symmetries or conservation laws. This has led to the identification of a small number of universality classes, among which the celebrated Edwards-Wilkinson and Kardar-Parisi-Zhang classes \cite{edwards1982surface,kardar1986dynamic}.

To characterize these scaling laws, one must introduce suitable observables. A natural distinction is between \emph{global} observables, which probe the roughness of the whole interface, and \emph{local} observables, obtained by restricting the measurement to a window much smaller than the full system. For a simple scale invariant interface, one would naively expect these two types of observables to lead to the same roughness exponent. In many systems, however, this is not the case: local and global measurements yield different scaling behaviors. This mismatch is referred to as \emph{anomalous scaling}.

The first part of the thesis then studies two models that illustrate this distinction, and in which \emph{anomalous} scaling has two distinct origins. In Chapter~\ref{sec:interfaces-elastic}, devoted to a heterogeneous elastic line, the apparent anomaly is a \emph{statistical} effect of disorder averaging. In Chapter~\ref{sec:interfaces-spme-main}, devoted to the stochastic porous medium equation, the anomaly is genuine: local and global fluctuations are governed by distinct mechanisms and therefore obey different scaling exponents.

The study of disordered elastic lines leads naturally to the second topic of this thesis: \textit{Anderson localization}. First introduced by Anderson in the context of electrons moving in disordered crystals \cite{anderson1958absence}, it describes how quenched disorder can qualitatively alter the spectral and transport properties of a wave equation. In a homogeneous medium, translational invariance implies that the eigenmodes are extended throughout the system. They take the form of delocalized waves, such as Bloch states in a crystal, and can therefore carry particles, energy, or information over macroscopic distances. Quenched disorder breaks this invariance and induces multiple scattering on the inhomogeneities of the medium. The resulting interference effects can suppress transport and produce eigenstates that are spatially localized, in the sense that their amplitudes are concentrated around a given center and decay exponentially away from it.

In this thesis, Anderson localization appears through the spectral properties of the random spring chain associated with the disordered elastic line in Chapter~\ref{sec:interfaces-elastic}. This is a natural one-dimensional localization problem, where one asks how disorder in the springs and masses modifies both the eigenvalues and the eigenvectors of the chain. Chapter~\ref{sec:interfaces-anderson} establishes these spectral and localization properties for chains with both random masses and random spring constants, with particular emphasis on the strong-disorder regime, where the underlying distributions are broad.

The third topic, developed in Part~II, is \textit{random multiplicative growth}, in which a quantity grows proportionally to its current size. A paradigmatic example is wealth accumulation: if an individual’s capital \(W\) grows at rate \(r\), then \(\Delta W \propto r W\). Because the increment is itself proportional to the current wealth, even small fluctuations in the growth rate are amplified over time, naturally generating broad distributions. This leads in particular to strong inequalities. This mechanism underlies the emergence of extreme heterogeneity in many contexts, including biological populations, city sizes, firm sizes, and wealth distributions.

A central question is then how such multiplicative amplification is modified by exchange or redistribution. In economics, this was put forward in the Bouchaud-M\'ezard model of economy \cite{bouchaud_mezard_2000_wealth}, where multiplicative growth is coupled to transfers between agents. In that setting, redistribution competes with the tendency of multiplicative growth to amplify differences, and the resulting dynamics may lead either to broadly spread wealth or to strong concentration on a few agents. Chapter~\ref{ch:population} begins by showing how this problem is connected to the KPZ equation discussed above, thus linking it back to interface growth models.

Of particular interest, and central to this thesis, is the case where the site-dependent growth rates can be decomposed into two distinct components: a quenched component, corresponding to a persistent advantage or disadvantage, and a temporal component, corresponding to transient fluctuations. The questions are then both to characterize the possible states of the system, whether it is delocalized or localized, and to understand the origin of the localization. When wealth concentrates on a few individuals, is this primarily due to persistent intrinsic advantage, to favorable temporal fluctuations, or to an interplay between the two? And how do these different mechanisms affect the overall growth rate of the system? These questions are studied successively in Chapter~\ref{sec:population-mf-growth}, which treats quenched heterogeneity without temporal noise, Chapter~\ref{sec:population-mf-growth-noise}, which adds multiplicative temporal noise, and Chapter~\ref{sec:population-resetting}, which replaces that noise by stochastic resetting.

\medskip
\noindent\textbf{Outline of the thesis.} Chapter~\ref{chap:introduction} introduces fluctuating interfaces, kinetic roughening, and anomalous scaling. Chapter~\ref{sec:interfaces-elastic} studies heterogeneous elastic lines and is based on \Pone. Chapter~\ref{sec:interfaces-spme-main} is devoted to the stochastic porous medium equation and is based on \Ptwo. Chapter~\ref{sec:interfaces-anderson} develops the corresponding one-dimensional localization problem in random chains and is based on \Pthree. Chapter~\ref{ch:population} introduces random multiplicative growth models and their connection with KPZ. Chapter~\ref{sec:population-mf-growth} treats the deterministic heterogeneous case, studied in \Pfour. Chapter~\ref{sec:population-mf-growth-noise} analyzes the effect of multiplicative temporal noise and is based in part on \Pfour. Chapter~\ref{sec:population-resetting} studies multiplicative growth with resetting.

\clearpage

 \chapter*{Résumé en français}
\addcontentsline{toc}{part}{Résumé en français}

Cette thèse étudie plusieurs problèmes de physique statistique des systèmes désordonnés : les interfaces rugueuses, la localisation d'Anderson dans des chaînes aléatoires, et la croissance multiplicative aléatoire. Ces trois sujets correspondent à des situations physiques différentes, mais ils posent une même question générale : comment le désordre modifie-t-il le comportement macroscopique d'un système ?

Dans les modèles étudiés ici, le désordre prend plusieurs formes : constantes de ressort ou masses aléatoires, diffusivité dépendant de la hauteur, ou taux de croissance hétérogènes. L'objectif est d'étudier comment ces hétérogénéités modifient les observables pertinentes de chaque problème : par exemple, les exposants de rugosité pour les interfaces, la densité d'états et la localisation des modes pour les chaînes aléatoires, et la répartition de la population dans les modèles de croissance.

\section*{Interfaces rugueuses et scaling anomal}

La première partie de la thèse est consacrée à la croissance d'interfaces. Une interface est décrite par un champ de hauteur \(h(x,t)\), et sa rugosité par les fluctuations de cette hauteur. Dans le cadre du scaling de Family-Vicsek, la rugosité est caractérisée par un exposant de rugosité, qui décrit la dépendance en taille du système, et par un exposant dynamique, qui décrit la propagation des corrélations dans le temps.

Une difficulté vient du fait que la rugosité peut être mesurée par des observables différentes. Les observables globales décrivent les fluctuations de l'interface entière, tandis que les observables locales mesurent les différences de hauteur sur une distance finie. Dans les cas les plus simples, ces deux descriptions donnent le même exposant de rugosité. Lorsque ce n'est pas le cas, on parle de scaling anomal.

\subsection*{Lignes élastiques hétérogènes}

Le premier modèle étudié est une ligne élastique hétérogène. Il s'agit d'une chaîne unidimensionnelle dont les hauteurs voisines sont reliées par des ressorts de raideurs aléatoires \(k_i\). La dynamique discrète est
\begin{equation*}
\partial_t h_x(t)
=
-k_x\bigl[h_x(t)-h_{x-1}(t)\bigr]
-k_{x+1}\bigl[h_x(t)-h_{x+1}(t)\bigr]
+\eta_x(t),
\end{equation*}
où \(\eta_x(t)\) est un bruit blanc gaussien. Le bruit est donc additif, alors que les constantes \(k_i\) forment un désordre gelé.

Deux observables sont utilisées. La première est une largeur globale \(D(L,t)\), qui mesure le déplacement quadratique moyen d'une interface de taille \(L\). La seconde est une corrélation locale \(G(x,t)\), qui mesure les fluctuations entre deux points séparés par une distance \(x\).

Lorsque la distribution des raideurs n'est pas trop large, le modèle retrouve le scaling standard d'Edwards-Wilkinson. En revanche, si la distribution donne un poids important aux très petites raideurs, des liens très faibles apparaissent. Ces liens se comportent presque comme des coupures de la chaîne : ils produisent de grands sauts de hauteur et dominent les moyennes sur le désordre.

Le scaling anomal obtenu dans ce modèle a donc une origine statistique. Il ne traduit pas nécessairement une différence entre le comportement local et global d'une interface typique. Il vient plutôt du fait que les observables moyennées sur le désordre sont dominées par des réalisations rares contenant des ressorts exceptionnellement faibles. Il faut donc distinguer le comportement typique du comportement moyen.

\subsection*{Équation stochastique des milieux poreux}

Le second modèle d'interface est l'équation stochastique des milieux poreux. Dans ce cas, la diffusivité locale dépend de la hauteur elle-même :
\begin{equation*}
\partial_t h(x,t)
=
\partial_x\!\left[D(h(x,t))\,\partial_x h(x,t)\right]
+\eta(x,t),
\end{equation*}
où \(D(h)\) est une diffusivité dépendant de la hauteur et \(\eta(x,t)\) un bruit additif. La dynamique est non linéaire : l'interface crée sa propre hétérogénéité effective.

Dans ce modèle, le scaling anomal a une origine différente. L'état stationnaire peut être décrit par une marche aléatoire effective dont la variance des incréments dépend de la hauteur locale. Cette description permet de comprendre pourquoi les fluctuations locales et globales ne sont pas contrôlées par le même mécanisme.

L'anomalie est donc intrinsèque. Elle ne provient pas seulement d'une moyenne dominée par des événements rares, mais d'une différence réelle entre les observables locales et globales. Cette analyse est complétée par une approche de groupe de renormalisation fondée sur le flot fonctionnel de la diffusivité \(D(h)\), nécessaire pour identifier les comportements asymptotiques sélectionnés par les grandes hauteurs.

\section*{Localisation d'Anderson dans des chaînes aléatoires}

La ligne élastique hétérogène conduit naturellement à un problème spectral. La même chaîne de ressorts aléatoires définit un système mécanique désordonné, dont les modes normaux sont obtenus à partir d'une équation aux valeurs propres. Cette observation motive ce chapitre consacrée à la localisation d'Anderson en dimension un.

Dans un milieu homogène, les modes propres sont étendus et peuvent transporter de l'énergie à grande distance. Dans un milieu désordonné, les diffusions multiples et les interférences peuvent au contraire produire des modes localisés, dont l'amplitude est concentrée autour d'une région finie et décroît loin de celle-ci.

Le modèle étudié est une chaîne unidimensionnelle avec masses aléatoires \(m_x\) et constantes de ressort aléatoires \(k_x\). Pour un mode propre de fréquence \(\omega\), le problème spectral s'écrit
\begin{equation*}
-\omega^2 m_x u_x
=
k_x(u_{x-1}-u_x)
+k_{x+1}(u_{x+1}-u_x).
\end{equation*}
Les deux quantités principales sont la densité d'états et la longueur de localisation. La première compte le nombre de modes à une fréquence donnée, tandis que la seconde mesure l'extension spatiale des modes propres.

L'analyse porte principalement sur le régime de basse fréquence et sur les distributions larges de masses et de raideurs. Lorsque les distributions possèdent des moments finies, on retrouve les résultats perturbatifs usuels. Lorsque le désordre devient plus large, de nouvelles lois de puissance apparaissent. Les masses très grandes et les ressorts très faibles contrôlent alors le comportement spectral à basse fréquence.

Cette partie précise aussi le lien avec les interfaces. Les liens faibles qui dominent certaines moyennes de rugosité sont les mêmes objets qui contrôlent les modes de basse fréquence de la chaîne associée.

\section*{Croissance multiplicative aléatoire}

La seconde partie de la thèse porte sur la croissance multiplicative aléatoire. Dans un tel modèle, l'accroissement d'une quantité est proportionnel à sa valeur courante. C'est une description naturelle pour des populations, des tailles de villes ou d'entreprises, ou encore des richesses, où les variations sont souvent relatives plutôt qu'absolues.

Le modèle principal est un modèle de croissance en champ moyen avec redistribution :
\begin{equation*}
\frac{d x_i}{dt}
=
\bigl(m_i(t)-\varphi\bigr)x_i
+
\varphi \overline{x},
\qquad
\overline{x}=\frac1N\sum_{j=1}^N x_j .
\end{equation*}
La variable \(x_i(t)\) représente la population ou la richesse du site \(i\), \(m_i(t)\) son taux de croissance, et \(\varphi\) le taux de redistribution. La redistribution tend à homogénéiser les fractions de population, tandis que l'hétérogénéité des taux de croissance favorise la concentration sur les sites les plus favorables.

Deux observables jouent un rôle central. La première est le taux de croissance asymptotique
\begin{equation*}
\gamma
=
\lim_{t\to\infty}
\frac1t \log \overline{x}(t),
\end{equation*}
qui mesure la croissance globale du système. La seconde est la famille des fractions
\begin{equation*}
p_i(t)=\frac{x_i(t)}{\sum_j x_j(t)} ,
\end{equation*}
qui décrit la répartition de la population totale. Une phase est dite délocalisée lorsque toutes les fractions s'annulent lorsque \(N\to\infty\). Elle est localisée lorsqu'un site porte une fraction finie. Elle est partiellement localisée lorsque plusieurs sites favorables peuvent porter des fractions importantes, avec des identités qui changent au cours du temps.

\subsection*{Hétérogénéité gelée sans bruit temporel}

Le premier cas est déterministe en temps : \(m_i(t)=m_i\), où les \(m_i\) sont des variables aléatoires gelées. Le modèle devient
\begin{equation*}
\frac{d x_i}{dt}
=
(m_i-\varphi)x_i
+
\varphi\overline{x}.
\end{equation*}
Il décrit la compétition entre redistribution et avantages persistants.

Pour des distributions à support compact, la structure de phase dépend du comportement de la distribution près de son bord supérieur. Si les sites presque optimaux sont suffisamment nombreux, le système peut rester délocalisé. S'ils sont trop rares, le système se localise sur le meilleur site lorsque la redistribution est faible.

Un cas particulièrement important est celui des distributions gaussiennes généralisées, dont la variance dépend de \(N\) afin que le maximum reste fini. Dans le cas gaussien,
\begin{equation*}
m_i\sim \mathcal N(0,\Sigma_N^2),
\qquad
\Sigma_N=\frac{\Sigma_0}{\sqrt{2\log N}},
\end{equation*}
de sorte que \(\max_i m_i\to \Sigma_0\). On obtient alors une transition au point
\begin{equation*}
\varphi_c=\Sigma_0 .
\end{equation*}
Pour \(\varphi>\Sigma_0\), le système est délocalisé et \(\gamma\to0\) lorsque \(N\to\infty\). Pour \(\varphi<\Sigma_0\), le système se localise sur le site le plus favorable et
\begin{equation*}
\gamma=\Sigma_0-\varphi .
\end{equation*}
La fraction portée par ce site est finie, avec \(p_{\max}\simeq 1-\varphi/\Sigma_0\). Dans ce régime, la localisation est donc due à un avantage persistant.

\subsection*{Bruit multiplicatif temporel}

Le second cas ajoute un bruit multiplicatif temporel,
\begin{equation*}
m_i(t)=m_i+\sigma \eta_i(t),
\end{equation*}
ce qui donne
\begin{equation*}
\frac{d x_i}{dt}
=
(m_i-\varphi)x_i
+
\varphi\overline{x}
+
\sigma x_i\eta_i(t).
\end{equation*}
Les \(\eta_i(t)\) sont des bruits blancs gaussiens indépendants. Le taux de croissance contient alors une partie gelée \(m_i\), qui représente un avantage persistant, et une partie fluctuante, qui représente des variations temporelles.

Dans le cas gaussien, le diagramme de phase contient trois phases : délocalisée, localisée et partiellement localisée. Dans la phase délocalisée, toutes les fractions sont d'ordre \(1/N\), et \(\gamma=0\) dans la limite \(N\to\infty\). Dans la phase localisée, le système est concentré à un instant donné sur l'un des meilleurs sites, avec \(m_i\simeq\Sigma_0\). Le taux de croissance est alors
\begin{equation*}
\gamma=\Sigma_0-\varphi-\frac{\sigma^2}{2},
\end{equation*}
où le terme \(-\sigma^2/2\) est la correction d'Ito associée au bruit multiplicatif.

La phase nouvelle est la phase partiellement localisée. Dans cette phase, la population n'est pas concentrée de façon permanente sur un unique site optimal. Plusieurs sites favorables, mais pas nécessairement optimaux, peuvent porter des fractions importantes. Un site dont le taux de croissance gelé est sous-optimal peut devenir dominant s'il bénéficie d'une fluctuation favorable pendant un temps suffisamment long. Cette phase est donc contrôlée par l'interaction entre hétérogénéité persistante et fluctuations temporelles.

L'analyse auto-moyennante décrit correctement la phase délocalisée et les lignes de transition qui l'impliquent, mais elle n'est pas suffisante dans les régions localisée et partiellement localisée. La frontière entre ces deux phases est en particulier déplacée par les fluctuations du site localisé. La thèse développe alors une description effective en termes de la fraction maximale \(p_{\max}\), qui permet d'obtenir un diagramme de phase corrigé.

\subsection*{Réinitialisation stochastique}

Le dernier modèle remplace le bruit multiplicatif continu par une réinitialisation stochastique. Entre deux réinitialisations, les sites suivent la dynamique déterministe
\begin{equation*}
\frac{d x_i}{dt}
=
(m_i-\varphi)x_i
+
\varphi\overline{x}.
\end{equation*}
Chaque site est en plus réinitialisé à une valeur \(x_0\) à des temps de Poisson de taux \(r\),
\begin{equation*}
x_i(t^+)=x_0.
\end{equation*}
Dans un contexte de populations, cela peut représenter une extinction locale suivie d'une recolonisation. Dans un contexte économique, cela peut représenter une faillite suivie d'un redémarrage.

La réinitialisation modifie le mécanisme de localisation. Un site ne peut devenir dominant que s'il évite d'être réinitialisé pendant un temps assez long pour passer d'une fraction petite à une fraction d'ordre un. Ce temps est d'ordre \(\log N\). Lorsque les réinitialisations ont lieu à un taux d'ordre un, la localisation complète sur un nombre fini de sites est donc supprimée.

Une phase partiellement localisée peut cependant subsister. Un ensemble sous-extensif de sites favorables contribue alors à la population totale, et certains d'entre eux peuvent porter de grandes fractions à un instant donné. Ce résultat montre que la localisation partielle ne dépend pas du détail précis du bruit temporel : elle apparaît dès que l'hétérogénéité persistante est combinée à un mécanisme qui empêche une condensation permanente sur un seul site.

\section*{Conclusion}

Les trois parties de la thèse illustrent des effets distincts du désordre. Dans les interfaces rugueuses, il peut modifier les lois d'échelle observées et rendre les moyennes dominées par des événements rares. Dans les chaînes aléatoires, il localise les modes propres et change les propriétés spectrales à basse fréquence. Dans les modèles de croissance multiplicative, il produit des phases localisées ou partiellement localisées, selon la façon dont les avantages persistants se combinent aux fluctuations temporelles.

Un point commun est l'importance du choix des observables. Les observables locales et globales d'une interface ne mesurent pas toujours le même mécanisme. Dans une chaîne désordonnée, la densité d'états et la longueur de localisation donnent des informations complémentaires. Dans les modèles de croissance, le taux global \(\gamma\) et les fractions \(p_i\) décrivent deux aspects distincts du problème.

 \part{Rough interfaces and anomalous scaling}
 \chapter{Generalities on Interfaces}
\label{chap:introduction}

An interface is typically described by a height field \(h(\mathbf{x},t)\) defined over a \(d\)-dimensional substrate, although in this thesis we restrict attention to the case \(d=1\). Its fluctuations are commonly quantified by the global width \(W(L,t)\), defined as the standard deviation of the height field in a system of linear size \(L\):
\begin{equation}
  W^2(L,t)=
  \left\langle
  \frac{1}{L}\int_0^{L}dy\,
  [h(y,t)-\bar h(t)]^2
  \right\rangle,
  \label{eq:intro-globalwidth}
\end{equation}
where
\begin{equation}
\bar h(t)=\frac{1}{L}\int_0^{L}dy\,h(y,t)
\end{equation}
denotes the spatial average of the height.

The evolution of this width is captured by the \textbf{Family--Vicsek scaling} picture, which distinguishes two asymptotic regimes. At early times, before correlations have extended across the whole system, the width grows algebraically in time:
\begin{equation}
    W(L,t) \sim t^{\beta},
\end{equation}
where \(\beta\) is the growth exponent. As the correlation length,
\begin{equation}
\xi(t)\sim t^{1/z},
\end{equation}
becomes comparable to the system size \(L\), the growth saturates and the width reaches a stationary value
\begin{equation}
    W(L) \sim L^{\alpha},
\end{equation}
where \(\alpha\) is the roughness exponent. The crossover between these two regimes occurs at a characteristic time \(t_{\times}(L)\sim L^{z}\), which defines the dynamical exponent \(z\). Equivalently, the exponents satisfy the scaling relation
\begin{equation}
\beta=\frac{\alpha}{z}.
\end{equation}

These regimes are unified by the Family-Vicsek Ansatz,
\begin{equation}
    W(L,t) \sim L^{\alpha} f\left( \frac{t}{L^{z}} \right),
\end{equation}
where the scaling function \(f(u)\) behaves as \(u^{\beta}\) for \(u\ll 1\) and approaches a constant for \(u\gg 1\). In this sense, the interface is \textit{self-affine}: its large-scale statistical properties are invariant under the rescaling
\begin{equation}
    \mathbf{x} \mapsto b\mathbf{x}, \qquad
    t \mapsto b^{z}t, \qquad
    h \mapsto b^{\alpha}h.
\end{equation}

The power of this scaling approach lies in its \textit{universality}: despite large differences in microscopic dynamics, many systems share the same large-scale behavior, determined only by broad features such as symmetries and conservation laws. %

Besides the global width, it is often useful to probe the interface on smaller spatial scales. This leads to the definition of the \textbf{local width}, measured in a window of size \(\ell \ll L\):
\begin{equation}
  w^2(\ell,t)=
  \left\langle
  \frac{1}{\ell}\int_x^{x+\ell}dy\,
  [h(y,t)-\bar h_{[x,x+\ell]}(t)]^2
  \right\rangle,
  \label{eq:intro-localwidth}
\end{equation}
where \(\bar h_{[x,x+\ell]}(t)\) is the average height over the interval \([x,x+\ell]\), and the brackets include an average over the position of the window.

A complementary characterization is obtained in Fourier space. Denoting by \(\hat h(k,t)\) the Fourier transform of \(h(x,t)\), one defines the \textbf{structure factor} as
\begin{equation}
  S(k,t)=\langle \hat h(k,t)\hat h(-k,t)\rangle.
  \label{eq:intro-S}
\end{equation}

For a self-affine interface, these different observables are expected to be governed by the same scaling exponents, and therefore to provide a consistent characterization of the same universality class. This is not always the case, as we explain below.

Since stationary properties will be discussed repeatedly throughout this thesis, we will often omit the explicit time dependence from the notation. Accordingly, we write
\begin{equation}
W(L)\equiv \lim_{t\to\infty} W(L,t),\qquad
w(\ell)\equiv \lim_{t\to\infty} w(\ell,t),\qquad
S(k)\equiv \lim_{t\to\infty} S(k,t).
\end{equation}

\section{Standard roughening and the Edwards-Wilkinson equation}
\label{sec:intro-ew}

The purpose of this section is to establish the standard picture of interface scaling. The model considered here was introduced by Edwards and Wilkinson in their study of the rough surface formed by a growing aggregate of particles. In their model, particles are deposited randomly onto an interface and are then allowed to relax locally before coming to rest. More precisely, one considers integer heights \(h_i(t)\) defined on lattice sites \(i\). At each deposition event, a site is chosen at random. If it is a local minimum, the particle remains there, whereas otherwise it relaxes to the lowest neighboring site. At large length and time scales, this discrete dynamics admits a coarse-grained continuum description in which the local relaxation gives rise to a diffusive smoothing term, while the randomness of the deposition process becomes a stochastic noise term. In one spatial dimension, this yields the Edwards-Wilkinson (EW) equation
\begin{equation}
  \partial_t h(x,t)=\nu\,\partial_x^2 h(x,t)+\eta(x,t),
  \label{eq:intro-ew}
\end{equation}
where \(\nu>0\) measures the strength of local relaxation and \(\eta(x,t)\) is a Gaussian white noise representing the fluctuations generated by random deposition:
\begin{equation}
  \langle \eta(x,t)\rangle =0,
  \qquad
  \langle \eta(x,t)\eta(x',t')\rangle
  =2D\,\delta(x-x')\delta(t-t').
  \label{eq:intro-ew-noise}
\end{equation}

Although it was introduced in the context of the statistics of granular aggregates \cite{EdwardsWilkinson1982}, its application is much broader, as it defines one of the universality classes of nonequilibrium interface growth \cite{KrugSpohn1991,FamilyVicsek1991,BarabasiStanley1995,HalpinHealyZhang1995,Krug1997}. 

\begin{figure}[!ht]
\centering
\includegraphics[width=0.99\textwidth]{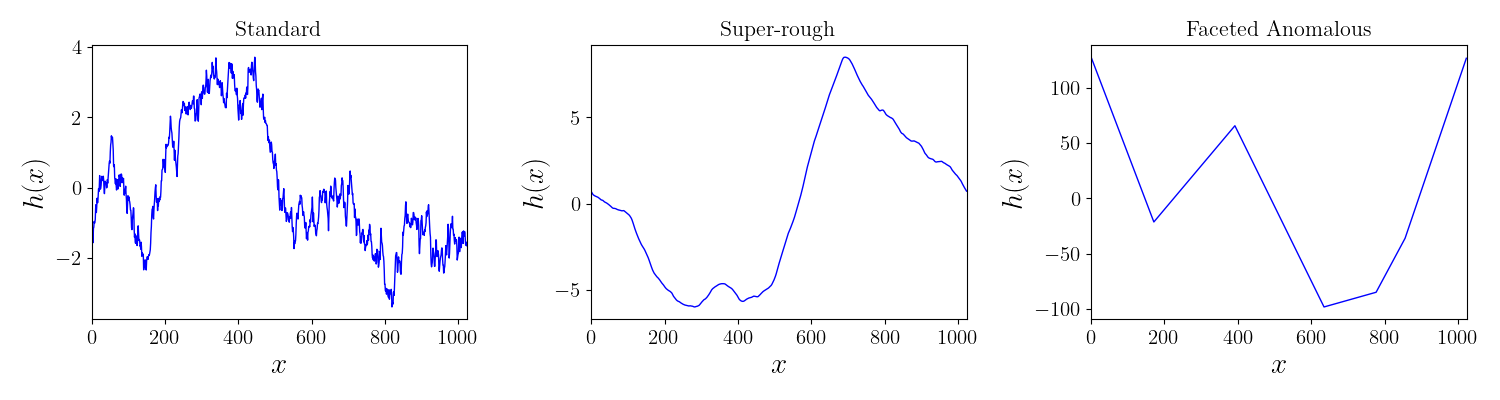}
\caption{Profiles of stationary interfaces: an Edwards-Wilkinson interface (standard scaling), a molecular-beam-epitaxy super-rough interface, and an anomalous faceted interface.}
\label{fig:intro-interface-taxonomy}
\end{figure}

While the noise term \(\eta(x,t)\) tends to generate independent fluctuations at each point of the interface, the Laplacian term smooths the profile and thus builds spatial correlations. This immediately suggests the existence of a time-dependent correlation length \(\xi(t)\). On scales smaller than \(\xi(t)\), fluctuations have already become correlated through relaxation. 0n larger scales, correlations have not yet spread and the interface is still out of equilibrium.
An illustration of a stationary EW interface profile is provided in Fig.~\ref{fig:intro-interface-taxonomy} (left).

For the EW equation, this behavior is most easily obtained in Fourier space. 
The Fourier modes evolve independently according to
\begin{equation}
  \partial_t \hat h(k,t)=-\nu k^2 \hat h(k,t)+\hat\eta(k,t),
\end{equation}
.%
For a flat initial condition, one finds
\begin{equation}
  S(k,t)=\frac{D}{\nu k^2}\left(1-e^{-2\nu k^2 t}\right).
  \label{eq:intro-ew-sk}
\end{equation}
This expression shows directly that the crossover between equilibrated and nonequilibrated modes occurs at \(k\sim t^{-1/2}\), hence
\begin{equation}
  \xi(t)\sim t^{1/2},
  \qquad
  z=2,
  \label{eq:intro-ew-xi}
\end{equation}
as long as \(\xi(t)<L\), at which point the system is equilibrated.
More generally, the structure factor obeys the scaling form
\begin{equation}
  S(k,t)\sim
  \begin{cases}
    t^{(2\alpha+1)/z}, & k\ll \xi(t)^{-1},\\[1mm]
    k^{-(2\alpha+1)}, & k\gg \xi(t)^{-1},
  \end{cases}
  \label{eq:intro-FV-S}
\end{equation}
where the first regime corresponds to modes that have not yet relaxed, while the second one describes the stationary self-affine fluctuations at scales smaller than the correlation length. In the EW case, Eq.~\eqref{eq:intro-ew-sk} yields \(z=2\) and  \(\alpha=1/2\).

The global width follows from the structure factor through
\begin{equation}
  W^2(L,t)\sim \int_{2\pi/L}^{\Lambda}\frac{dk}{2\pi}\,S(k,t),
  \label{eq:intro-WSk}
\end{equation}
with \(\Lambda\) a microscopic ultraviolet cutoff. Using Eq.~\eqref{eq:intro-FV-S}, one obtains the standard Family-Vicsek form
\begin{equation}
  W(L,t)\sim
  \begin{cases}
    t^\beta, & t\ll L^z,\\[1mm]
    L^\alpha, & L^z \ll t,
  \end{cases}
  \qquad
  \beta=\frac{\alpha}{z}.
  \label{eq:intro-FV-W}
\end{equation}
For the EW equation in \(1+1\) dimensions, this gives
\begin{equation}
\alpha=\frac{1}{2},\qquad \beta=\frac{1}{4}, \qquad z=\frac{\alpha}{\beta}=2
\end{equation}

Likewise, the local width satisfies $w^2(\ell,t)=2\int_{2\pi/L}^{\Lambda}\frac{dk}{2\pi}\,
  \left[1-\cos(k\ell)\right] S(k,t)$, hence
\begin{equation}
  w(\ell,t)\sim
  \begin{cases}
    \xi(t)^\alpha\sim t^\beta, & t \ll \ell^z, \\[1mm]
    \ell^\alpha, & \ell^z \ll t,
  \end{cases}
  \label{eq:intro-FV-w}
\end{equation}
so that local, global, and spectral observables are all governed by the same scaling exponents, as expected for an ordinary self-affine interface.%

\section{Beyond standard scaling}
\label{sec:intro-exponents}

We now turn to situations in which the standard Family-Vicsek picture is no longer satisfied. 
In the ordinary self-affine case, the global width, the local width, and the structure factor are all governed by the same roughness exponent. 
However, this need not remain true in general. 

We first consider the stationary regime and distinguish the roughness exponents extracted from the three main observables. 
The \emph{global roughness exponent} \(\alpha\) is defined through the width,
\begin{equation}
  W(L)\sim L^\alpha.
\end{equation}
The \emph{local roughness exponent} \(\alpha_{\rm loc}\) is defined from the local width in a window of size \(\ell\),
\begin{equation}
  w(\ell)\sim \ell^{\alpha_{\rm loc}}L^{\alpha-\alpha_{\rm loc}},
  \qquad 1\ll \ell\ll L,
\end{equation}
where homogeneity requires an additional system-size dependence, which allows one to recover the global scaling as \(\ell \to L\).
The \emph{spectral roughness exponent} \(\alpha_s\) is defined from the stationary structure factor,
\begin{equation}
  S(k)\sim k^{-(2\alpha_s+1)}L^{2(\alpha-\alpha_s)}.
\end{equation}
In the standard case, one has
\begin{equation}
\alpha=\alpha_{\rm loc}=\alpha_s,
\end{equation}
and the explicit system-size dependence disappears from both \(w(\ell)\) and \(S(k)\). 

Different scenarios have been reported in the literature:
\begin{align}
  &\text{standard:}
  && \alpha_s=\alpha<1,
  \qquad \alpha_{\rm loc}=\alpha,
  \nonumber\\
  &\text{anomalous:}
  && \alpha_s=\alpha_{\rm loc}<1,
  \qquad \alpha>\alpha_{\rm loc},
  \nonumber\\
  &\text{super-rough:}
  && \alpha_s=\alpha>1,
  \qquad \alpha_{\rm loc}=1,
  \nonumber\\
  &\text{faceted anomalous:}
  && \alpha_s\neq \alpha,
  \qquad \alpha_{\rm loc}=1,
  \label{eq:intro-taxonomy}
\end{align}
and will be discussed in more detail in the following sections.

The evolution of the global width remains unchanged:
\begin{equation}
  W(L,t)\sim
  \begin{cases}
    t^\beta, & t\ll L^z,\\[1mm]
    L^\alpha, & t\gg L^z,
  \end{cases}
  \qquad
  \beta=\frac{\alpha}{z},
  \label{eq:intro-summary-W}
\end{equation}
suggesting that the correlation length still scales as \(\xi(t)\sim t^{1/z}\). Since the local width and the structure factor have different exponents, continuity requires a new intermediate time-dependent regime.
One gets
\begin{equation}
  S(k,t)\sim
  \begin{cases}
    t^{(2\alpha+1)/z}, & k\ll \xi(t)^{-1},\\[1mm]
    k^{-(2\alpha_s+1)}\,t^{2(\alpha-\alpha_s)/z}, & 1/L \gg k\gg \xi(t)^{-1}, \\
     k^{-(2\alpha_s+1)}L^{2(\alpha-\alpha_s)} &  \xi(t)=L.
  \end{cases}
  \label{eq:intro-generic-S}
\end{equation}
and for the local width:
\begin{equation}
  w(\ell,t)\sim
  \begin{cases}
    t^\beta, & t\ll \ell^z,\\[1mm]
    \ell^{\alpha_{\rm loc}}\,t^{(\alpha-\alpha_{\rm loc})/z}, & \ell^z\ll t\ll L^z,\\[1mm]
    \ell^{\alpha_{\rm loc}}\,L^{\alpha-\alpha_{\rm loc}}, & t\gg L^z.
  \end{cases}
  \label{eq:intro-anomalous-w}
\end{equation}
where $(\alpha-\alpha_{\rm loc})/z$ is a new intermediate growth exponent.

The next sections discuss these different nonstandard cases one by one, beginning with super-roughness, then the faceted case, and finally anomalous scaling.

\section{Super-roughness}
\label{sec:intro-superrough}

The purpose of this section is to discuss the best understood case of a local-global exponent mismatch. 
Super-roughness is defined by the condition
\begin{equation}
  \alpha=\alpha_s>1,
  \qquad
  \alpha_{\rm loc}=1.
\end{equation}
For short distances, the height difference \(h(x+\ell,t)-h(x,t)\) cannot grow faster than linearly with the distance \(\ell\). Consequently, the local roughness exponent cannot be greater than \(1\) and is forced to saturate at \(1\), even though the global width grows faster than linearly with system size. The amplitude of local fluctuations then depends explicitly on the largest scale available in the problem \cite{LopezRodriguezCuerno1997,Ramasco2000,LopezCastroGallego2005}.

A standard example is provided by idealized models of molecular-beam epitaxy. In molecular-beam epitaxy, atoms are deposited onto a crystalline substrate and may diffuse on the surface before being incorporated into the growing film. At the microscopic level, the interface is represented by discrete heights \(h_i(t)\) on a lattice, and particles are deposited at random sites. Each particle relaxes within a local neighborhood toward a position where it has more lateral bonds to the crystal, and is therefore energetically more stable. This local relaxation tends to smooth the surface by reducing sharp height variations. In a coarse-grained description, the corresponding smoothing mechanism is governed by a fourth-order derivative, which penalizes curvature variations. This leads to the linear molecular-beam epitaxy (MBE) equation, also known as the Mullins-Herring (MH) equation:
\begin{equation}
  \partial_t h(x,t)=-\kappa\,\partial_x^4 h(x,t)+\eta(x,t),
  \label{eq:intro-mh}
\end{equation}
with Gaussian white noise
\begin{equation}
  \langle \eta(x,t)\eta(x',t')\rangle
  =2D\,\delta(x-x')\delta(t-t').
\end{equation}

An illustration of a stationary linear MBE interface profile is provided in Fig.~\ref{fig:intro-interface-taxonomy} (middle). The profile is visibly much smoother than in the EW case.

In Fourier space, the MBE equation reads
\begin{equation}
  \partial_t \hat h(k,t)=-\kappa k^4 \hat h(k,t)+\hat\eta(k,t),
\end{equation}
so it differs from the EW equation only by the replacement \(k^2\to k^4\). The same analysis therefore applies, yielding the exponents
\begin{equation}
\alpha=\frac32,\qquad \beta=\frac38,\qquad z=4,
\end{equation}
valid for both the structure factor and the global width.
The important difference appears for the local width. In the stationary regime, the structure factor behaves as \(S(k)\sim k^{-4}\), and the stationary local width is obtained from
\begin{equation}
  w^2(\ell)\sim \int_{1/L}^{\Lambda} dk\,\bigl[1-\cos(k\ell)\bigr]\,k^{-4}.
  \label{eq:eqwsuperroug}
\end{equation}
For \(1\ll \ell\ll L\), the integral is dominated by the infrared cutoff \(k \sim 1/L \ll 1/\ell\) , where
\begin{equation}
1-\cos(k\ell)\simeq \frac{(k\ell)^2}{2},
\end{equation}
which gives
\begin{equation}
w^2(\ell)\sim \ell^2\int_{1/L}^{1/\ell} dk\,k^{-2}\sim \ell^2 L.
\end{equation}
Hence
\begin{equation}
  w(\ell)\sim \ell\,L^{1/2}.
\end{equation}
The dependence on \(\ell\) is therefore linear, so that
\begin{equation}
\alpha_{\rm loc}=1,
\end{equation}
while the extra factor \(L^{1/2}\) shows that the amplitude of local fluctuations is set by the largest scale in the system. The global exponent satisfies \(\alpha>1\), but the local exponent saturates at its maximal geometric value, \(\alpha_{\rm loc}=1\).

More generally, if the stationary structure factor scales as \(S(k)\sim k^{-(2\alpha+1)}\) with \(\alpha>1\), the same argument gives
\begin{equation}
  w(\ell)\sim \ell\,L^{\alpha-1}.
\end{equation}
Super-roughness is therefore not specific to the MBE equation, it is the generic consequence of a global roughness exponent larger than one \cite{LopezRodriguezCuerno1997,Ramasco2000}.

Beyond the linear MBE equation, super-rough behavior has also been reported in discrete limited-mobility molecular-beam epitaxy models such as Wolf-Villain and Das Sarma-Tamborenea \cite{WolfVillain1990,DasSarmaTamborenea1991,TamboreneaDasSarma1993}. 
In these systems, however, the numerical extraction of asymptotic exponents can be made difficult by long crossover regimes and significant finite-size corrections, as emphasized in several simulation studies \cite{Schroeder1993,SiegertPlischke1994,PunyinduDasSarma1998,DasSarmaEtAl2002}. 
Super-roughness also arises in driven disordered elastic systems at depinning, for example in the one-dimensional short-range quenched Edwards-Wilkinson universality class\cite{RossoKrauth2001,DuemmerKrauth2005,FerreroEtAl2013}. 
Experimentally, super-rough scaling has been reported in various systems, such as tumor-growth interfaces \cite{BruEtAl1998}, reactive-sputtered AlN films \cite{AugerEtAl2005}, CdTe films grown at sufficiently high substrate temperature \cite{MataEtAl2008}, and transient regimes of electrodeposited Ni/NiW films \cite{OrrilloEtAl2017}.

\section{Faceted anomalous scaling}
\label{sec:intro-faceted}

The purpose of this section is to discuss interfaces made of facets, as illustrated in Fig.~\ref{fig:intro-interface-taxonomy} (right).
Faceted anomalous scaling is characterized in terms of exponents by
\begin{equation}
  \alpha_{\rm loc}=1,
  \qquad
  \alpha_s\neq \alpha.
  \label{eq:intro-faceted-def}
\end{equation}

A faceted profile is made of long segments of nearly constant slope, separated by isolated points where the slope changes abruptly. At distances smaller than the typical facet length, the interface is therefore essentially linear, and one necessarily has
\begin{equation}
  w(\ell)\sim \ell,
  \qquad
  \alpha_{\rm loc}=1.
\end{equation}
Thus, as in the super-rough case, the local roughness exponent saturates at its maximal geometric value, $\alpha_{\rm loc}=1$. The difference lies in the structure factor: in a faceted interface, the dominant contribution comes from the slope changes, and this produces a spectral exponent \(\alpha_s\) that differs from the global one.

To make this mechanism explicit, it is useful to consider an idealized piecewise-linear stationary profile.

Consider a stationary interface profile \(h(x)\) of a system of size \(L\), assumed to be piecewise linear with a finite number \(N\) of slope flips at positions \(X_n\), \(n=1,\dots,N\). On each interval between two successive flips, the slope is constant. If the slope jump at \(X_n\) is \(a_n\), then the second derivative of the height is concentrated at the flip positions:
\begin{equation}
  q(x) = \partial_x^2 h(x)=\sum_{n=1}^{N} a_n\,\delta(x-X_n).
  \label{eq:intro-qx-faceted}
\end{equation}

Let \(\hat h(k)\) and \(\hat q(k)\) denote the Fourier transforms of \(h(x)\) and \(q(x)\). 
For \(k\neq 0\),
\begin{equation}
  \hat q(k)=-k^2 \hat h(k),
\end{equation}

The structure factor is therefore given by
\begin{equation}
  S(k)=\left\langle |\hat h(k)|^2 \right\rangle
  =\frac{1}{k^4}\,\left\langle |\hat q(k)|^2 \right\rangle.
  \label{eq:intro-Sk-from-q}
\end{equation}

Consider \(N\) independent uniformly distributed flips. Then the spectrum \(\hat q(k)\) is flat and, in particular,
\begin{equation}
  \left\langle |\hat q(k)|^2 \right\rangle \sim \frac{N}{L},
\end{equation}
see Appendix~\ref{app:faceted-q-spectrum}. In the case relevant here, \(N=O(1)\), and
\begin{equation}
  S(k)\sim \frac{N}{L}\,\frac{1}{k^4}
  \label{eq:intro-kminus4}
\end{equation}
which, using \(S(k)\sim k^{-(2\alpha_s+1)}\), gives
\begin{equation}
\alpha_s=\frac32.
\end{equation}
Moreover, since there are \(O(1)\) flips, \(\alpha_{\rm loc}=\alpha=1\). Hence,
\begin{equation}
\alpha_{\rm loc}=\alpha=1,
\qquad
\alpha_s=\frac32.
\end{equation}

As in the super-rough case, the local exponent saturates at \(\alpha_{\rm loc}=1\), but for faceted interfaces the spectral behavior is controlled by slope discontinuities.

Faceted anomalous scaling has been reported in several distinct settings. Purrello \emph{et al.} found this behavior in the anharmonic Larkin model \cite{Purrello2019}. Related faceted scaling was also identified by Al\'es and L\'opez in KPZ growth driven by long-range temporally correlated noise, as well as in anharmonic elastic-interface models in correlated random media \cite{Ales2019,Ales2021}. Experimentally,  it was found in the dissolving polycrystalline iron \cite{CordobaTorres2009} and in the epitaxial growth of CdTe films \cite{Nascimento2011}.

\section{Anomalous scaling}
\label{sec:intro-anomalous}

We now turn to the least understood form of anomalous roughening, generically denoted anomalous scaling. This is the case studied in the rest of this thesis. It is characterized by
\begin{equation}
  \alpha_{\rm loc}=\alpha_s,
  \qquad
  \alpha>\alpha_{\rm loc}.
  \label{eq:intro-anom-def}
\end{equation}

\begin{figure}[!ht]
\centering
\includegraphics[width=0.72\textwidth]{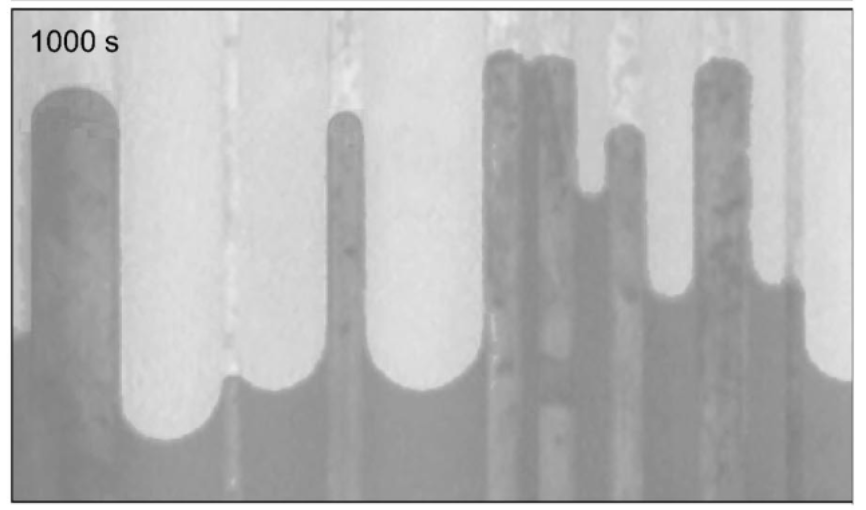}
\caption{Experimental interface profiles displaying anomalous roughening in a Hele-Shaw cell with quenched disorder. Taken from \cite{soriano2003anomalous}}
\label{fig:intro-imbibition-front}
\end{figure}

This phenomenology has been reported in several contexts. An experimental example is provided by roughening fluid fronts in Hele-Shaw cells with quenched disorder \cite{SorianoEtAl2002,soriano2003anomalous,SorianoEtAl2005}, see Fig.~\ref{fig:intro-imbibition-front} for an example of an interface profile in this case. It has also recently been discussed in the context of urban expansion fronts \cite{MarquisArtimeGallottiBarthelemy2025}. Other examples include electrodeposition, thin-film growth, and fracture \cite{HuoSchwarzacher2001,Schwarzacher2004,CuernoVazquez2004,AugerEtAl2006,SchmittbuhlEtAl1997,LopezEtAl1998}.

Historically, much of the early evidence came from  MBE models such as the Wolf-Villain and Das Sarma-Tamborenea models \cite{WolfVillain1990,TamboreneaDasSarma1993,Schroeder1993,Krug1994,DasSarmaGhaisasKim1994}. Later work showed that these models display long crossovers and strong finite-size effects, so one must distinguish carefully between genuine asymptotic scaling and effective exponent mismatches measured at finite system sizes \cite{Krug1997,PunyinduDasSarma1998,DasSarmaEtAl2002}. 

A first interpretation was proposed by L\'opez and collaborators \cite{lopez1995growth,Lopez1999,Ramasco2000,LopezCastroGallego2005}. They introduced the term ``intrinsic'' anomalous scaling to distinguish this case from super-rough and faceted scaling, and argued that the mismatch between local and global exponents reflects genuinely different local and global roughness properties of the interface. In particular, they proposed a heterogeneous elastic line model displaying anomalous scaling. We will revisit this model in Chapter~\ref{sec:interfaces-elastic}, where we argue against this ``intrinsic'' interpretation.

It has been argued that this anomalous scaling is related to the presence of large slopes \cite{Krug1994,Lopez1999}. When slope fluctuations are broad, second moments are often insufficient, and one should examine all moments, leading to \emph{multiscaling}. Instead of characterizing local roughness only through the quadratic width, one defines the \(q\)-dependent structure functions
\begin{equation}
  w_q(\ell,t)=\left\langle \left|h(x+\ell,t)-h(x,t)\right|^q \right\rangle^{1/q}.
\end{equation}
In the stationary state, one may then have
\begin{equation}
  w_q(\ell)\sim \ell^{\alpha_{\rm loc}(q)} L^{\alpha-\alpha_{\rm loc}(q)},
\end{equation}
where \(\alpha_{\rm loc}(q)\) is independent of \(q\) in the standard case, but may depend on \(q\) when multiscaling is present.

\section{Plan of Part I}
\label{sec:intro-roadmap}

Chapter~\ref{sec:interfaces-elastic} is devoted to the heterogeneous elastic line model introduced by L\'opez and collaborators \cite{lopez1995growth}. In their interpretation, the local and global fluctuations are genuinely distinct, so that the system should be characterized by two independent roughness exponents, one local and one global. We revisit this picture by deriving, in particular, the full distribution of stationary height differences. The main result of that chapter is that the discrepancy between local and global roughness can be understood as a statistical effect of disorder averaging and rare events, without invoking a new local roughness exponent. In particular, the model exhibits multiscaling as a consequence of broad disorder-induced fluctuations.

Chapter~\ref{sec:interfaces-spme-main} then turns to the one-dimensional stochastic porous medium equation. Despite the nonlinearity of the model, we again obtain the stationary distribution of height differences. In this case, by contrast, the local exponent is genuinely distinct from the global one. The interface exhibits multiscaling of local increments due to the broad distribution of slopes. This corresponds to a genuine case of ``intrinsic'' anomalous scaling.

Chapter~\ref{sec:interfaces-anderson} is devoted to the spectral properties of the heterogeneous elastic line, and provides results that will be used in Chapter~\ref{sec:interfaces-elastic}.

\chapter{Heterogeneous Elastic Lines}
\label{sec:interfaces-elastic}

\providecommand{\be}{\begin{equation}}
\providecommand{\ee}{\end{equation}}
\providecommand{\bea}{\begin{align}}
\providecommand{\eea}{\end{align}}
\providecommand{\nn}{\nonumber\\}
\providecommand{\D}{\mathrm{d}}
\providecommand{\eqlaw}{\stackrel{\mathrm{law}}{=}}
\providecommand{\free}{\mathrm{free}}
\providecommand{\fixed}{\mathrm{fixed}}
\providecommand{\lambdaMinTyp}{\lambda_{\min}^{\mathrm{typ}}}

The aim of this chapter is to study a solvable model in which anomalous roughening has a \emph{statistical} origin. When the interface contains large jumps, disorder averages need not reflect the typical behavior and may instead be dominated by rare realizations.

The model we consider is a variant of the Edwards-Wilkinson interface with quenched spatially varying diffusivity, introduced in \cite{lopez1995growth}. Because it is linear, the stationary measure can be obtained exactly and the finite-time behavior can also be analyzed in detail. This provides a useful setting in which to separate \emph{typical} roughness from \emph{disorder-averaged} roughness. Previous works had identified anomalous scaling in this model, but their interpretation was incorrect. Our contribution is to provide an exact description of the stationary state, which allows us to determine both the global and local exponents and to clarify their origin. Contrary to earlier claims, we find that these anomalies have a purely statistical original. They also lead to a form of multiscaling consistent with related models and with experimental observations that had remained unexplained \cite{soriano2003anomalous}.

\section{Heterogeneous Edwards-Wilkinson model}

The particular model we study is an interface with spatially inhomogeneous diffusivity defined by
\be
\label{eq:rdcont}
\partial_t h(x,t)=\partial_x\!\big[k(x)\partial_x h(x,t)\big]+\eta(x,t),
\ee
where \(k(x)\) is the quenched random diffusivity. The usual EW equation is recovered for \(k(x)=k\). 

\begin{figure}[!ht]
\centering
\includegraphics[width=0.7\textwidth]{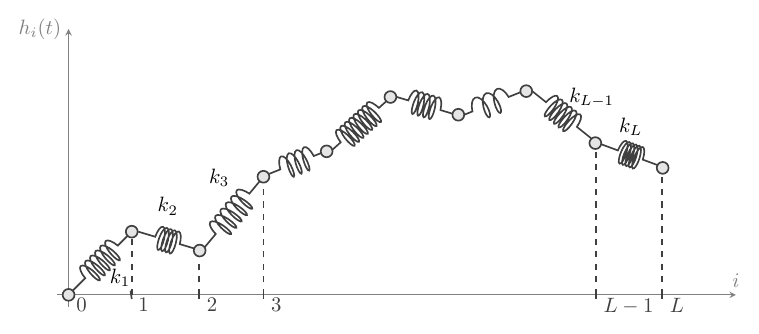}
\caption{Illustration of a random spring chain with one end fixed and the other end free. This will be referred to as the \emph{free} case.}
\label{schema}
\end{figure}

More precisely, we work with a discrete-space version. 
On sites \(n=0,1,\dots,L\), the heights evolve according to
\begin{equation}
  \partial_t h_n(t)
  =
  -k_n\big(h_n(t)-h_{n-1}(t)\big)
  -k_{n+1}\big(h_n(t)-h_{n+1}(t)\big)
  +\eta_n(t),
  \label{eq:el-eom}
\end{equation}
with independent Gaussian white noises
\begin{equation}
  \langle \eta_n(t)\rangle=0,
  \qquad
  \langle \eta_n(t)\eta_m(t')\rangle=2\,\delta_{nm}\delta(t-t').
\end{equation}
where Eq.~\eqref{eq:el-eom} also describes a random spring chain where \(k_n\) is interpreted as a spring constant subject to thermal fluctuations, as illustrated in Fig.~\ref{schema}.

The spring constants are taken to be i.i.d.\ quenched variables drawn from
\begin{equation}
  p(k)=\mu\,k^{\mu-1},
  \qquad k\in(0,1),
  \label{eq:el-pk}
\end{equation}
with \(\mu>0\). 
Small spring constants are therefore common when \(\mu<1\) and rare when \(\mu>1\). 
It is convenient to work with the inverse spring constants
\begin{equation}
  X_i=\frac{1}{k_i},
\end{equation}
since both stationary height differences and several finite-time observables will be expressed directly in terms of sums of these variables. 
The \(X_i\) are Pareto distributed,
\begin{equation}
  q(X)=\mu\,X^{-1-\mu},
  \qquad X\ge1.
\end{equation}
The heavy tail of \(X_i\) will be the origin of the deviation from standard Edwards-Wilkinson scaling.

\paragraph{Boundary conditions}
We impose a Dirichlet condition at the left end,
\begin{equation}
  h_0(t)=0,
  \label{eq:el-left-dirichlet}
\end{equation}
and at the right end we consider two cases:
\begin{enumerate}[label={(\arabic*)},leftmargin=*,align=left,itemsep=0.1cm]
  \item the \emph{free case}, in which the right end is unconstrained, corresponding to \(k_{L+1}=0\);
  \item the \emph{fixed case}, in which the right end is also pinned, corresponding to a second Dirichlet condition \(h_{L+1}=0\).
\end{enumerate}

The equation \eqref{eq:el-eom} may be rewritten as
\begin{equation}
\label{eqmotion2}
\partial_t h_i(t)
=
-\sum_{j=1}^{L}\Lambda_{ij}h_j(t)+\eta_i(t),
\qquad i=1,\dots,L,
\end{equation}
where \(\Lambda\) is the real symmetric tridiagonal matrix
\begin{align} \label{eq:eqlambda}
  \Lambda_{i,i}&=k_i+k_{i+1},
  \qquad i=1,\dots,L,
  \nonumber\\
  \Lambda_{i,i+1}&=\Lambda_{i+1,i}=-k_{i+1},
  \qquad i=1,\dots,L-1.
\end{align}
In both the free and fixed cases, pinning the left end removes the center-of-mass fluctuation (the zero mode) and makes \(\Lambda\) invertible.
While boundary conditions do not affect the typical behavior of the interface in the bulk, they might change the nature of the rare events. For instance, consider an interface with large ``rips''. In the free case, one can have a single rip, while one needs at least two in the fixed case to match the boundary conditions. As a consequence, disorder averages dominated by rare events may depend strongly on the choice of boundary conditions.

In the following, we will {\bf focus on the free case}, and unless stated otherwise all results below refer to that case.

\paragraph{Thermal averages, disorder averages, and observables}

The model contains two distinct sources of randomness: quenched disorder, through the random spring constants, and thermal noise. For a fixed disorder realization, we write \(\langle\cdot\rangle\) the average over the thermal noise, and \(\overline{(\cdot)}\) the average over the quenched disorder. Since the model is linear, the thermal measure is Gaussian at fixed disorder. The disorder average, on the other hand, probes the fluctuations of the springs, which can lead to broad distributions, and the average can then be dominated by rare realizations.

We use two equal-time observables. While in the introduction we used the local and global width, it will be more convenient here to work, equivalently, with the mean-square displacement
\begin{equation}
  D(L,t)=\frac{1}{L}\sum_{n=1}^L \langle h_n(t)^2\rangle,
  \label{eq:el-D}
\end{equation}
for the global observable and the squared height-difference
\begin{equation}
  G_n(x,t)=\langle (h_{n+x}(t)-h_n(t))^2\rangle,
  \qquad 1\ll x\ll L,
  \label{eq:el-G}
\end{equation}
for the local one.

\section{Summary of results}
\label{sec:el-observables}
The central point of this chapter is the distinction between \emph{typical} roughness and \emph{disorder-averaged} roughness. 
For narrow disorder distributions, the two notions are effectively equivalent and one recovers standard Edwards-Wilkinson behavior. 
For the heavy-tailed distribution \eqref{eq:el-pk}, however, these two notions can differ significantly as the average is dominated by rare events.

The behavior of the model is controlled by the parameter \(\mu\), which determines in particular the density of weak springs. 
Two principal regimes must be distinguished.
\paragraph{Regime \(\bm{\mu>1}\).}
In this case the disorder is effectively weak, in the sense that \(\overline{X_i}=\overline{1/k_i}\) is finite. 
The model then displays standard Edwards--Wilkinson scaling, with
\begin{equation}
  \alpha_{\mathrm{EW}}=\frac12,
  \qquad
  z_{\mathrm{EW}}=2,
  \qquad
  \beta_{\mathrm{EW}}=\frac14.
\end{equation}
One finds
\begin{align}
\label{w2}
\overline{D(L,t)}
&\sim
\begin{cases}
t^{2\beta_{\mathrm{EW}}}, & t\ll L^{z_{\mathrm{EW}}},
\\[1mm]
L^{2\alpha_{\mathrm{EW}}}, & t\gg L^{z_{\mathrm{EW}}},
\end{cases}
\\
\label{g2}
\overline{G(x,t)}
&\sim
\begin{cases}
t^{2\beta_{\mathrm{EW}}}, & t\ll x^{z_{\mathrm{EW}}},
\\[1mm]
x^{2\alpha_{\mathrm{EW}}}, & t\gg x^{z_{\mathrm{EW}}},
\end{cases}
\end{align}
for \(x\ll L\). 
Thus the quenched heterogeneity does not modify the standard EW scaling and exponents.

\begin{figure}
\hspace*{-2cm}
\includegraphics[width=1.2\textwidth]{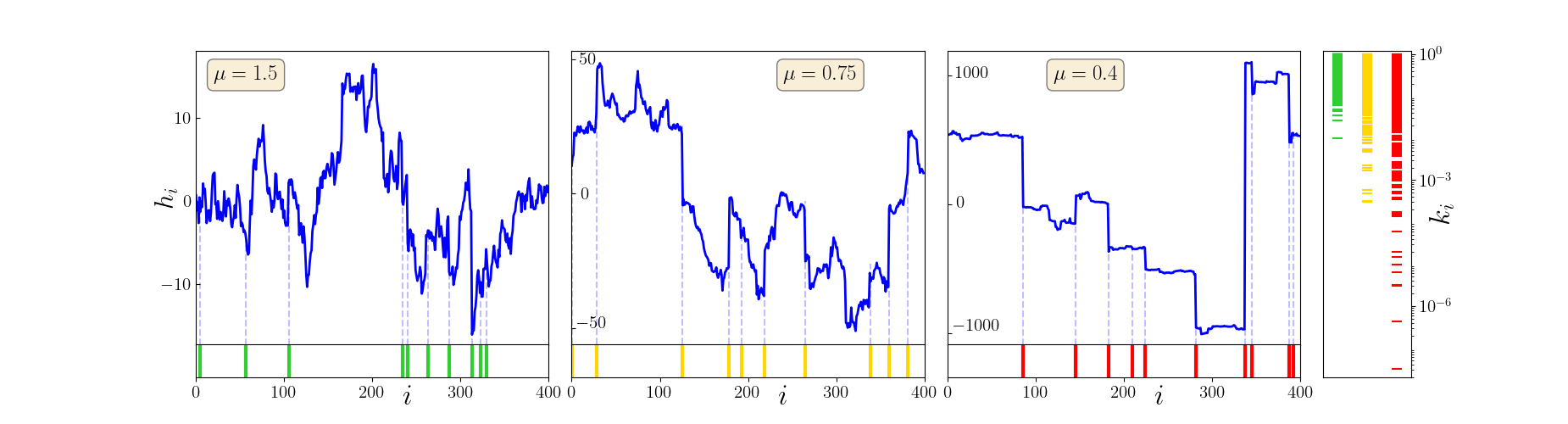}
\caption{Examples of stationary interfaces for \(\mu=1.5\), \(0.75\), and \(0.4\), with Dirichlet boundary conditions and zero mean. The ten weakest springs are indicated below each interface. As \(\mu\) decreases, weak springs become more frequent and the interfaces develop larger jumps. These large jumps are the origin of the anomalous disorder-averaged behavior.}
\label{fig:el-figint}
\end{figure}

\paragraph{Regime \(\bm{\mu<1}\).}
In this case \(\overline{1/k_i}\) diverges and very weak springs occur frequently enough to alter the shape of the interface and the disorder average. 
Typical stationary interfaces display large jumps, as illustrated in Fig.~\ref{fig:el-figint}. 
The roughness and dynamical exponents become
\begin{equation}
  \alpha=\frac{1}{2\mu},
  \qquad
  z=\frac{1+\mu}{\mu},
  \qquad
  \beta=\frac{1}{2(1+\mu)}.
\end{equation}
Because of the large jumps, one must distinguish typical observables from disorder-averaged ones. The typical observables, here denoted by the superscript ``typ.'', satisfy the standard scaling:
\begin{align}
\label{scaleDmu05freetyp}
D^{{\rm typ.}}(L,t)
&\sim
\begin{cases}
t^{2\beta}, & t\ll L^z,
\\[1mm]
L^{2 \alpha}, & t\gg L^z,
\end{cases}
\\
\label{eq:anomscalingtyp}
G^{{\rm typ.}}(x,t)
&\sim
\begin{cases}
t^{2\beta}, & t\ll x^z,
\\[1mm]
x^{2 \alpha}, & t\gg L^z.
\end{cases}
\end{align}

This is not the case for disorder-averaged observables. In particular, the scaling behavior is anomalous and depends on the boundary conditions. For the free case, one obtains%
\begin{align}
\label{scaleDmu05free}
\overline{D^\free(L,t)}
&\sim
\begin{cases}
t^{2\beta}, & t\ll L^z,
\\[1mm]
L^\mu t^{1-\mu}, & t\gg L^z,
\end{cases}
\\
\label{eq:anomscaling}
\overline{G^\free(x,t)}
&\sim
\begin{cases}
t^{2\beta}, & t\ll x^z,
\\[1mm]
x\,t^{(2\alpha-1)/z}, & x^z\ll t\ll L^z,
\\[1mm]
x\,L^{-1+\mu}t^{1-\mu}, & t\gg L^z.
\end{cases}
\end{align}
For the fixed case, the results are presented and derived in \Pone.
Two features deserve to be emphasized.
First, neither disorder-averaged observable saturates at long times, even though every \emph{individual} interface reaches thermal equilibrium after a finite relaxation time. Indeed, the disorder averages are dominated by the rare interfaces that have not yet equilibrated. These rare interfaces contain very weak springs, which take a long time to equilibrate.
Second, the local disorder average scales, for times large enough, linearly in \(x\), which may look like a local roughness exponent \(1/2\), but in fact has a different origin. 

\paragraph{Origin of the anomalous scaling}
Consider a window of size \(x\). Either it does not contain a large jump, in which case the local height difference is of the typical scale, or it does, in which case the height difference is much larger. The probability that a window contains such a jump is proportional to its size \(x\). In the strong disorder case \(\mu<1\),  the disorder average is dominated by these rare windows and therefore scales linearly in \(x\).

In that sense, the apparent anomalous scaling of \(\overline{G(x,t)}\) is purely statistical. In experiments or numerical simulations, one typically averages over many windows or many realizations, which in the present setting is equivalent to averaging over both disorder and thermal noise. Therefore, standard analysis of experiments does not correspond to a measure of typical behavior.%

The remainder of the chapter derives these results, first in the stationary regime and then at finite time.

\section{Stationary regime}
\label{sec:el-stationary}

For a given realization of the springs, the thermal process is Gaussian and the equal-time covariance is known explicitly \cite{gupta2013dynamics}:
\begin{equation}
  \langle h_i(t)h_j(t)\rangle
  =
  \left(\frac{1-e^{-2\Lambda t}}{\Lambda}\right)_{ij}.
  \label{eq:el-cov-time}
\end{equation}
where $\Lambda$ is the matrix introduced above, \eqref{eq:eqlambda}.
At long times this converges to
\begin{equation}
  \langle h_i h_j\rangle=(\Lambda^{-1})_{ij}.
  \label{eq:el-cov-stat}
\end{equation}
Once \(\Lambda^{-1}\) is known, both stationary observables can be deduced straightforwardly:
\begin{align}
  G_i(x)
  &=
  (\Lambda^{-1})_{i+x,i+x}+(\Lambda^{-1})_{i,i}-2(\Lambda^{-1})_{i,i+x},
  \label{eq:el-Gstat}
  \\
  D(L)
  &=
  \frac{1}{L}\operatorname{Tr}\Lambda^{-1}.
  \label{eq:el-Dstat}
\end{align}

The inversion of the tridiagonal matrix \(\Lambda\) is carried out in Appendix A of our paper \Pone. 
In the free case, the result is particularly simple:
\begin{equation}
  (\Lambda^{-1})_{ij}
  =
  \sum_{k=1}^{\min(i,j)}X_k,
  \qquad
  X_k=\frac{1}{k_k}.
  \label{Lambdafree}
\end{equation}
This formula is exact. It is also possible to find an explicit formula for the "fixed" case \Pone.

For fixed disorder, the stationary interface is distributed as a random walk with independent Gaussian increments of variances \(X_k\):
\begin{equation}
  h_x \eqlaw \sum_{j=1}^x \eta_j,
  \qquad
  \eta_j\sim\mathcal N(0,X_j),
\end{equation}
where \(\eqlaw\) denotes equality in law. For a given realization of disorder, the stationary free interface is a random walk in space with independent but non-identically distributed increments, and the broad distribution of those increment variances comes from the quenched disorder.

Using Eq.~\eqref{eq:el-Gstat}, one obtains
\begin{align}
G_i^\free(x)
&=
\sum_{j=i+1}^{i+x} X_j
\;\eqlaw\;
\xi_x,
\label{Gavg}
\\
D^\free(L)
&=
\sum_{j=1}^L \frac{L-j+1}{L}X_j,
\label{Davg}
\end{align}
where \(\xi_x\) denotes the sum of \(x\) independent Pareto variables. 
In particular, the one-site distribution of \(G_i^\free(x)\) is independent of \(i\), and we denote it by \(P^\free_x(G)\).

\paragraph{Typical scale and tail of the stationary distributions}

The distribution \(P^\free_x(G)\) coincides with the distribution \(\pi_x(\xi)\) of the sum \(\xi_x\), and has the following general properties:
\begin{itemize}
\item[(i)] its support is \((x,\infty)\);
\item[(ii)] it is peaked around the \emph{typical} scale \(G\sim x^{2\alpha}\);
\item[(iii)] for large \(G\) it decays as
\begin{equation}
P^\free_x(G)\sim x\,G^{-1-\mu}.
\end{equation}
\end{itemize}

The value of the roughness exponent \(\alpha\) depends on \(\mu\). 
For \(\mu>1\), the law of large numbers applies and one has \(2\alpha=1\), \emph{i.e.}
\begin{equation}
G^{\free,{\rm typ.}}(x)\sim x.
\end{equation}
For \(\mu<1\), the sum \(\xi_x\) is in the domain of attraction of a one-sided stable law and is dominated by its largest term, so that
\begin{equation}
G^{\free,{\rm typ.}}(x)\sim x^{1/\mu},
\qquad \text{thus} \qquad
\alpha=\frac{1}{2\mu}.
\end{equation}

One must distinguish the \emph{typical scale} from the \emph{mean value}.
For \(\mu>1\), the disorder average is finite and
\begin{equation}
\overline{G^\free(x)}=\overline{\xi_x}=x\,\frac{\mu}{\mu-1},
\end{equation}
so that the mean and the typical scale coincide up to constants. 
For \(\mu<1\), however, the stationary mean \(\overline{G^\free(x)}\) diverges. 
Thus the roughness exponent \(\alpha=1/(2\mu)\) characterizes the \emph{typical} local geometry, while the disorder average is dominated by the tail \(P^\free_x(G)\sim x\,G^{-1-\mu}\).

The distribution \(Q^\free_L(D)\) of the global observable \(D^\free(L)\) may be analyzed in the same way. 
It has support on \((L,\infty)\), is peaked around the typical scale \(D\sim L^{2\alpha}\), and for large \(D\) decays as
\begin{equation}
Q^{\free,{\rm typ.}}_L(D)\sim L\,D^{-1-\mu}.
\end{equation}
In particular, \(\overline{D^\free(L)}=\infty\) for \(\mu<1\). 
This divergence is caused by rare samples containing an exceptionally weak spring and therefore an exceptionally large jump.

\section{Finite-time regime}
\label{sec:el-finite}

We now turn to finite times.  
We begin with a general scaling relation that follows from the conservative nature of the drift. We then study the global observable \(\overline{D(L,t)}\), for which an exact spectral representation is available. 
The second part focuses on the local observable \(\overline{G(x,t)}\), whose distribution can be recovered thanks to the insight gained from the study of the global observable.%

\subsection{A general scaling relation}
\label{sec:scalrel}

Before turning to the spectral analysis, it is useful to recall a general scaling relation that holds for growth equations whose drift is a gradient $\partial_x F(x,t,h)$:
\begin{equation}
\partial_t h(x,t)=\partial_x F(x,t,h)+\eta(x,t).
\end{equation}
Integrating over the whole system \(x\in[0,L]\) gives
\begin{equation}
\frac{d}{dt}\int_0^L h(x,t)\,\D x
=
F(L,t,h)-F(0,t,h)+\int_0^L \eta(x,t)\,\D x.
\end{equation}
$F(L,t,h)-F(0,t,h)$ is a boundary term which can be neglected. Integrating in time from \(0\) to the saturation time \(L^z\), the white noise term scales as
\begin{equation}
\int_0^{L^z}\D t\int_0^L \D x \, \eta(x,t)
\sim
L^{(1+z)/2}.
\end{equation}
On the other hand, at saturation one expects
\begin{equation}
\int_0^L h(x,t=L^z)\,\D x \sim L\,L^\alpha=L^{1+\alpha}.
\end{equation}
Matching the two scalings yields
\begin{equation}
z=2\alpha+1
\end{equation}
in \(d=1\). In general dimension, the same argument leads to \(z=2\alpha+d\). Together with the scaling relation \(\beta=\alpha/z\), this already gives for $\mu>1$
\begin{equation}
z=\frac{1+\mu}{\mu},
\qquad
\alpha=\frac{1}{2\mu},
\qquad
\beta=\frac{1}{2(1+\mu)}.
\end{equation}
while one recovers the standard Edwards-Wilkinson exponents $\beta=\frac14$ and $z=2$ for $\mu>1$.
The spectral analysis below recovers these exponents and completes the picture by yielding the late-time behavior as well.

\subsection{Spectral density and scaling of \texorpdfstring{$\overline{D(L,t)}$}{D(L,t)}}

\begin{figure}[!ht]
\centering
\includegraphics[width=0.45\textwidth]{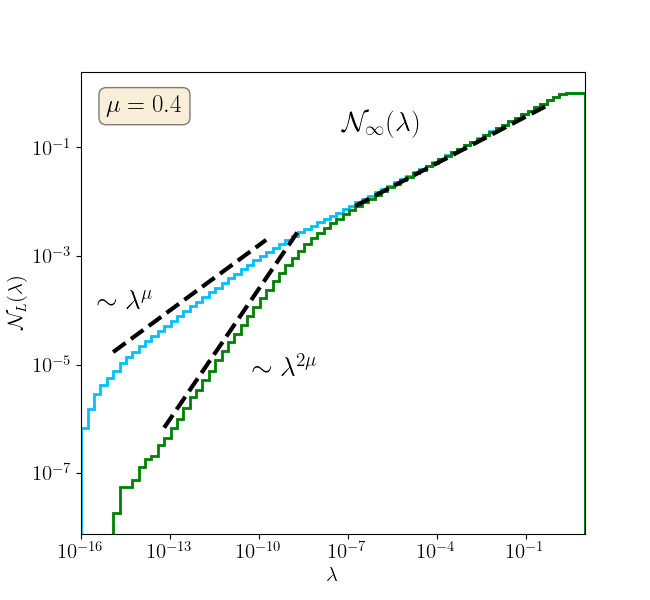}
\includegraphics[width=0.45\textwidth]{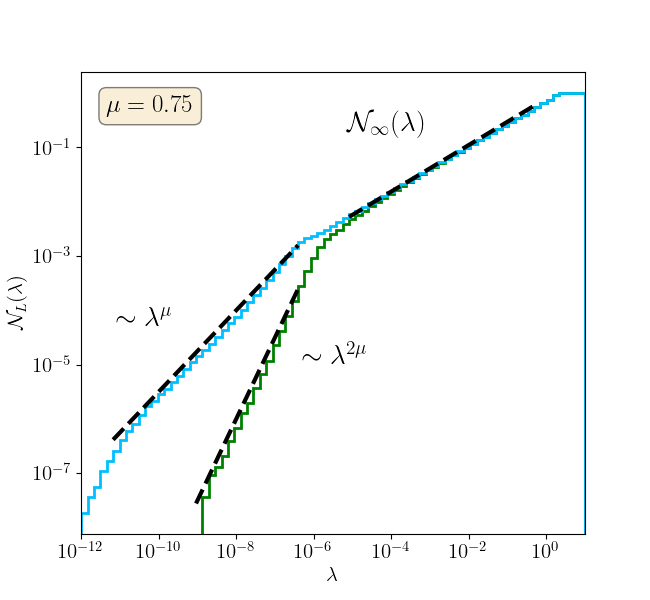}
\caption{Cumulative spectral density \(\mathcal N_L(\lambda)\) for \(\mu=0.4\) and \(\mu=0.75\), with \(L=500\) and \(N_s=10^8\) realizations, for free (blue) and fixed (green) boundary conditions. The dashed black curves are the analytical predictions obtained from the small-\(\lambda\) asymptotics and the infinite-volume density \(\rho(\lambda)\). The low-\(\lambda\) tail is controlled by the weakest spring in the free case and by the second-weakest spring in the fixed case.}
\label{DoS2}
\end{figure}

Taking the trace of Eq.~\eqref{eq:el-cov-time} gives
\begin{equation}
\overline{D(L,t)}
=
\frac{1}{L}\,
\overline{\operatorname{Tr}\!\left(\frac{1-e^{-2\Lambda t}}{\Lambda}\right)}
=
\frac{1}{L}\,
\overline{\sum_{\alpha=1}^{L}\frac{1-e^{-2\lambda_\alpha t}}{\lambda_\alpha}},
\end{equation}
where \(\lambda_\alpha\) are the eigenvalues of \(\Lambda\). 
Since the trace is basis invariant, the disorder average depends only on the spectral density
\begin{equation}
\rho_L(\lambda)=\frac{1}{L}\overline{\sum_{\alpha=1}^{L}\delta(\lambda-\lambda_\alpha)}.
\end{equation}
Therefore
\begin{equation}
\overline{D(L,t)}
=
\int \D\lambda\,\rho_L(\lambda)\,
\frac{1-e^{-2\lambda t}}{\lambda}.
\label{Dt}
\end{equation}
We also denote by
\begin{equation}
\rho(\lambda)=\lim_{L\to\infty}\rho_L(\lambda)
\end{equation}
the spectral density. This quantity can be analyzed by various means: in our paper \Pone, we used the standard Dyson-Schmidt method combined with more recent results on products of random $2 \times 2$ matrices using group theoretical considerations \cite{Tex20}. Another method is introduced in Chapter~\ref{sec:interfaces-anderson}, which allows one to obtain the DOS for more general disorder models of spring chains.
This quantity is studied in detail in Chapter~\ref{sec:interfaces-anderson}.
The low-\(\lambda\) behavior of \(\rho(\lambda)\) controls the long-time dynamics. 
For \(\mu>1\), one has, as in the homogeneous case,
\begin{equation}
\rho(\lambda)\simeq \frac{1}{2\pi\sqrt{k_a\lambda}},
\qquad
\frac{1}{k_a}=\overline{\frac{1}{k_i}},
\end{equation}
and one retrieves the EW scaling.
For \(\mu<1\), by contrast, the disorder strongly modifies the small-\(\lambda\) density of states. One obtains the exact asymptotic form
\begin{equation}
\label{BernasconiDoS}
\rho(\lambda)
\underset{\lambda\to0}{\simeq}
\frac{\mu\, \Gamma(1-\mu)^{\frac{1}{1+\mu}}}
{(1+\mu)^{\frac{2}{1+\mu}}\Gamma(\frac{1}{1+\mu})^2}
\,\, \lambda^{-\frac{1}{1+\mu}},
\qquad \mu<1.
\end{equation}
The power law was also obtained earlier by different means in \cite{BerSchWys80,AleBerSchOrb81}, and the prefactor in \cite{stephen1982diffusion}. Figure~\ref{DoS2} compares this asymptotic behavior with numerical data for the cumulative density of states
\begin{equation}
\mathcal N_L(\lambda)=\int_0^\lambda \rho_L(s)\,\D s.
\end{equation}

A finite system of size \(L\) has a smallest eigenvalue \(\lambda_1\), whose typical scale is determined by
\begin{equation}
\mathcal N_L(\lambdaMinTyp)\sim \frac{1}{L},
\qquad
\lambdaMinTyp\sim L^{-(1+\mu)/\mu}.
\end{equation}
This immediately yields the dynamical exponent
\begin{equation}
z=\frac{1+\mu}{\mu}.
\end{equation}
For \(\lambda\gtrsim \lambdaMinTyp\), one expects \(\rho_L(\lambda)\approx \rho(\lambda)\), independently of the boundary conditions. 
Inserting the asymptotic form \eqref{BernasconiDoS} into Eq.~\eqref{Dt} then yields
\begin{align}
\overline{D(t)}
&=
\int_0^\infty \D\lambda\,\rho(\lambda)\,\frac{1-e^{-2\lambda t}}{\lambda}
\sim C_\mu\,t^{2\beta},
\qquad
t\ll L^z,
\\
2\beta&=\frac{1}{1+\mu},
\qquad
\beta=\frac{1}{2(1+\mu)},
\end{align}
with $C_\mu=
\left(
  \frac{2\,\Gamma(1-\mu)}{(1+\mu)^\mu}
\right)^{\frac{1}{1+\mu}}
\frac{\mu\,\Gamma(\frac{\mu}{1+\mu})}
{\Gamma(\frac{1}{1+\mu})^2}$.

\paragraph{Rare samples and the late-time behavior of \texorpdfstring{$\overline{D(L,t)}$}{D(L,t)}}

The preceding argument applies for \(t\ll L^z\), when the disorder average is controlled by the bulk  density of states (\(L\to+\infty\)). The large-time regime must be treated separately, because it is dominated by rare samples with anomalously small eigenvalues. In the free case, this occurs when the interface contains an exceptionally weak spring: such a spring creates a large rip and therefore a very long equilibration time.
Thus, the lowest mode of a free interface is localized around the weakest spring, and one has \(\lambda_1 \sim k_{\min}\). 
The distribution of the weakest spring is given by standard order statistics \cite{schehr2014exact}
\begin{equation}
f_{1,L}(k)=\mu L\,k^{-1+\mu}e^{-Lk^\mu}.
\label{weakest}
\end{equation}
By matching the rare-event tail to the bulk density of states ( $L \to +\infty$) at \(\lambda\sim L^{-z}\), one obtains
\begin{equation}
\label{rhoper}
\rho^\free_L(\lambda)\simeq
\begin{cases}
C^\free L^\mu \lambda^{\mu-1},
& \lambda<L^{-z},
\\[1mm]
\rho(\lambda),
& \lambda>L^{-z},
\end{cases}
\end{equation}
where \(C^\free\) is a constant of order one. 
Inserting this form into Eq.~\eqref{Dt} yields the late-time scaling announced in Eq.~\eqref{scaleDmu05free}:
\begin{equation}
\overline{D^\free(L,t)}
\sim
L^\mu t^{1-\mu},
\qquad
t\gg L^z.
\end{equation}
Thus the disorder average does not saturate, even though each individual sample has already reached equilibrium after a finite time. 
Indeed, at time \(t\), the average is still dominated by the small fraction of samples whose weakest spring is so weak that their relaxation time exceeds \(t\).

\paragraph{Dependence on the boundary condition}

This scaling also depends on the boundary conditions. If the interface is fixed at both ends, a single weak spring is no longer sufficient to create a large rip: one needs two weak springs. The relevant rare-event tail is then controlled by the distribution of the second-weakest spring, \(f_{2,L}(\lambda)\sim \lambda^{2\mu-1}\).

\subsection{Scaling of \texorpdfstring{$\overline{G(x,t)}$}{G(x,t)}}

The local observable \(G_i(x,t)\) is not basis invariant, so its disorder average depends on the eigenvector distribution and not simply the eigenvalues.
One can nonetheless obtain its behavior from physical arguments, supported by numerical data.

For \(1\ll x\ll L\) and away from the boundaries, we write
\begin{align}
G_i(x,t)
&=
\langle (h_{i+x}(t)-h_i(t))^2\rangle
\nonumber\\
&=
\langle h_{i+x}(t)^2\rangle+\langle h_i(t)^2\rangle
-2\langle h_{i+x}(t)h_i(t)\rangle.
\end{align}
We distinguish three time regimes in the free case.

\paragraph{(i) Short times: \(\bm{t\ll x^z}\).}
The growing length \(\ell(t)\sim t^{1/z}\) is still smaller than the observation window. 
The heights \(h_{i+x}(t)\) and \(h_i(t)\) are effectively uncorrelated, and each has variance of order \(t^{2\beta}\). 
Therefore
\begin{equation}
\overline{G(x,t)}\approx 2\,\overline{D(t)}\sim t^{2\beta},
\end{equation}
as stated in Eq.~\eqref{eq:anomscaling}.

\begin{figure}[!ht]
\centering
\includegraphics[width=0.48\textwidth]{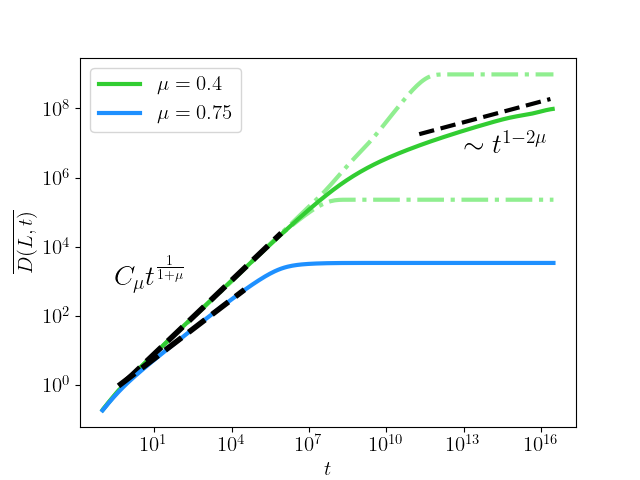}
\includegraphics[width=0.48\textwidth]{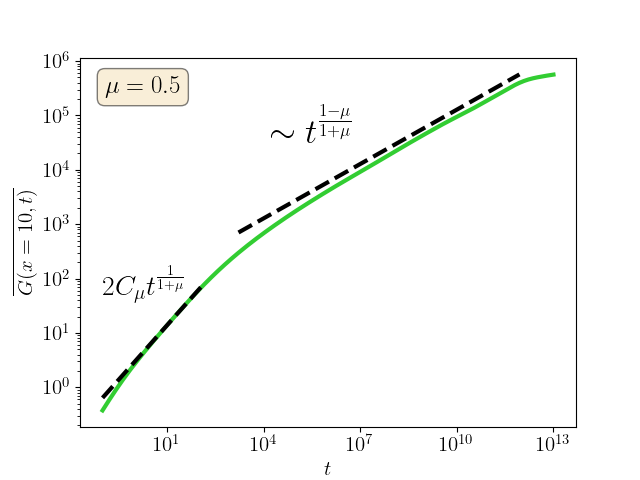}
\caption{Time-dependent scaling in the fixed case. Compared with the free case, the main difference appears in the regime \(t\gg L^z\) as shown in \Pone: for \(1/2<\mu<1\), the disorder-averaged width saturates, whereas for \(\mu<1/2\) it grows as \(t^{1-2\mu}\). Left: \(\overline{D(L,t)}\) for \(L=500\), \(N_s=10^5\), with \(\mu=0.75\) (lower blue curve) and \(\mu=0.4\) (upper green curve). Dashed lines show \(D(L,t)\) for two individual disorder realizations at \(\mu=0.4\). Right: time evolution of \(\overline{G(x=10,t)}\) for \(\mu=0.5\), \(L=5000\), and \(N_s=10^6\).}
\label{DtGx3}
\end{figure}

\paragraph{(ii) Intermediate times: \(\bm{x^z\ll t\ll L^z}\).}
In this regime the observation window is smaller than the correlated region, thus the distribution of jump amplitudes follows the same power-law tail as the stationary distribution,
\begin{equation}
P_x(G)\sim x\,G^{-1-\mu},
\end{equation}
but with a time-dependent cutoff that is given by the largest rip size squared. It is given by $\overline{D(L,t)} \sim t^{2 \beta}$, thus the cutoff is given by
\begin{equation}
G_{\max}(x,t)\sim t^{2\beta}.
\end{equation}
Integrating up to the time-dependent cutoff yields
\begin{equation}
\overline{G(x,t)}
\sim
\int_{x^{1/\mu}}^{t^{2\beta}}\D G\,
G\,
\frac{x}{G^{1+\mu}}
\sim
x\,t^{\frac{1-\mu}{1+\mu}}
=
x\,t^{(2\alpha-1)/z},
\end{equation}
which reproduces the intermediate-time anomalous scaling in Eq.~\eqref{eq:anomscaling}. This time dependence is confirmed numerically in Fig.~\ref{DtGx3}. 

\paragraph{(iii) Late times: \(\bm{t\gg L^z}\).}
At these times most samples have equilibrated, but the disorder average is still dominated by the rare non-equilibrated realizations containing an exceptionally weak spring. 
The scaling form may be deduced from the exact identity
\begin{equation}
\overline{D(L,t)}
=
\frac{1}{L}\sum_{x=1}^{L}\overline{G(x,t)},
\end{equation}
together with the probabilistic observation that the factor \(x\) originates from the probability that the observation window contains the dominant jump. 
This leads to
\begin{equation}
\overline{G^\free(x,t)}
\sim
x\,L^{-1+\mu}t^{1-\mu},
\qquad
t\gg L^z,
\end{equation}
as stated in Eq.~\eqref{eq:anomscaling}.

\section{Discussion and interpretation}
\label{sec:el-discussion}

This chapter has studied a random elastic line with spring constants drawn from the distribution \eqref{eq:el-pk}. 
The behavior is governed by the exponent \(\mu\), which controls the density of weak springs. 
For \(\mu>1\), the first inverse moment \(\overline{1/k_i}\) is finite, and the system remains in the ordinary Edwards--Wilkinson universality class. 
For \(\mu<1\), by contrast, the heavy tail of \(X_i=1/k_i\) produces a qualitatively different regime characterized by
\begin{equation}
\alpha=\frac{1}{2\mu},
\qquad
z=\frac{1+\mu}{\mu},
\qquad
\beta=\frac{1}{2(1+\mu)},
\end{equation}
together with nonstandard disorder-averaged scaling.

The key point is that the anomalous form of the disorder-averaged observables has a purely \emph{statistical} origin. Typical local fluctuations scale as \(x^{2\alpha}\), whereas disorder means are dominated by the rare windows that contain the largest jump of a weak-link sample. The linear dependence on \(x\) in \(\overline{G(x,t)}\) therefore measures the probability of encountering such a jump, not the roughness of a typical local fluctuation. 
This interpretation differs from the one often adopted in the literature \cite{lopez1996lack,lopez1997superroughening,lopez1999scaling,lopez1997power}. 
Those works correctly identified the presence of anomalous scaling, but interpreted the linear dependence \(\overline{G(x,t)}\propto x\) in terms of a local roughness exponent \(\alpha_{\mathrm{loc}}=1/2\), and argued that the local behavior differs from the global one. 

From the distributions, one can also derive the multiscaling properties
\begin{equation}
\overline{G(x,t)^{q/2}}
\sim
x^{\alpha_{\mathrm{loc}}(q)q},
\end{equation}
with
\begin{equation}
\alpha_{\mathrm{loc}}(q)=
\begin{cases}
\alpha, & q<1/\alpha,
\\[1mm]
1/q, & q>1/\alpha.
\end{cases}
\end{equation}
Small moments remain controlled by typical fluctuations while larger moments are dominated by rare jump-dominated samples. Indeed, from the probabilistic argument, any sufficiently large moment of the height difference scales as \(\propto x\), leading to \(q \alpha_{\mathrm{loc}}(q)=1\), hence \(\alpha_{\mathrm{loc}}(q)=1/q\).
This multiscaling is consistent with recent numerical simulations \cite{Rodriguez-Fernandez_2025} and experimental measurements reported in \cite{soriano2002anomalous,soriano2003anomalous,soriano2005anomalous}, where its origin had remained unclear. Our approach shows that it follows naturally from the statistical mechanism described above, thereby providing both a prediction and an interpretation.

The next chapter provides a complementary example in which the anomalous scaling is genuinely intrinsic, with a local exponent that differs from the global one. It also exhibits multiscaling, due to the presence of large slopes.

\clearpage

\chapter{Stochastic Porous Medium Equation}
\label{sec:interfaces-spme-main}

\renewcommand{\be}{\begin{equation}}
\renewcommand{\ee}{\end{equation}}
\renewcommand{\bea}{\begin{eqnarray}}
\renewcommand{\eea}{\end{eqnarray}}
\renewcommand{\nn}{\nonumber\\}
\providecommand{\red}[1]{#1}
\providecommand{\blue}[1]{#1}
\providecommand{\green}[1]{#1}
\providecommand{\cyan}[1]{#1}

\providecommand{\eqrefMT}{\eqref}
\providecommand{\generalab}{generalab}%
\providecommand{\thetaxZERO}{thetax0}%
\providecommand{\varianceNddimHO}{variance_N_ddimHO}%
\providecommand{\Ad}{Ad}%
\providecommand{\defNab}{defNab}%
\providecommand{\eqrelationKH}{eq:relation_K_H}%
\providecommand{\resWKBTWO}{resWKB2}%
\providecommand{\aTWO}{a2}%
\providecommand{\mrONE}{mr1}%
\providecommand{\FTWO}{F2}%
\providecommand{\DMGeneral}{DMGeneral}%
\providecommand{\Haa}{Haa}%
\providecommand{\varBessel}{var_Bessel }%
\providecommand{\Vl}{Vl}%
\providecommand{\eqsumofcumulantsangularsectors}{eq:sum_of_cumulants_angular_sectors}%
\providecommand{\integrate}{integrate}%
\providecommand{\final}{final}%
\providecommand{\cumulevenGUEgen}{cumuleven_GUE_gen}%
\providecommand{\cumulfree}{cumul_free}%
\providecommand{\entropy}{entropy}%

\providecommand{\x}{{\bf x}}
\providecommand{\y}{{\bf y}}
\providecommand{\p}{{\bf p}}
\providecommand{\z}{{\bf z}}
\providecommand{\tf}{\tilde{f}}
\providecommand{\tF}{\tilde{F}}
\providecommand{\tZ}{\tilde{Z}}
\providecommand{\tX}{\tilde{X}}
\providecommand{\tY}{\tilde{Y}}
\providecommand{\beq}{\begin{equation}}
\providecommand{\eeq}{\end{equation}}
\providecommand{\beqn}{\begin{eqnarray}}
\providecommand{\eeqn}{\end{eqnarray}}

\providecommand{\sinc}{\operatorname{sinc}}
\providecommand{\Ai}{\operatorname{Ai}}
\providecommand{\Bi}{\operatorname{Bi}}
\providecommand{\Li}{\operatorname{Li}}
\providecommand{\sgn}{\operatorname{sgn}}
\providecommand{\cotan}{\operatorname{cotan}}
\providecommand{\arccotan}{\operatorname{arccotan}}
\providecommand{\J}{\operatorname{J}}
\providecommand{\Det}{\operatorname{Det}}
\providecommand{\Tr}{\operatorname{Tr}}
\providecommand{\q}{\frac{\hbar^2}{2m}}
\providecommand{\dep}[2]{\ensuremath{\frac{\partial #1}{\partial #2}}}
\providecommand{\dd}{\ensuremath{\mathrm d}}
\providecommand{\tr}{\ensuremath{\text{tr}}}
\providecommand{\dt}[2]{\ensuremath{\frac{\dd #1}{\dd #2}}}
\providecommand{\dtn}[3]{\ensuremath{\frac{\dd^{#3} #1}{\dd #2^{#3}}}}
\providecommand{\ket}[1]{\ensuremath{|#1\rangle}}
\providecommand{\ketb}[2]{\ensuremath{|#1\rangle_{#2}}}
\providecommand{\bra}[1]{\ensuremath{\langle #1|}}
\providecommand{\braket}[2]{\ensuremath{\langle #1| #2 \rangle}}
\providecommand{\brab}[2]{{}_{#2}\ensuremath{\langle #1|}}
\providecommand{\moy}[1]{\ensuremath{\langle #1 \rangle}}
\providecommand{\abs}[1]{\ensuremath{\left| #1 \right|}}
\providecommand{\pFq}[5]{{}_{#1}\mathrm{F}_{#2} \left( \begin{array}{c} #3
\\ #4 \end{array} ; #5 \right)}

The previous chapter presented a model in which the local-global mismatch was \emph{statistical}: the apparent local exponent extracted from disorder-averaged observables did not characterize the typical local scaling of the interface. 
The purpose of the present chapter is to analyze a different mechanism. 
Here the local fluctuations are \emph{genuinely} governed by an exponent that differs from the global one, and can be broad enough to generate multiscaling.

Two complementary approaches will be used. 
The first is a random walk description of the stationary interface in \(d=1\), which provides a physical interpretation of the local and global exponents and gives access to the stationary increment distribution. 
The second is a functional renormalization-group (FRG) analysis of the stochastic equation, which yields the global scaling and the associated fixed points. It does not, however, allow one to obtain the local exponents.

The organization of the chapter is as follows. 
Section~\ref{sec:spme-model} introduces the continuum and discrete models and summarizes the main results. 
Section~\ref{sec:spme-mapping} derives the effective random walk description of the stationary interfaces in \(d=1\), from which one obtains the global roughness exponent and the local second-moment exponent. 
Section~\ref{sec:spme-cont} develops the continuum random walk limit, the mapping to a Bessel process, and the increment distributions that underlie the multiscaling behavior. 
Section~\ref{sec:spme-FRG} focuses on deriving the functional renormalization-group (FRG) analysis and studying its flow.

\section{The stochastic Porous Medium equation}
\label{sec:spme-model}

\begin{figure}[htbp]
    \centering
    \includegraphics[width=0.7\textwidth]{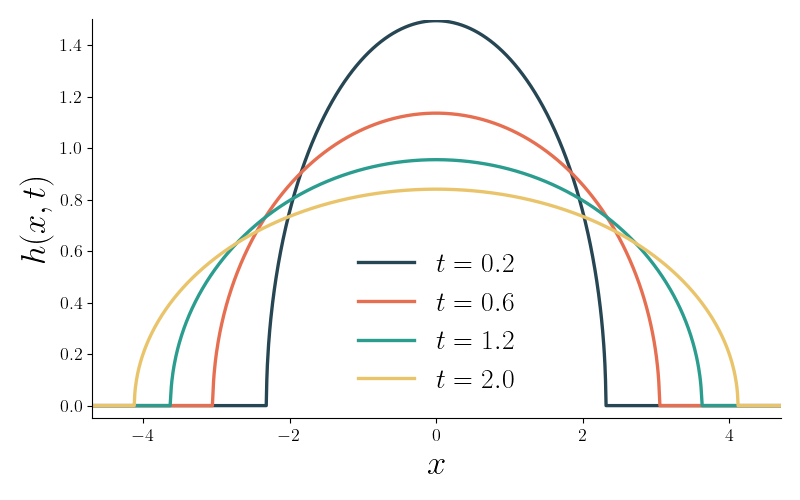}
    \caption{Example of a Barenblatt solution of the one-dimensional porous medium equation for $s=3$ shown at several times. The solution has compact support and finite speed propagation: its support is bounded by moving free boundaries.}
    \label{fig:barenblatt_pme_free_boundary}
\end{figure}

Before introducing the stochastic porous medium equation, let us recall its deterministic version:
\begin{equation}
\partial_t h = \nabla \left[ |h|^{s-1} \nabla h \right],
\end{equation}
it describes a situation in which the effective diffusivity depends on the field itself: diffusion is weaker where \(h\) is small and stronger where \(h\) is large when \(s>1\). The regime \(s>1\) arises, for instance, in gas flow through porous media \cite{Muskat1946}, in groundwater infiltration  \cite{Boussinesq1904}, and in certain nonlinear heat-transfer problems \cite{ZeldovichRaizer1967}. Since the diffusion coefficient \(|h|^{s-1}\) vanishes at \(h=0\), it leads to features absent in linear diffusion, such as moving free boundaries, namely interfaces separating the region where \(h>0\) from the region where \(h=0\), as illustrated in Fig.\ref{fig:barenblatt_pme_free_boundary}. By contrast, when \(s<1\) one enters the fast-diffusion regime, where diffusion is enhanced at low density. For a complete mathematical review, including applications, see for instance \cite{Vazquez2006Smoothing,vazquez}.

In geomorphology, nonlinear diffusion emerges naturally when the sediment-transport coefficient depends on the topography. While Culling's foundational model \cite{culling1960analytical} introduces the role of curvature-driven smoothing, its linear, isotropic, deterministic nature cannot account for the self-affine complexity observed in natural landscapes. The non-linearity and anisotropy of real slopes can be explained as follow: as overland water flow drains downhill, it accumulates causing the shear stress exerted on the soil bed to grow in the direction of the flow \cite{scheidegger2012theoretical,rodriguez1997fractal}. 

Driven by this observation, Pastor-Satorras and Rothman \cite{pastor1998stochastic,pastor1998scalingJStat} formulated an anisotropic stochastic description of erosion. By applying a local conservation law to the height field $h$,
\begin{equation}
\partial_t h = -\nabla\cdot \mathbf{J}+\eta,
\end{equation}
they postulated a current $\mathbf{J}$ contribution:
\begin{equation}
\mathbf{J}=-\nu\nabla h-D(h)\nabla_\parallel h,
\end{equation}
where $\nabla_\parallel$ denotes the gradient in the direction of the slope. In this formulation, $D(h)$ represents an effective diffusivity that encodes the nonlinearity caused by the accumulated erosive flux. The stochastic term $\eta$ accounts for the inherent randomness of the environment, such as the intermittent nature of rainfall or variations in soil erodibility.

Here, we will focus on the isotropic version, and especially on its one-dimensional version ($d=2$ is the upper critical dimension). The stochastic porous medium equation with additive, non-conservative white noise reads
\begin{align}
  \partial_t h(x,t)
  &= \partial_x^2 V(h(x,t))+\eta(x,t)
  \nonumber\\
  &= \partial_x\!\left[D(h(x,t))\,\partial_x h(x,t)\right]+\eta(x,t),
  \label{eq:spme-cont}
\end{align}
where \(h(x,t)\) is the interface height, \(D(h)=V'(h)\) is an even function, and \(\eta\) is a Gaussian white noise with correlations
\begin{equation}
  \langle \eta(x,t)\rangle=0,
  \qquad
  \langle \eta(x,t)\eta(x',t')\rangle=2\sigma\,\delta(x-x')\delta(t-t').
  \label{eq:spme-noise}
\end{equation}
The class of models of interest is characterized by the large-height behavior
\begin{equation}
  D(h)\sim c\,|h|^{s-1},
  \qquad
  |h|\to\infty,
  \label{eq:spme-power}
\end{equation}
with \(s>0\) and \(c>0\). When \(s=1\), one recovers the linear Edwards-Wilkinson case. For \(s>1\), the diffusivity increases with the local height, so the interface tends to flatten at large \(h\) and its slopes become smaller. By contrast, for \(0<s<1\), the diffusivity decreases with \(h\), so the slopes become steeper at large height. Examples of such interfaces are shown in Fig.~\ref{fig:figint}.

Previous work on related stochastic nonlinear-diffusion equations mostly focused on the anisotropic erosion problem introduced by Pastor-Satorras and Rothman \cite{pastor1998stochastic,pastor1998scalingJStat}, and on its renormalization-group analysis \cite{antonov1995quantum,antonov2017scaling,duclut2017nonuniversality,DuclutThesis}, although none of these works succeeded in deriving the critical exponents, nor did they identify the presence of anomalous scaling. The RG analysis of this model was first performed in \cite{antonov1995quantum} within a polynomial expansion of the function \(V(h)\), keeping only the first few couplings.
Here, we will use two complementary approaches. The first consists in describing the stationary state in terms of a random walk, and can also be applied directly to the anisotropic problem. The second consists in performing the RG analysis while working directly with the full functional flow, which is necessary to identify the fixed points selected by the large \(h\) behavior of \(V(h)\).

{\bf Discrete model and boundary conditions}

For numerical work and for the connection with the random walk description, it is useful to work on a lattice. 
On sites \(n=0,1,\dots,L\), we consider
\begin{equation}
  \partial_t h_n
  =
  D(h_{n-1})(h_{n-1}-h_n)
  +
  D(h_n)(h_{n+1}-h_n)
  +
  \eta_n(t),
  \label{eq:spme-discrete}
\end{equation}
with independent Gaussian noises satisfying
\begin{equation}
  \langle \eta_n(t)\eta_m(t')\rangle
  =2\,\delta_{nm}\delta(t-t').
\end{equation}
The left boundary is chosen fixed, \(h_0=0\), while the right boundary is free. 
In the discrete model, the free boundary is implemented by the condition
\begin{equation}
h_{L+1}-h_L=0.
\end{equation}

Thus, at a given time, the interface may be viewed as a spring chain whose local spring constants are not quenched, but depend on the local height: the spring connecting sites \(n\) and \(n+1\) has stiffness \(D(h_n)\). 
In that sense, the stochastic porous medium equation resembles the heterogeneous elastic line of the previous chapter, but with a field-dependent rather than quenched inhomogeneity.

\section{Main results}
\label{sec:spme-three-layers}

In \(d=1\) (with \(d=2\) the upper critical dimension), the scaling exponents depend continuously on $s$ and are given by
\begin{equation}
  \alpha=\frac{1}{1+s},
  \qquad
  z=\frac{s+3}{s+1},
  \qquad
  \beta=\frac{\alpha}{z}=\frac{1}{s+3},
  \label{eq:spme-exponents-d1}
\end{equation}
and satisfy the scaling relation $z=2\alpha+1$.

The interface exhibits \emph{intrinsic} anomalous scaling. In particular, the second-order local roughness exponent is
\begin{equation}
  \alpha_{\rm loc}(q=2)=\frac12.
\end{equation}
This can be understood heuristically as follows. On sufficiently small scales, the interface varies slowly, so that the coefficient \(D(h)\) in the operator
\begin{equation}
\partial_x\!\left[D(h(x,t))\,\partial_x h(x,t)\right]
\end{equation}
can be treated as approximately constant. Locally, the dynamics therefore reduce to an effective linear equation, which yields the local roughness exponent \(1/2\).

For \(0<s<1\), this argument remains valid for all moments \(q\), so that
\begin{equation}
  \alpha_{\rm loc}(q)=\frac12,
  \qquad 0<s<1,\ \forall q.
\end{equation}

For \(s>1\), the picture changes for sufficiently high moments. The previous argument captures the typical behavior, and therefore remains valid for low-order moments. However, rare excursions toward \(h\approx 0\), where \(D(h)\sim |h|^{s-1}\) becomes small, give rise to unusually large height increments and hence to a broad increment distribution. As a result, sufficiently high moments are dominated by these rare events rather than by the typical local fluctuations. The local roughness exponent is then given by
\begin{equation}
  \alpha_{\rm loc}(q)
  =
  \begin{cases}
    1/2, & \, q<q_c,
    \\[2mm]
    \alpha+\dfrac{1}{q}\dfrac{s}{1+s}, & \, q>q_c,
  \end{cases}
  \label{eq:alphalocqs1}
\end{equation}
with threshold
\begin{equation}
  q_c=2+\frac{2}{s-1}.
  \label{eq:spme-qc}
\end{equation}

\section{Stationary regime: the random walk mapping}
\label{sec:spme-mapping}

Before studying the random walk, let us recall that the drift term is a gradient, \\
$\partial_x\!\left[D(h(x,t))\,\partial_x h(x,t)\right]$, and therefore, as shown in Section~\ref{sec:scalrel}, implies the scaling relation \(z=2\alpha+1\). The dynamical exponents can thus be deduced from the roughness exponent that we will obtain here in the stationary state.

The stationary state in \(d=1\) can be understood thanks to an effective random walk description. 
If the stiffness were uniform, as in the Edwards-Wilkinson case, the stationary measure would be Gaussian and the increments would be independent centered Gaussians.
For the stochastic porous medium equation, the stiffness is not uniform, but when the local height is large it varies slowly compared to the slope.
One is then naturally led to an inhomogeneous random walk in which the local variance depends on the local height itself.

This reasoning leads to the effective random walk
\begin{equation}
  \tilde h_{n+1}
  =
  \tilde h_n+\frac{\chi_n}{\sqrt{D(\tilde h_n)}},
  \label{eq:RW}
\end{equation}
where the \(\chi_n\) are i.i.d.\ centered Gaussian variables of unit variance. 
This mapping is not an exact identity; it is an asymptotically valid description of the stationary \emph{bulk} in large systems, where the local height is large and \(D(h)\) varies weakly over one step.

Indeed, expanding \(D(h)\) to first order over one lattice spacing gives the condition
\begin{equation}
  |h_{n+1}-h_n|
  \ll
  \frac{D(h_n)}{|D'(h_n)|}
  \sim |h_n|,
  \label{eq:condgauss1}
\end{equation}
where the last estimate uses the asymptotic form \(D(h)\sim c|h|^{s-1}\). 
Since a typical increment scales as
\begin{equation}
|h_{n+1}-h_n|\sim \frac{1}{\sqrt{D(h_n)}}\sim |h_n|^{(1-s)/2},
\end{equation}
the condition~\eqref{eq:condgauss1} is automatically satisfied for \(|h_n|\gg1\), for all \(s>0\).

\begin{figure}
\centering
\includegraphics[width=0.8\textwidth]{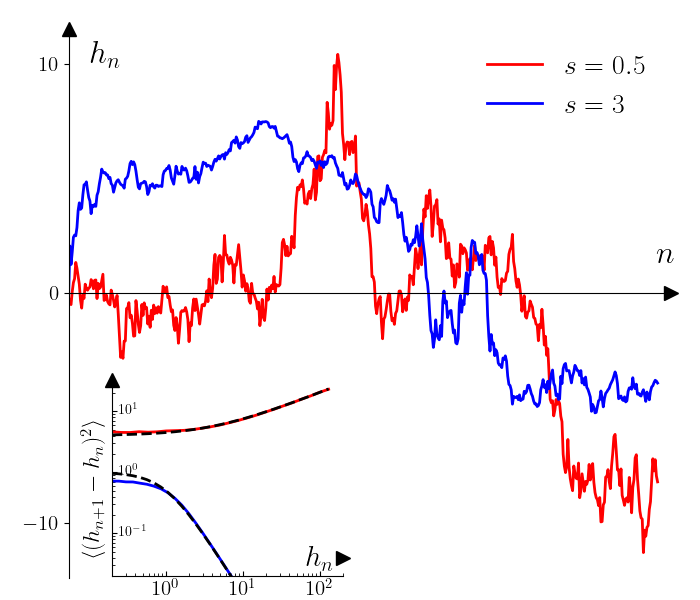}
\caption{Examples of stationary interfaces of the stochastic porous medium equation for two values of \(s\). For \(s>1\), the interface flattens at large \(h\), whereas for \(s<1\) it displays larger increments. Inset: variance of the local height increment \(h_{n+1}-h_n\). The random walk description becomes accurate at large enough height, where the variance is well approximated by \(1/D(h_n)\) (black dashed line).
}
\label{fig:figint}
\end{figure}

At a bulk point \(n=O(L)\), one expects \(h_n\sim L^\alpha\), and for any \(s>0\) this diverges with \(L\). 
The random walk description~\eqref{eq:RW} should therefore become asymptotically correct in the stationary bulk.
This interpretation is well supported numerically. 
In the inset of Fig.~\ref{fig:figint}, we plot the variance of the increment $h_{n+1}-h_n$ as a function of $h_n$, and show that it matches very well the prediction $1/D(h_n)$ for $h_n$ large enough. 
In the inset of the left panel of Fig.~\ref{fig:compRWPME}, the distribution $P(h_{n+1}|h_n)$ is plotted and compared to a Gaussian of mean $h_n$ and variance $1/D(h_n)$. The agreement is very good, and increases as $h_n$ increases. 

{\bf Derivation of the global roughness exponent}

We now study the discrete random walk
\begin{equation}
h_{n+1}
=
h_n+\frac{\chi_n}{\sqrt{D(h_n)}},
\label{RWmodelSM}
\end{equation}
where the \(\chi_n\) are i.i.d.\ centered Gaussian variables of unit variance. 
The trajectory of the walker corresponds to a stationary interface profile, and the walker position corresponds to the interface height.

Because the increments are independent, one has the exact relation
\begin{equation}
\langle (h_{n+\ell}-h_n)^2\rangle
=
\sum_{m=n}^{n+\ell-1}
\left\langle \frac{1}{D(h_m)}\right\rangle.
\label{variancedifference}
\end{equation}

We now assume the scaling form
\begin{equation}
h_n\sim n^\alpha \tilde h
\qquad (n\to\infty),
\end{equation}
where \(\tilde h\) is a rescaled random variable independent of \(n\). 
Then
\begin{equation}
\langle h_n^2\rangle
=
\sum_{m=0}^{n-1}
\left\langle \frac{1}{D(h_m)}\right\rangle
\simeq
\frac{1}{c}
\sum_{m=0}^{n-1}
\langle |h_m|^{1-s}\rangle.
\end{equation}
Using \(h_m\sim m^\alpha \tilde h\), one obtains
\begin{equation}
\langle h_n^2\rangle
\simeq
\frac{1}{c(1+\alpha(1-s))}
n^{1+\alpha(1-s)}
\langle |\tilde h|^{1-s}\rangle,
\end{equation}
whereas the left-hand side scales as \(\langle \tilde h^2\rangle n^{2\alpha}\). 
Matching the powers of \(n\) yields
\begin{equation}
\alpha=\frac{1}{1+s},
\qquad
\langle \tilde h^2\rangle
=
\frac{s+1}{2c}\,\langle |\tilde h|^{1-s}\rangle.
\label{momentrelationRW}
\end{equation}
Thus the random walk description reproduces the global roughness exponent quoted in Eq.~\eqref{eq:spme-exponents-d1}.

{\bf Derivation of the local anomalous exponent ($q=2$)}

The same exact relation \eqref{variancedifference} gives the scaling of height differences. 
For \(n\gg1\) and \(1\ll \ell\ll n\), one obtains
\begin{equation}
\langle (h_{n+\ell}-h_n)^2\rangle
\simeq
\frac{1}{c}\langle |\tilde h|^{1-s}\rangle
\sum_{p=0}^{\ell-1}(n+p)^{\frac{1-s}{1+s}}.
\label{sumscalemom2}
\end{equation}
There are two regimes.

\emph{Local regime: \(\ell\ll n\).}
In this regime, the dependence on \(p\) may be neglected in the sum, and one finds
\begin{equation}
\langle (h_{n+\ell}-h_n)^2\rangle
\simeq
\frac{1}{c}\langle |\tilde h|^{1-s}\rangle
n^{2\alpha-1}\,\ell.
\label{eq:scalemom2}
\end{equation}
Thus 
\begin{equation}
\alpha_{\rm loc}(2)=\frac12.
\label{q2anomdisc}
\end{equation}

\emph{Global regime: \(\ell=O(n)\).}
If \(\ell\sim n\), the same formula yields
\begin{equation}
\langle (h_{n+\ell}-h_n)^2\rangle
\simeq
\frac{1}{c}\langle |\tilde h|^{1-s}\rangle
n^{2\alpha} f(\lambda=\ell/n),
\end{equation}
with
\begin{equation}
f(\lambda)=\frac{s+1}{2}\left[(1+\lambda)^{2/(s+1)}-1\right].
\end{equation}
The two regimes match smoothly when \(\lambda\to0\).

In conclusion, the local second-moment roughness exponent is always \(1/2\), while the global roughness exponent is \(\alpha=1/(1+s)\). 
For \(s\neq 1\), the nonlinearity of the equation thus leads to a genuine intrinsic anomalous scaling.

\section{Beyond the second moment: limiting distributions and multiscaling}
\label{sec:spme-cont}

The discrete random walk description already gives the correct global exponent and the local second-moment exponent. 
To obtain the full stationary height difference distribution and the multiscaling exponents, however, it is more convenient to pass to a continuum description.

At large scales, the random walk is described by the It\^o process
\begin{equation}
dh(x)=\frac{1}{\sqrt{D(h(x))}}\,dB(x),
\end{equation}
where \(B(x)\) is a Brownian motion in the spatial variable \(x\). 
The one-point probability density \(P(h,x)\) satisfies the Fokker-Planck equation
\begin{equation}
\partial_x P(h,x)=\frac12\,\partial_h^2\!\left(\frac{P(h,x)}{D(h)}\right).
\label{FKPh}
\end{equation}
For \(D(h)\simeq c|h|^{s-1}\), the large \(x\) solution takes the scaling form
\begin{equation}
P(h,x)
\simeq
\frac12\,x^{-1/(1+s)}\,
\tilde P(|h|/x^{1/(1+s)}).
\end{equation}
where \(1/(1+s)\) is the roughness exponent. Inserting this ansatz into Eq.~\eqref{FKPh}, one finds the limiting distribution of the rescaled height
\begin{equation}
\tilde P(\tilde h)
=
\mathcal N\,|\tilde h|^{s-1}
\exp\!\left[
-\frac{2c|\tilde h|^{s+1}}{(s+1)^2}
\right],
\label{eq:onepointh}
\end{equation}
where \(\tilde P\) is normalized on \([0,\infty)\). 

{\bf Mapping to the Bessel process}

A convenient way to analyze the continuum process is to transform it to a process with additive noise. 
Introduce
\begin{equation}
{\sf Y}(h)=\int_0^h dh'\,\sqrt{D(h')},
\qquad
Y(x)={\sf Y}(h(x)).
\end{equation}
Applying It\^o's formula gives a Langevin equation of the form
\begin{equation}
dY=dB(x)-U'(Y)\,dx,
\label{eq:LangevinUp}
\end{equation}
where \(U(Y)\) is the effective potential generated by the height dependence of \(D(h)\). 
For the power-law case \(D(h)=c|h|^{s-1}\), one finds
\begin{equation}
U(Y)=-\frac{s-1}{2(s+1)}\log|Y|.
\end{equation}
Thus \(|Y(x)|\) is a Bessel process \cite{WikipediaBesselProcess} of effective dimension
\begin{equation}
d_{\mathrm{B}}=\frac{2s}{s+1}<2.
\qquad (s>0).
\end{equation}
The propagator of this process is known exactly (See \cite{katori2011besselprocess} for instance).

\begin{figure}
\centering
\includegraphics[width=0.48\textwidth]{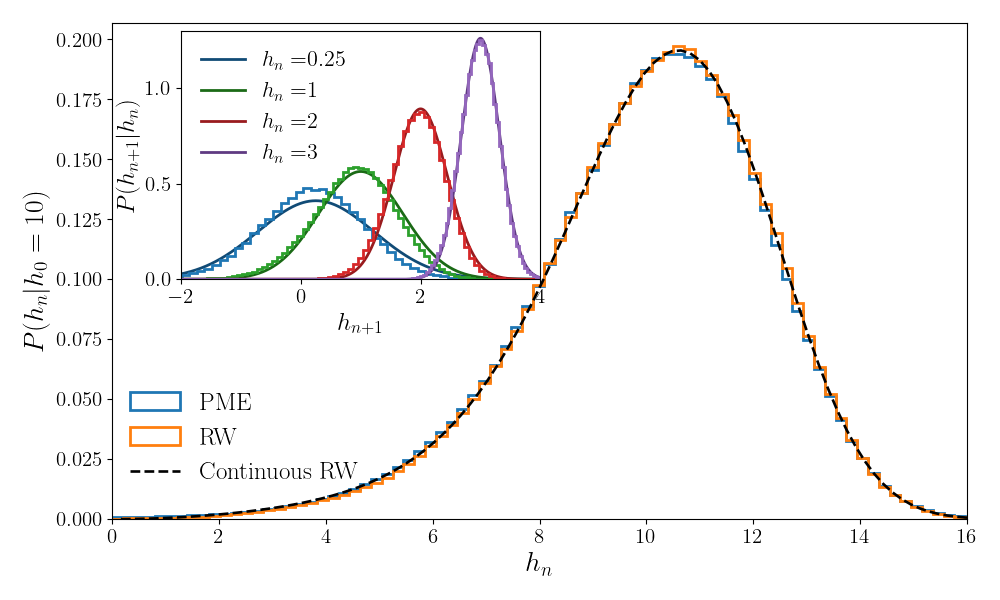}
\includegraphics[width=0.48\textwidth]{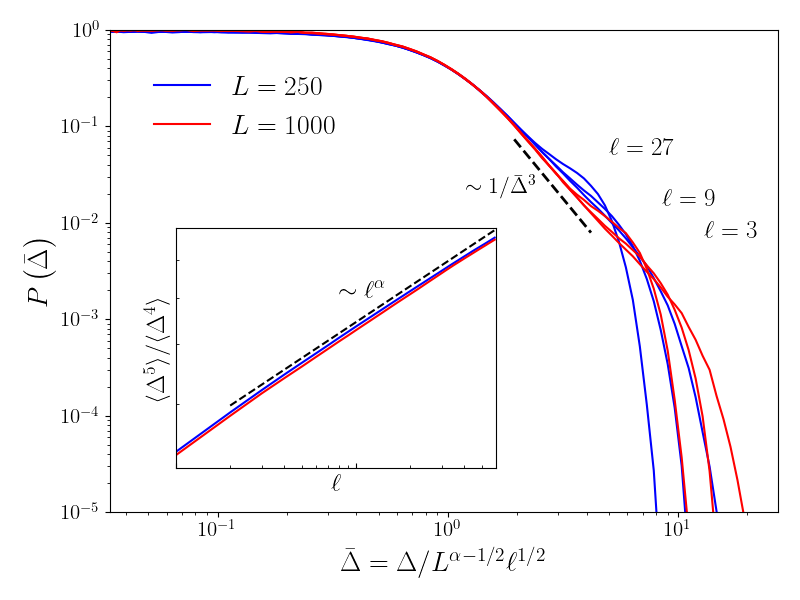}
\caption{Left: comparison between stationary interfaces of the stochastic porous medium equation, the discrete random walk model, and the continuum prediction for the one-point distribution. The inset shows the distribution $P(h_{n+1}|h_n)$ for the stochastic porous medium equation (histogram), compared with the Gaussian prediction with variance \(1/D(h_n)\) (continuous line). 
Right: distribution of $P(\bar \Delta)$, with $\bar \Delta= \Delta/( L^{\alpha - \frac{1}{2}} \ell^{\frac{1}{2}})$, for $s=3$, $\alpha=1/4$ (blue line: $L=250$, red line: $L=1000$) and for three different values of $\ell$. The collapse at small $\bar \Delta$ shows the scaling of the typical values $\Delta_\text{typ}(\ell) \sim \sqrt{\ell} L^{\alpha -1/2}$. The plot also shows a power-law decay with the predicted exponent $3$ for $L=1000$. Inset: evaluation of the cutoff $\Delta_\text{max}(\ell) \sim \langle \Delta^5 \rangle/\langle \Delta^4 \rangle$, which indeed scales as $\ell^\alpha$.
}
\label{fig:compRWPME}
\end{figure}

For \(D(h)=c|h|^{s-1}\), the propagator of \(|h|\), i.e. the probability density for the random walk to be at \(|h|\) at position \(x\), given that it started from \(h_0\) at \(x=0\), reads
\begin{equation}
\begin{split}
P(|h|,x||h_0|,0)=\;&
\frac{2c}{(s+1)x}\sqrt{|h_0|}\,|h|^{s-\frac12}
\exp\!\left[
  -\frac{2c(|h|^{s+1}+|h_0|^{s+1})}{(s+1)^2x}
\right]
\\
&\times
I_{-\frac{1}{s+1}}\!\left(
  \frac{4c|h|^{\frac{s+1}{2}}|h_0|^{\frac{s+1}{2}}}{(s+1)^2x}
\right),
\end{split}
\label{eq:propagatorAbsh}
\end{equation}
where \(I_\nu\) is the modified Bessel function. This distribution is obtained and compared, for $h_0$, to stationary interfaces of the stochastic porous medium equation in the left panel of Fig.~\ref{fig:compRWPME}. The agreement is excellent.

While we are interested in the distribution of the height increment \(h(x+y)-h(x)\) for \(y \ll x\), it is more convenient to consider instead
\begin{equation}
\Delta = |h(x+y)|-|h(x)|.
\end{equation}
This is expected to display the same behavior, since for \(y \ll x\) the interface rarely crosses \(0\). Because the conditional distribution of \(|h(x+y)|\) given \(|h(x)|\) is known, the distribution of \(\Delta\) follows by conditioning on \(|h(x)|\) and integrating over it:
\begin{equation}
P_{x,y}(\Delta)
=
\int_{\max(0,-\Delta)}^\infty
dh\,
P(h+\Delta,y|h,0)\,P(h,x|0,0).
\label{eq:expprobdiff-summary}
\end{equation}
whose behavior can be extracted exactly since both quantities are known. 
The main results are as follows.
\paragraph{Case \(s>1\).}
For \(y\ll x\), the height-difference distribution takes the form
\begin{align}
  P_{x,y}(\Delta)
  \sim
  \begin{cases}
    x^{\frac12-\alpha}y^{-\frac12},
    & |\Delta|\ll x^{\alpha-\frac12}y^{1/2},
    \\[2mm]
    x^{-\frac{s}{s+1}}\,y^{\frac{s}{s-1}}
    |\Delta|^{-\left(3+\frac{2}{s-1}\right)},
    & x^{\alpha-\frac12}y^{1/2}\ll |\Delta|\ll y^\alpha,
  \end{cases}
  \label{eq:distincrRW-summary}
\end{align}
followed by a stretched-exponential cutoff at \(|\Delta|\sim y^\alpha\). The power-law comes from the large increments when $h$ is small, and is observed numerically, with $x \sim L, y \sim \ell$, in the right panel of Fig.~\ref{fig:compRWPME}, as well as the large $\Delta$ cutoff $\sim \ell^\alpha$ (Inset).

\paragraph{Case \(0<s<1\).}
The large $\Delta$ behavior is now given by a stretched exponential:
\begin{align}
  P_{x,y}(\Delta)
  \sim
  \begin{cases}
    x^{\frac12-\alpha}y^{-\frac12},
    & |\Delta|\ll x^{\alpha-\frac12}y^{1/2},
    \\[2mm]
    x^{-\frac{s(1-s)}{2(s+1)}}\,y^{-s/2}\,
    |\Delta|^{-(1-s)}
    \exp\!\left[
      -b_s
      \left|\frac{\Delta}{x^{\alpha-\frac12}y^{1/2}}\right|^{s+1}
    \right],
    & |\Delta|\gg x^{\alpha-\frac12}y^{1/2},
  \end{cases}
  \label{eq:distincRW2-summary}
\end{align}
with
\begin{equation}
b_s=\frac{2^{1-s}c}{(s+1)(1-s^2)^{(1-s)/2}}.
\end{equation}
Thus the increment distribution always has a central region, which is responsible for \(\alpha_{\rm loc}(2)=1/2\), but only for \(s>1\) does it develop a power-law tail. For \(0<s<1\), the stretched-exponential decay suppresses large enough fluctuations for all moments to remain controlled by the typical region, and there is therefore no multiscaling.

{\bf Moments and multiscaling}

The multiscaling exponents follow directly from the height-difference distributions.

\emph{Case \(0<s<1\).}

Since the increment distribution has a stretched-exponential tail, all moments are controlled by the central region:
\begin{equation}
\langle |\Delta|^q\rangle
\simeq
c_q\,y^{q/2}x^{q(\alpha-\frac12)},
\end{equation}
and, using the definition $\langle |\Delta|^q\rangle^{1/q}
\sim
y^{\alpha_{\rm loc}(q)}x^{\alpha-\alpha_{\rm loc}(q)}$, one has
\begin{equation}
\alpha_{\rm loc}(q)=\frac12
\qquad
\text{for all } q>0.
\end{equation}
There is therefore no multiscaling for \(0<s<1\).

\emph{Case \(s>1\).}

There are two regimes in \(q\).

For \(q<q_c\), the moments are dominated by the central region of the distribution, and one finds
\begin{equation}
\langle |\Delta|^q\rangle
\simeq
c_q\,y^{q/2}x^{q(\alpha-\frac12)}.
\qquad
q<q_c=2+\frac{2}{s-1},
\end{equation}
Since the tail exponent of the distribution is \(3+2/(s-1)\), this moment diverges as \(q\to q_c^{-}\).

For \(q>q_c\), the moment is controlled by the upper cutoff \(|\Delta|\sim y^\alpha\), and one obtains
\begin{equation}
\langle |\Delta|^q\rangle
\sim
y^{q\alpha+\frac{s}{s+1}}x^{-\frac{s}{s+1}},
\qquad
q>q_c.
\end{equation}
This yields
\begin{equation}
\alpha_{\rm loc}(q)
=
\begin{cases}
\frac12, & q<q_c,
\\[2mm]
\alpha+\dfrac{1}{q}\dfrac{s}{1+s}, & q>q_c,
\end{cases}
\qquad
s>1.
\end{equation}

\section{Functional RG approach}
\label{sec:spme-FRG}

We now turn to the functional RG approach, which provides an alternative characterization of the stochastic porous medium equation. In order to perform the RG approach, we consider the stochastic porous medium equation in general dimension $d$. We begin by recalling the standard method and the main previous results. The derivation of the action functional, the analysis of divergences, and the resulting RG treatment were pioneered in \cite{antonov1995quantum}. We therefore only sketch the main ideas here, and refer the interested reader to that work for details, as well as to the later paper~\cite{antonov2017scaling}, which considers an anisotropic version of the model in the context of erosion.

In \cite{antonov1995quantum}, the one-loop correction was used to study the RG flow only for the first few coefficients in the Taylor expansion of $V(h)$. However, identifying all possible fixed points requires treating the full functional flow of $V(h)$, which we derive here.

We then analyze the fixed point relevant for the case considered here, $D(h)\sim |h|^{s-1}$, and find an associated roughness exponent
\begin{equation}
\alpha = \frac{2-d}{1+s},
\end{equation}
which in $d=1$ reduces to
\begin{equation}
\alpha = \frac{1}{1+s},
\end{equation}
in agreement with the result obtained from the random walk mapping. Finally, we study the stability of these fixed points and find that they are attractive.

\begin{figure}  \label{fig:flowz}
    \centering
    \includegraphics[width=0.58\textwidth]{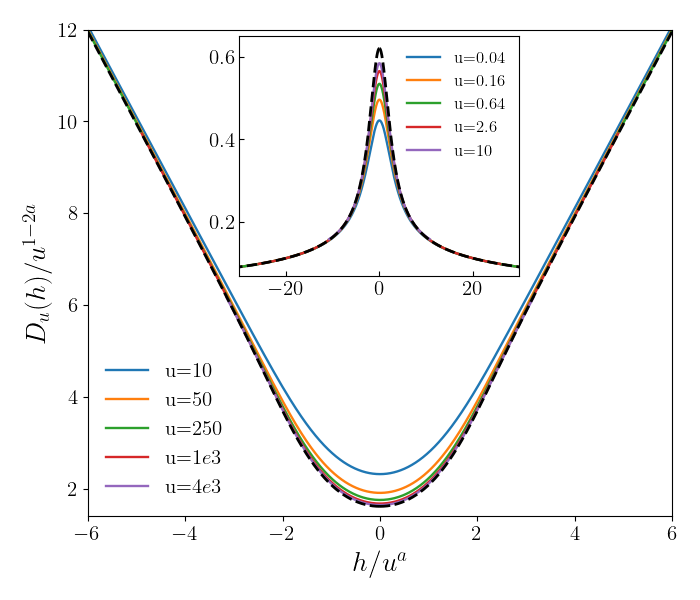}
    \caption{Numerical solution of Eq.~\eqref{RG2} illustrating the collapse of $D_u(h)$. Main panel, $s>1$: starting from $D_0(h) =2 \sqrt{4 +h^2}$, which corresponds to $s=2$ and $a=1/(1+s)=1/3$, the flow converges toward a self-similar fixed point, shown by the black dashed lines, which corresponds to $F(H)$ (with $b^2=1.612$) calculated numerically from \eqref{eqn:FF2-2}. %
    Inset, $s<1$: Starting from $D_0(h) = \frac{1}{2}(1/4^4+h^2)^{-1/4}$%
    corresponding to $s=0.5$ and $a=2/3$. The dotted line corresponds to the predicted fixed point (with $b^2=0.62$).%
    }
    \label{fig:figRG}
\end{figure}

\subsection{Derivation of the RG flow of $V(h)$}

We study the isotropic stochastic porous medium equation in space dimension $d$ 
\be \label{eqmo1}
\partial_t h(x,t) =  \nabla_x^2  V_0(h(x,t)) + \eta(x,t), %
\ee
where $\eta(x,t)$ is a Gaussian white noise with zero mean and variance
\be
\left\langle \eta(x,t) \eta(x',t')\right\rangle  = 2\sigma_0 \delta(x-x')\delta(t-t').
\ee
Here we use the subscript zero to denote the bare quantities.
We use the dynamical action, obtained by enforcing
the equation of motion using an auxiliary response field $\hat h=- i \tilde h$, and averaging over the noise.
As a result, the average of any observable
${\cal O}[h_s]$ of the solution $h=h_s$ of \eqref{eqmo1} over the noise can be written as a path integral
\be
\langle {\cal O}[h_s] \rangle = \int {\cal D} h {\cal D} \tilde h \, {\cal O}[h] \, e^{- S[\tilde h, h] }
\ee
where the dynamical action reads
\be \label{dynaction}
S[\tilde h, h] = \int dt d^d x  [ \tilde h (\partial_t h - \nabla_x^2 V_0(h))  - \sigma_0 \tilde h^2 ]
\ee
and one may include the initial conditions in the definition of the path integral if needed. An infrared cutoff $\mu$
is implicit everywhere to regularize the system at large distance (alternatively one may use a finite system size $L \sim 1/\mu$).
Splitting $V_0(h)=\nu_0 h + U_0(h)$, where $\nu_0$ is the bare diffusivity  and $U_0(h)$ can be treated as the non-linear interaction, one sees from the quadratic part of the action
that the bare dimensions of the fields are $h=L^{(2-d)/2}$ and $\tilde h=L^{- (d+2)/2}$, with scaling $t \sim x^{z_0}$ with $z_0=2$, and that
the non-linear part $U_0(h)$ becomes relevant for $d \leq 2$. One thus expects logarithmic divergences in
perturbation theory at the upper critical dimension $d=d_c=2$. Since the field $h$ is dimensionless
there one expects the need for functional RG to describe the flow of the full function $V_0(h)$.

One proceeds by computing the effective action functional $\Gamma[\phi] = \Gamma[\tilde h,h]$ associated
to the action $S[\phi] = S[\tilde h, h]$, using the two component field notation $\phi=(\tilde h,h)$
whenever convenient. Let us recall that one first defines
the partition sum in presence of sources, $Z[j]= \int {\cal D}\phi e^{- S[\phi] + \int dt d^d x j(x) \cdot \phi(x)}$.
Then $W[j]=\log Z[j]$ is the generating function of the connected correlations of the field (obtained by taking successive derivatives w.r.t $j$).
$\Gamma[\phi]$ is then defined as the Legendre transform of $W[j]$,
i.e. $\Gamma[\phi]= \int dt d^d x j(x) \cdot \phi(x) - W[j]|_{W'[j]=\phi}$.
Its important property is that
connected correlations of $\phi$ are now tree diagrams obtained from $\Gamma[\phi]$. Hence
the vertices of $\Gamma[\phi]$ are the dressed (or renormalized) vertices which sum up all loops.
Decomposing $S=S_0+S_{\rm int}$ where $S_0$ is the part of the action which is quadratic in the fields,
it can be computed from the perturbative formula
\be \label{eq:loop-expansion}
\Gamma[\phi]= S_0[\phi] + \sum_{p \geq 1} \frac{(-1)^{p+1}}{p!} \langle \left( S_{\rm int}[\phi + \delta \phi] \right)^p \rangle^{1PI}_{\delta \phi, S_0}
\ee
where $\langle \dots \rangle^{1PI}_{\delta \phi, S_0}$ denotes an average over the field $\delta \phi$ (at fixed background field $\phi$)
keeping only 1-particle irreducible diagrams (i.e. connected diagrams in the vertices $S_{\rm int}$ which
cannot be disconnected by cutting a propagator from $S_0$). The RG analysis then proceeds by
considering the leading dependence of $\Gamma[\phi]$ on the IR cutoff $\mu$ as $\mu \to 0$,
which is related to the analysis of the divergences in the 1PI diagrams (loop diagrams).

In \cite{antonov1995quantum}, it was found that to one loop it takes the same form as the action \eqref{dynaction}
\be
\Gamma[\tilde h, h] = \int dt d^d x  [ \tilde h (\partial_t h - \nabla_x^2 V(h))  - \sigma \tilde h^2 ] + \dots
\ee

To see this, it is convenient to rewrite the interaction term by integration by parts:
\begin{equation}
\int dt\,d^dx\, \tilde h\,\nabla_x^2 V_0(h)
=
\int dt\,d^dx\, V_0(h)\,\nabla_x^2 \tilde h ,
\label{eq:ibpV}
\end{equation}
up to boundary terms. In this form, one sees immediately that the nonlinear part of the action always involves spatial derivatives acting on the response field \(\tilde h\). Therefore the interaction does not couple to the uniform mode of \(\tilde h\), and loop corrections cannot generate terms proportional to \(\tilde h\) or \(\tilde h^2\) without derivatives. In particular, the coefficients of \(\tilde h\,\partial_t h\) and \(\tilde h^2\) are not renormalized. Hence no field renormalization is needed, \(\sigma=\sigma_0\) (which we set to \(1\) below), and the only quantity that flows is the potential \(V(h)\).

As a result, to one loop the effective action keeps the same form as the bare action, up to higher-gradient terms which are irrelevant near \(d=2\):
\begin{equation}
\Gamma[\tilde h,h]
=
\int dt\,d^dx\,
\left[
\tilde h\bigl(\partial_t h-\nabla_x^2 V(h)\bigr)- \tilde h^2
\right]
+\ldots
\label{eq:renorgamma}
\end{equation}

Near \(d=2-\epsilon\), it has been found \cite{antonov1995quantum}
\begin{equation}
\tilde\Gamma_1(\tilde h,h)
=
- A_d\frac{\mu^{-\epsilon}}{\epsilon}
\int dt\,d^dx\,
\frac{V_0''(h)}{V_0'(h)}\,\nabla_x^2 \tilde h,
\label{eq:Gamma1}
\end{equation}
where \(\mu\) is the renormalization scale and \(A_2=1/(4\pi)\).

In \cite{antonov1995quantum}, the correction \eqref{eq:Gamma1} was used to study the flow of the first few coefficients in the Taylor expansion of \(V(h)\). In the present work, by contrast, we keep the full functional dependence of \(V(h)\), which is necessary in order to identify all possible fixed points.

Since only $V(h)$ is renormalized, the divergent part of the one-loop correction is therefore directly identified with a renormalization of \(V(h)\).
The one-loop counterterm may thus be written as
\begin{equation}
\delta V(h)
=
\frac12 \int \frac{d\omega}{2\pi}\frac{d^dk}{(2\pi)^d}
\frac{2V_0''(h)}{D_0^2(h)(k^2+\mu^2)^2+\omega^2}
=
A_d\frac{\mu^{-\epsilon}}{\epsilon}\frac{V_0''(h)}{V_0'(h)}
+O(\epsilon^0),
\label{eq:deltaV1loop}
\end{equation}

Taking a derivative w.r.t. $\mu$ at fixed $V_0$, and then replacing $V_0 \to V$, we obtain
the RG flow for the full potential $V(h)$ to leading order in $\epsilon$ as
\begin{equation}\label{aa2}
- \mu \partial_\mu V(h)  = \frac{1}{4 \pi} \mu^{-\epsilon} \frac{V''(h)}{V'(h)}.
\end{equation}
The general exact RG method was developed in \cite{wetterich1993exact,morris1994exact} and
leads to an exact RG equation for the functional $\Gamma[\phi]$ (see also Section 3 in \cite{balents2005thermal}
for a simplified summary).
Arriving at \eqref{aa2} requires
truncations, which can be performed in a systematic way in a multilocal expansion \cite{schehr2003exact,le2010exact},
and within a controlled dimensional expansion near $d=2$. These truncations can be performed more
generally within the non-perturbative FRG approach, which was applied to the anisotropic version of the model (erosion)
in
\cite{duclut2017nonuniversality,DuclutThesis}. Here we consider perturbative FRG applied to the isotropic model.
\\

\paragraph{Critical exponents}
The critical exponents are associated to scale invariant solutions of the RG flow.
Let us first examine the form that these solutions can take. To analyze \eqref{aa2} let us replace the
variable $\mu$ by the variable
\be
u=\frac{1}{4 \pi} \int_{\mu}^{\mu_0} d\mu'/(\mu')^{1+\epsilon},
\ee
where $\mu_0$ is the bare IR cutoff, and we are interested in the limit $\mu \to 0$. In that
limit we see that $u \sim \mu^{-\epsilon}/\epsilon$ for $\epsilon>0$,
and $u= \frac{1}{4 \pi} \log(\mu_0/\mu)$ in $d=2$. As we pointed out in the main text,
the {running} derivative of the potential, which we denote $D_u(h)=V'(h)$ (indicating the explicit $u$ dependence)
satisfies (taking $\partial_h$ in \eqref{aa2}) a logarithmic PME
\be \label{RG2}
\partial_u D_u(h) = \partial_h^2 \log D_u(h)
\ee
This equation admits  self-similar solutions of the form
\be \label{eqn:scaleW}
D_u(h) = u^{1- 2a} F( u^{-a} h),
\ee
where $F$ is a fixed function with $F(0)$ finite, and the exponent $a$ takes some given value (depending on $s$, see below).

Recalling that $u \sim \mu^{-\epsilon}$ where $\mu \sim 1/L$ is the IR cutoff, we have $u \sim L^\epsilon$
where $L$ is the system size. %
{The form of the fixed point solution~\eqref{eqn:scaleW} implies the scaling behavior}
\be   \label{eq:roughness-1}
h \sim u^a \sim L^\alpha \quad   \text{with} \quad \alpha = a \epsilon
\ee

Note that the relation $z = 2 - \epsilon(1- 2 a) = d + 2 \alpha$ can also be recovered from the FRG upon studying the renormalization of the two point height correlator. (See Sup.Mat. I of \Ptwo)

\subsection{Functional RG flow: fixed points, linear stability, and basin of attraction}
\label{sec:RG-flow}

In this section, we identify the RG fixed point relevant to the stochastic porous medium equation when
\begin{equation}
D(h)\sim |h|^{s-1},
\label{eq:RG-tail-D}
\end{equation}
and show, both analytically and numerically, that it is attractive. The corresponding fixed point is characterized by $a=\frac{1}{1+s}$,
which, in \(d=1\), reproduces the random walk result
\begin{equation}
\alpha=\frac{1}{1+s}.
\label{eq:RG-alpha-s}
\end{equation}
The other critical exponents then follow from the scaling relation \(z=d+2\alpha\). The full classification of fixed points, together with the analysis of their eigenvalues, eigenvectors, and basins of attraction, is given in \Ptwo. Here we only summarize the solutions relevant to the stochastic PME with \(D(h)\sim |h|^{s-1}\), together with the main ideas of the stability/basin of attraction analysis.

\paragraph{Fixed point equation.}
We focus on self-similar solutions of the RG flow of the form \eqref{eqn:scaleW},
\begin{equation}
D_u(h)=u^{1-2a}F_u(H),
\qquad
H=u^{-a}h,
\end{equation}
and look for profiles \(F_u(H)\) that converge to a \(u\)-independent limit \(F(H)\) as \(u\to\infty\). In terms of \(F_u\), the RG flow becomes
\begin{equation}
u\,\partial_u F_u(H)
=
aH F_u'(H)
-(1-2a)F_u(H)
+\frac{F_u''(H)}{F_u(H)}
-\frac{F_u'(H)^2}{F_u(H)^2},
\label{eqn:FF2}
\end{equation}
where primes denote derivatives with respect to \(H\). The fixed-point equation is therefore
\begin{equation}
aH F'(H)
-(1-2a)F(H)
+\frac{F''(H)}{F(H)}
-\frac{F'(H)^2}{F(H)^2}
=0.
\label{eqn:FF2-2}
\end{equation}
If \(F(H)\) is a solution, then so is \(b^2F(bH)\) for any \(b\), so this scaling freedom may be fixed by choosing \(F(0)=1\).

\paragraph{Fixed point relevant to the stochastic PME.}
The full classification of self-similar solutions of \eqref{eqn:FF2-2}, based on a phase-plane analysis, is given in \Ptwo. Here we focus on the branch relevant to the stochastic PME with
\begin{equation}
V(h)\sim h|h|^{s-1}.
\label{eq:RG-tail-V}
\end{equation}
For \(0<a<1\), there exists an even, positive fixed-point profile \(F(H)\), regular at the origin, whose large-\(|H|\) behavior is
\begin{equation}
F(H)\sim |H|^{-2+1/a}
\qquad (|H|\to\infty).
\label{eq:RG-F-tail}
\end{equation}
For
\begin{equation}
a=\frac{1}{1+s},
\label{eq:RG-a-s}
\end{equation}
this becomes
\begin{equation}
F(H)\sim |H|^{s-1},
\label{eq:RG-F-tail-s}
\end{equation}
which is precisely the asymptotic behavior required to match the initial condition \(D_0(h)\sim |h|^{s-1}\). This is therefore the candidate fixed point for the stochastic PME.

\paragraph{Basin of attraction.}
One must still show that this fixed point is not merely a formal self-similar solution of the RG flow, but a true attractor for the class of initial conditions \(V_0(h)\sim h|h|^{s-1}\). To test this numerically, one integrates the RG flow
\begin{equation}
\partial_u D_u(h)=\partial_h^2 \log D_u(h),
\label{eq:eqflowSM}
\end{equation}
on a large but finite grid in \(h\), using a discrete Laplacian and explicit time stepping in the RG time \(u\), starting from an initial profile such that \(D_0(h)\sim |h|^{s-1}\).

In the regime of interest, namely \(0<a<1\), or equivalently \(s>0\), the numerics show that the large-\(|h|\) tail of the initial condition is preserved by the flow, and that the rescaled profile \(F_u(H)\) converges to the fixed-point function \(F(H)\) with
\begin{equation}
a=\frac{1}{1+s}.
\label{eq:RG-a-s-basin}
\end{equation}
This convergence is illustrated in Fig.~\ref{fig:figRG}.

\paragraph{Linear stability.}
To characterize the approach to the fixed point more precisely, one considers a perturbation of the form
\begin{equation}
D_u(h)=u^{1-2a}\bigl(F(H)+\delta F_u(H)\bigr),
\qquad
H=u^{-a}h,
\qquad
\delta F_u(H)=u^\lambda \Phi(H),
\label{eq:pertF}
\end{equation}
and linearizes the RG equation around \(F(H)\). This yields the eigenvalue problem
\begin{align}
0&=({\cal L}_\lambda\Phi)(H)\\
&=
\Phi''(H)
-\Phi'(H)\left(2\frac{F'(H)}{F(H)}-aHF(H)\right)
+\Phi(H)\left[aHF'(H)-(2-4a+\lambda)F(H)+\frac{F'(H)^2}{F(H)^2}\right].
\label{eqd2H2}
\end{align}
This equation is analyzed in detail in \Ptwo, both analytically and numerically. For initial conditions of the form
\begin{equation}
D_0(h)\simeq C_1 h^{s-1}+C_2 h^{s-1-\gamma},
\qquad
a=\frac{1}{1+s},
\label{eq:RG-initial-correction}
\end{equation}
the leading correction is governed by the largest eigenvalue of the linearized flow. In the regime of interest, this eigenvalue is either
\begin{equation}
\lambda_{\rm exp}=a-1,
\qquad\text{or}\qquad
\lambda_\gamma=-a\gamma=-\frac{\gamma}{s+1}.
\label{eq:RG-eigenvalues}
\end{equation}
The first is associated with a rapidly decaying eigenvector, while the second corresponds to a power-law perturbation,
\begin{equation}
\Phi(H)\sim H^{s-1-\gamma}
\qquad (H\to\infty).
\label{eq:RG-powerlaw-eigenvector}
\end{equation}

\begin{figure}
    \centering
    \includegraphics[width=0.48\linewidth]{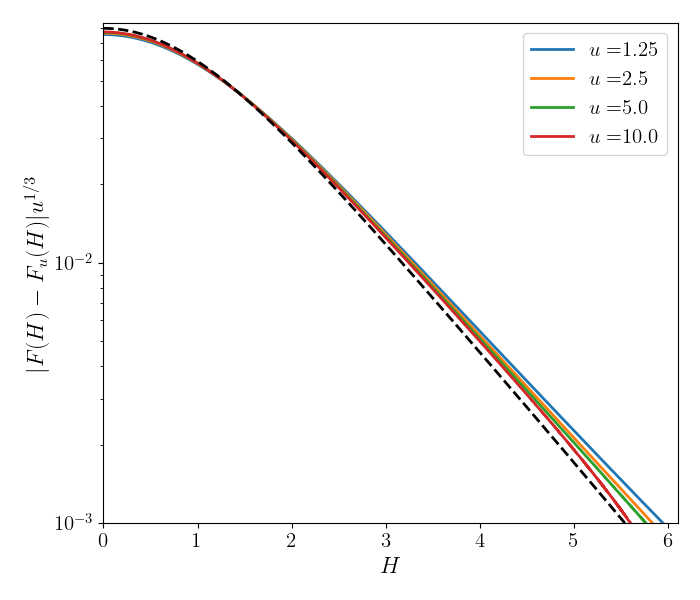}
    \includegraphics[width=0.48\linewidth]{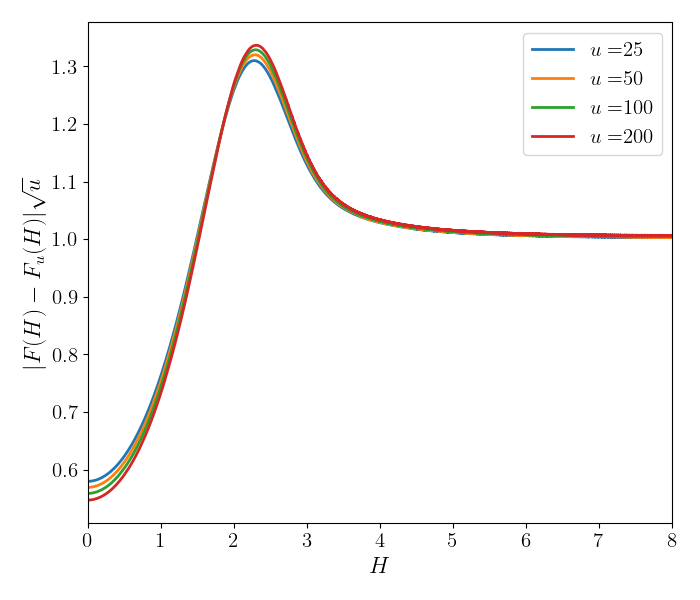}
    \caption{RG flow near the fixed point $|F(H)-F_u(H)|$ rescaled by $u^{\lambda}$, with $\lambda$ the predicted largest eigenvalue. %
    Left panel: starting from $D_0(h) = \left(1/4^4 + h^2 \right)^{-0.25}/2$ ($s=0.5$), with $\lambda=a-1=-1/3$, the corresponding eigenvector shown by the black dotted line decays exponentially. Right panel: starting from $D_0(h) = 1+h^2$ ($s=3$), one has $\lambda=-1/2$, with a corresponding eigenvector of order $O(1)$ when $H \to \infty$.}
    \label{fig:flowRGdiff}
\end{figure}

Since both eigenvalues are negative for \(0<a<1\), the fixed point is attractive. This is confirmed numerically in Fig.~\ref{fig:flowRGdiff}, which shows a clear collapse onto the scaling form. We thus conclude that, for initial conditions \(V_0(h)\sim h|h|^{s-1}\), the RG flow converges to the fixed point associated with
\begin{equation}
a=\frac{1}{1+s},
\label{eq:RG-a-s-conclusion}
\end{equation}
thereby recovering the roughness exponent \(\alpha=1/(1+s)\) obtained previously from the random walk mapping.

\clearpage

\chapter{Spinoff: One-Dimensional Localization in Random Chains}
\label{sec:interfaces-anderson}

\renewcommand{\K}{K}
\providecommand{\IDoS}{\mathcal{N}}
\providecommand{\EXP}[1]{\mathrm{e}^{#1}}
\providecommand{\eqdef}{\stackrel{\mathrm{def}}{=}}
\providecommand{\mean}[1]{\left\langle #1 \right\rangle}
\providecommand{\smean}[1]{\left\langle #1 \right\rangle}

\providecommand{\symmetriz}[1]{\left\lfloor #1 \right\rfloor}
\providecommand{\ssymmetriz}[1]{\lfloor #1\rfloor}

This chapter is devoted to the spectral problem that appeared earlier in Chapter~\ref{sec:interfaces-elastic}, when studying the heterogeneous elastic line. There, the dynamics could be rewritten in terms of a random tridiagonal matrix \(\Lambda\), and the interface observables were controlled by its lower spectrum. The latter, and more, can be obtained using tools from Anderson localization, as we detail below.
The results of this chapter are based on \Pthree.

\begin{figure}[!h]
\centering
\includegraphics[width=0.48\textwidth]{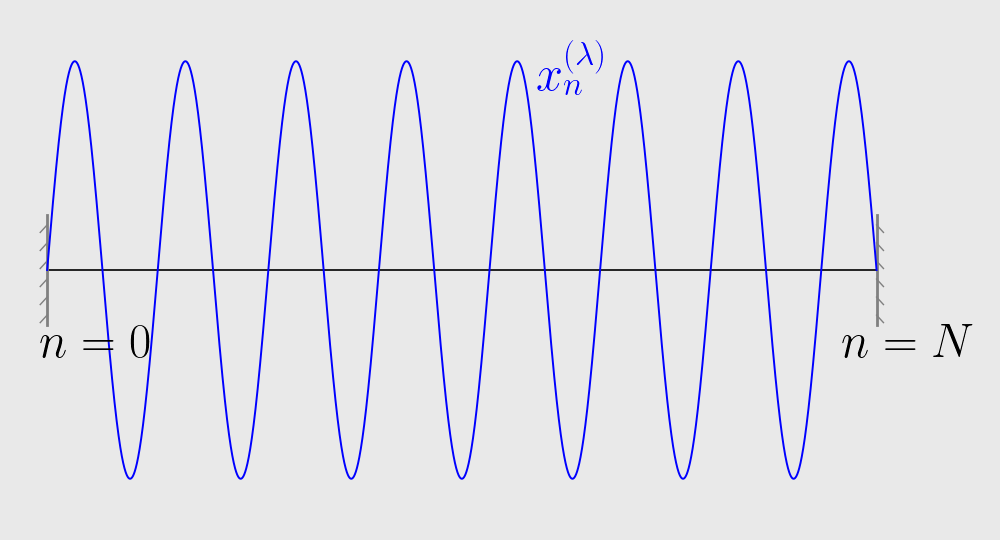}
\includegraphics[width=0.48\textwidth]{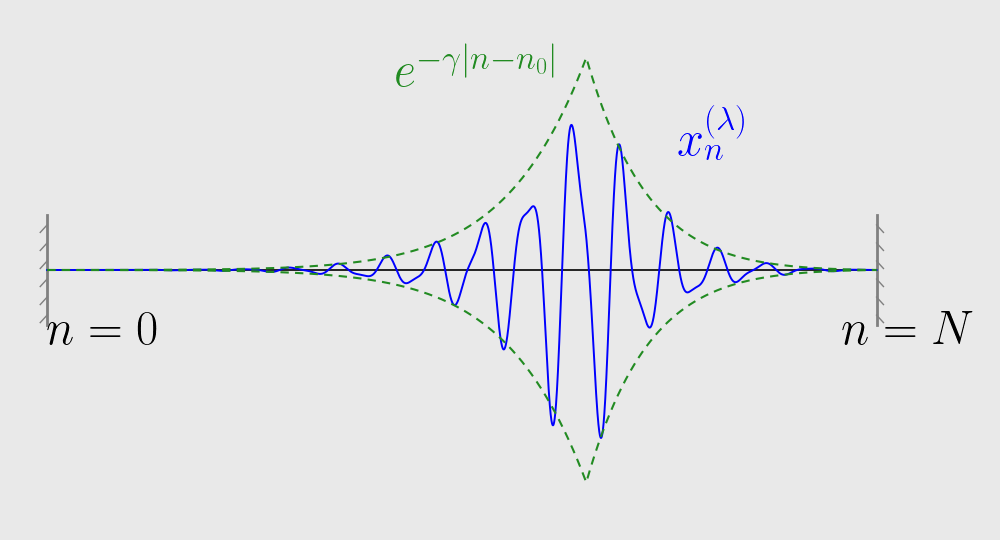}
\caption{Typical eigenmodes of the spring chain. Left: in the absence of disorder, translational invariance leads to extended modes. Right: with disorder, the eigenmodes become localized and decay exponentially fast away from a localization center.}
\label{fig:eigenvectors-anderson-spring}
\end{figure}

Anderson localization is the localization of wave excitations by static disorder. It was introduced by Anderson in the context of electrons moving in a crystal with impurities, where repeated scattering by a quenched random potential and the resulting interference may suppress diffusion \cite{anderson1958absence}. By contrast, a perfect crystal is translationally invariant and Bloch's theorem predicts extended eigenstates. Disorder breaks translation invariance and may instead produce states whose amplitude is concentrated in a finite region of space, with an exponential envelope characterized by a localization length. The two essential ingredients are therefore the wave nature of the excitation and quenched disorder. Typical examples of these two situations are shown in Fig.~\ref{fig:eigenvectors-anderson-spring}.

Consider, for instance, the Anderson model on a discrete lattice. The Schr\"odinger equation reads
\begin{equation}
E \psi_n = V_n \psi_n - \sum_m t_{nm}\psi_m,
\end{equation}
where \(\psi_n\) is the wavefunction amplitude at site \(n\), \(V_n\) the on-site potential, and \(t_{nm}\) the hopping amplitude between sites \(n\) and \(m\). In one dimension, with nearest-neighbor hopping, this becomes
\begin{equation}
E \psi_n = V_n \psi_n - t_{n} \psi_{n+1} - t_{n-1}\psi_{n-1}.
\end{equation}

The role of dimension is crucial: the lower the dimension, the easier it is to localize. In three dimensions, localized and extended states may coexist, separated by a mobility edge, and increasing the disorder strength can drive a metal-insulator transition \cite{mott1968mobility,kramer1993localization,evers2008anderson}. In contrast, for one-dimensional disordered systems, all states are localized. The main questions are then to determine the density of states and the localization length of the corresponding eigenmodes.

Another important setting for wave localization is provided by elastic systems such as random spring chains \cite{lieb_mattis_1966}. For the spring chain with random masses, one starts from Newton's equation of motion
\begin{equation}
\label{eq:NewtonSpringChain}
m_n \ddot X_n(t)
=
K_{n-1}\bigl(X_{n-1}(t)-X_n(t)\bigr)
+
K_n\bigl(X_{n+1}(t)-X_n(t)\bigr).
\end{equation}
Looking at the modes \(X_n(t)=x_n\,\EXP{-\I\omega t}\), and introducing
\begin{equation}
\lambda=\omega^2,
\end{equation}
one obtains the spectral equation
\begin{equation}
\label{eq:SpringEquationMasses}
-m_n \lambda x_n
=
K_{n-1}(x_{n-1}-x_n)+K_n(x_{n+1}-x_n).
\end{equation}
For uniform masses $m_n=1$, this reduces to the eigenvalue problem of the operator \(\Lambda\) encountered in the elastic-line chapter,
\begin{equation}
\label{eq:SpringEquation}
-\lambda x_n
=
K_{n-1}(x_{n-1}-x_n)+K_n(x_{n+1}-x_n).
\end{equation}
Equivalently,
\begin{equation}
(K_{n-1}+K_n-\lambda)x_n-K_nx_{n+1}-K_{n-1}x_{n-1}=0,
\end{equation}
which has exactly the form of a one-dimensional Anderson equation with random hoppings
\begin{equation}
t_n=K_n,
\end{equation}
and diagonal term
\begin{equation}
V_n-E=K_{n-1}+K_n-\lambda.
\end{equation}
The spring-chain problem is therefore an Anderson-type problem with diagonal and off-diagonal disorder (within this picture, note that the diagonal terms $V_n$ and $V_{n+1}$ are correlated).

The first quantity we study is the density of states
\begin{equation}
  \rho(\lambda)
  \eqdef
  \lim_{N\to\infty}\frac{1}{N}\sum_{\alpha=1}^N \delta(\lambda-\lambda_\alpha),
\end{equation}
and its integrated DoS
\begin{equation}
  \IDoS(\lambda)
  \eqdef
  \int_0^\lambda \D\lambda'\,\rho(\lambda').
\end{equation}
These are often written in terms of frequencies \(\lambda=\omega^2\),
\begin{equation}
  \varrho(\omega)=2\omega\,\rho(\omega^2).
\end{equation}

\begin{figure}[!h]
\centering
\includegraphics[width=0.65\textwidth]{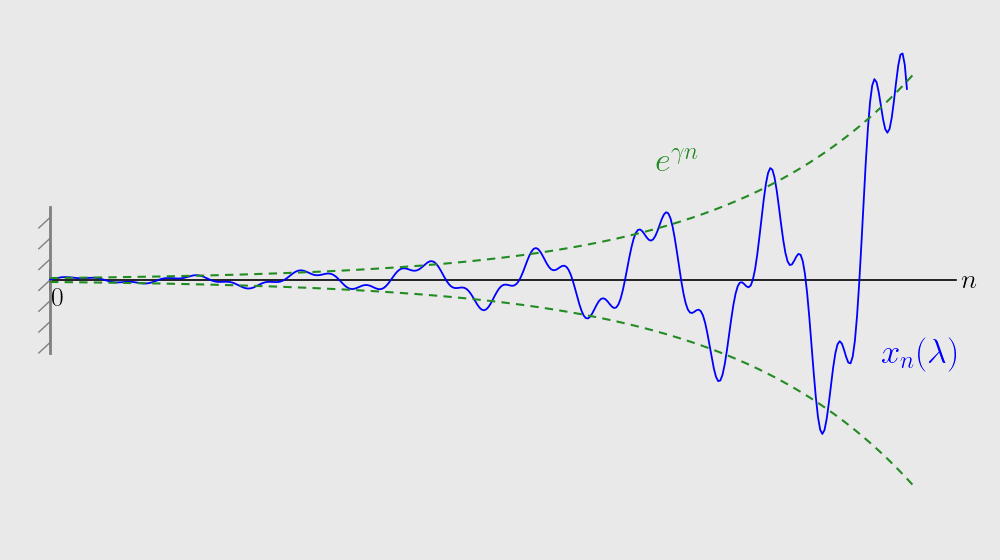}
\caption{A typical solution of the initial-value problem for a fixed \(\lambda\). The sequence \(x_n\) may change sign and cross zero, while its envelope grows exponentially as \(e^{\gamma(\lambda)n}\). This growth rate is the quantity used to define the Lyapunov exponent.}
\label{fig:gammaenv}
\end{figure}

The second central quantity is the localization length. While the eigenvectors are obtained by solving the spectral problem with boundary conditions, one usually accesses localization through the initial-value problem. Starting from fixed initial conditions, one studies the growth of the envelope of the corresponding solution \(x_n\), as illustrated in Fig.~\ref{fig:gammaenv}. If we define the Lyapunov exponent of the corresponding solution by
\begin{equation}
\gamma(\lambda)=\lim_{n\to\infty}\frac{1}{n}\ln |x_n(\lambda)|,
\end{equation}
then its inverse provides a measure of localization length
\begin{equation}
\xi_\lambda=\frac{1}{\gamma(\lambda)}.
\end{equation}

Note, however, that \(\xi_\lambda\) and \(\gamma(\lambda)\) characterize the solution of the initial-value problem rather than the solution of the spectral problem with Dirichlet boundary conditions. The assumption that \(\xi_\lambda\) also characterizes the true eigenmodes is known as the Borland conjecture \cite{Borland1963}. While standard in one-dimensional localization, it may fail in some singular situations \cite{Texier_2010}.

In this chapter, we study these two quantities for \eqref{eq:SpringEquationMasses}, where \(m_n\) and \(K_n\) are i.i.d.\ random variables. The results required for Chapter~\ref{sec:interfaces-elastic} on the heterogeneous elastic line are then recovered as the special case of uniform masses.

\section{Main Results}
\label{sec:anderson-main-results}

The central quantity in our approach is the complex Lyapunov exponent
\begin{equation}
  \Omega(\lambda)=\lim_{n\to\infty}\frac{1}{n}\ln x_n(\lambda).
\end{equation}
Once it is known, both observables of interest are obtained from
\begin{equation}
  \Omega(\lambda+\I0^+)=\gamma(\lambda)-\I\pi\,\IDoS(\lambda),
\end{equation}
so that its real part gives the Lyapunov exponent and its imaginary part gives the integrated density of states, as we show in Section~\ref{sec:anderson-ds}. The main result of the chapter is the following general formula for \(\Omega(\lambda)\):
\begin{equation}
  \boxed{
  \Omega(\lambda)
  =
  -\min_{\theta\in[0,1]}
  \int_0^\theta \D t\,
  \ln\!\left[
    \frac{4}{-\lambda}\,
    \frac{1-t}{1+t}\,
    \varphi_*^{(1/\K)}(t)\,
    \varphi_*^{(m)}(t)
  \right]}
  \label{eq:MainResultForOmega}
\end{equation}
where the two "fugacities" are defined implicitly by
\begin{equation}
  \varphi_*^{(1/\K)}(\theta)\,
  \mean{
    \frac{\K^{-1}}{1+\varphi_*^{(1/\K)}(\theta)\,\K^{-1}}
  }
  =
  \theta,
\end{equation}
and
\begin{equation}
  \varphi_*^{(m)}(\theta)\,
  \mean{
    \frac{m}{1+\varphi_*^{(m)}(\theta)\,m}
  }
  =
  \theta.
\end{equation}
where \(\langle \cdots \rangle\) denotes averaging with respect to the corresponding disorder distribution.

Equation~\eqref{eq:MainResultForOmega} is not exact, as it relies on the symmetrization procedure described below. It nevertheless turns out to be a very good approximation, as confirmed by comparison with the exact result obtained by Nieuwenhuizen in a special case \cite{Nie84} and by numerical simulations in the low-frequency regime; see Fig.~\ref{fig:gamma-exp-benchmark-chap} (Left). It is also completely general: it applies to arbitrary disorder distributions and to any value of \(\lambda\), and reduces the spectral problem to the study of two scalar transcendental equations, one for the masses and one for the inverse spring constants. This makes it particularly convenient for deriving asymptotic behaviors.

Of particular interest is the behavior at low frequency, which usually displays universal properties. Generally, in this regime, both the spectral density and the Lyapunov exponent show power-law behavior:
\begin{equation}
  \gamma(\lambda)\sim \lambda^\zeta,
  \qquad
  \rho(\lambda)\sim \lambda^{\eta-1},
  \qquad
  \lambda\to0^+,
  \label{eq:main-results-lowfreq-laws}
\end{equation}
or, equivalently, in frequency variables, $\varrho(\omega)\sim \omega^{2\eta-1}$.

To probe this low-frequency behavior, we will consider the power-law distributions
\begin{equation}
  p(\K)=\mu\,\K^{-1+\mu},
  \qquad 0<\K<1,
\end{equation}
and
\begin{equation}
  q(m)=\nu\,m^{-1-\nu},
  \qquad m>1,
\end{equation}

The exponent \(\eta\) is controlled by the first moments of \(m\) and \(1/\K\), whereas \(\zeta\) is sensitive to the second moments as well. The full classification is summarized in Tables~\ref{tab:Eta} and \ref{tab:Zeta}.

\begin{table}[!h]
\centering
\begin{tabular}{cc|cc}
$\eta$ & & $\mu<1$ & $\mu>1$ \\
& & $\mean{\K^{-1}}=\infty$ & $\mean{\K^{-1}}<\infty$ \\
\hline
$\nu<1$ & $\mean{m}=\infty$ & $\big(\frac{1}{\mu}+\frac{1}{\nu}\big)^{-1}$ & $\frac{\nu}{1+\nu}$ \\
$\nu>1$ & $\mean{m}<\infty$ & $\frac{\mu}{1+\mu}$ & $\frac12$
\end{tabular}
\caption{Exponent \(\eta\) controlling the low-frequency spectral density.}
\label{tab:Eta}
\end{table}

\begin{table}[!h]
\centering
\begin{tabular}{c|ccc}
$\zeta$ & $0<\mu<1$ & $1<\mu<2$ & $\mu>2$ \\
\hline
$0<\nu<1$ & $\big(\frac{1}{\mu}+\frac{1}{\nu}\big)^{-1}$ & $\big(1+\frac{1}{\nu}\big)^{-1}$ & $\big(1+\frac{1}{\nu}\big)^{-1}$ \\
$1<\nu<2$ & $\big(1+\frac{1}{\mu}\big)^{-1}$ & $\frac12\min(\mu,\nu)$ & $\nu/2$ \\
$\nu>2$ & $\big(1+\frac{1}{\mu}\big)^{-1}$ & $\mu/2$ & $1$
\end{tabular}
\caption{Exponent \(\zeta\) controlling the low-frequency Lyapunov exponent.}
\label{tab:Zeta}
\end{table}

The spectral density exhibits two first-order phase transition lines, at \(\mu=1\) and \(\nu=1\), that is, at the boundary of the region where \(\mean{m}<\infty\) and \(\mean{\K^{-1}}<\infty\). The Lyapunov exponent also exhibits first-order phase transitions on these lines, together with additional ones at the boundaries of the region where \(\mean{m^2}<\infty\) and \(\mean{\K^{-2}}<\infty\), namely on \(\mu=2\) and \(\nu=2\). The exponents remain continuous across the transition lines, but logarithmic corrections appear on the marginal boundaries.
In several related statistical-mechanics problems, the Lyapunov exponent plays the role of a free-energy density, which makes the interpretation of these discontinuities in terms of first-order phase transitions natural.

\begin{figure}[!h]
\centering
\includegraphics[width=0.65\textwidth]{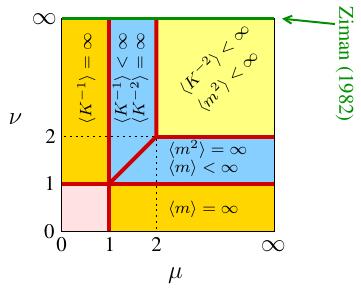}
\caption{Phase diagram for the low-frequency spectral and localization properties of the random mass-spring chain in the \((\mu,\nu)\) plane. The spectral exponent \(\eta\) changes when the first moments of \(m\) or \(1/\K\) diverge, whereas the localization exponent \(\zeta\) is additionally sensitive to the divergence of the second moments. On the marginal lines \(\mu=1\) and \(\nu=1\), the power laws are dressed by logarithmic corrections.}
\label{fig:lowfreq-summary-anderson}
\end{figure}

The phase diagram is shown in Fig.~\ref{fig:lowfreq-summary-anderson}. Weak-disorder expansions only describe the leading low-frequency behavior of the density of states and of the Lyapunov exponent in the region \(\mu>2\) and \(\nu>2\), where both second moments are finite. Outside this region, anomalous power laws appear. The green line corresponds to the special case studied by Ziman \cite{Zim82}, for which the exponent \(\zeta\) can be determined exactly.

Before presenting our approach, let us briefly summarize the existing results. Earlier works had established only partial aspects of the overall picture. For random masses with uniform springs, the low-frequency behavior was obtained only in the weak-disorder regime \cite{MatIsh70}. For random springs with uniform masses, Bernasconi and collaborators derived the low-frequency spectral behavior \cite{BerSchWys80,AleBerSchOrb81} while Ziman \cite{Zim82} obtained the behavior of the Lyapunov exponent in a related model. For the more difficult case in which both the spring and the masses are random, Nieuwenhuizen \cite{Nie84} identified an exactly solvable case in which both variables are exponentially distributed. Our method, by contrast, reveals the complete phase diagram: it recovers the power-law exponents in every regime and also the logarithmic corrections along the transition lines.

\section{The Spectral Problem and the Dyson-Schmidt Method}
\label{sec:anderson-ds}

There are two complementary routes to analyze this problem. On the one hand, one may use the standard one-dimensional localization machinery, based on transfer matrices and Riccati variables, which relates the density of states and the Lyapunov exponent to the stationary distribution of an associated Riccati process \cite{luck1992systemes,lifshitz_gredeskul_pastur_1988}. On the other hand, we follow a different route, starting from the finite-size spectral determinant and reorganizing it combinatorially. In order to make clear what is standard and what is specific to our contribution, we first recall the Dyson-Schmidt method, before turning to our new combinatorial approach.

We impose the initial conditions
\begin{equation}
  x_0 = 0,
  \qquad
  x_1 = 1,
\end{equation}
corresponding to an elastic line pinned at $n=0$. By linearity, the value of $x_1$ is arbitrary.
The relation \eqref{eq:SpringEquation} can be reorganized as a second-order recurrence
\begin{equation}
  \label{eq:TheRecurrence}
  x_{n+1}(\lambda)
  =
  \left(
    1+\frac{\K_{n-1}-m_n\lambda}{\K_n}
  \right)x_n(\lambda)
  -
  \frac{\K_{n-1}}{\K_n}\,x_{n-1}(\lambda).
\end{equation}
The eigenvalues are then fixed by the quantization condition, which for Dirichlet (pinned line) boundary conditions simply reads
\begin{equation}
  x_{N+1}(\lambda)=0.
\end{equation}
Hence the zeros of the polynomial $x_{N+1}(\lambda)$ are precisely the eigenvalues $\lambda_\alpha=w_\alpha^2$ of the disordered chain.
One can in particular rewrite
\begin{equation}
  x_{N+1}(\lambda)
  =
  \prod_{j=1}^N \frac{m_j}{\K_j}
  \prod_{\alpha=1}^N(\lambda_\alpha-\lambda),
\end{equation}
where the prefactor is obtained by inspection of \eqref{eq:TheRecurrence}, which shows that the highest-degree term is \(\prod_j (-m_j\lambda/K_j)\propto \lambda^N\). This shows that \(x_{N+1}(\lambda)\) has the structure of a spectral determinant.
For \(\lambda\notin\mathrm{Spectrum}\), one introduces the complex Lyapunov exponent
\begin{equation}
  \Omega(\lambda)
  =
  \lim_{N\to\infty}\frac{1}{N}\ln x_{N+1}(\lambda)
\end{equation}
which obeys
\begin{equation}
  \Omega(\lambda)
  =
  \mean{\ln(m_j/\K_j)}
  +
  \int \D\lambda'\,\rho(\lambda')\ln(\lambda'-\lambda).
\end{equation}
Approaching the real axis from the upper half-plane gives
\begin{equation}
  \Omega(\lambda+\I0^+)=\gamma(\lambda)-\I\pi\,\IDoS(\lambda),
\end{equation}

so that the imaginary part of \(\Omega\) is the integrated density of states, while its real part, \(\gamma(\lambda)=\lim_{N \to +\infty}\frac{1}{N}\ln|x_N|\), is the Lyapunov exponent.

The standard way to extract these observables is the Dyson-Schmidt method, also known in the mathematical literature as the Furstenberg approach to products of random matrices \cite{Dys53,Sch57,luck1992systemes}. One introduces the Riccati variable
\begin{equation}
  z_n \eqdef \frac{x_n}{x_{n-1}},
\end{equation}
which transforms the second-order recurrence into a non-linear first-order one:
\begin{equation}
  \label{eq:BadRiccati}
  z_{n+1}
  =
  1+\frac{\K_{n-1}-m_n\lambda}{\K_n}
  -
  \frac{\K_{n-1}}{\K_n}\,\frac{1}{z_n}.
\end{equation}
This is a M\"obius transformation,
\begin{equation}
  z_{n+1}=\mathcal{M}_n(z_n),
  \qquad
  \mathcal{M}_n(z)=\frac{a_n z+b_n}{c_n z+d_n},
\end{equation}
with matrix representative
\begin{equation}
  M_n=
  \begin{pmatrix}
    1+\dfrac{\K_{n-1}-m_n\lambda}{\K_n} & -\dfrac{\K_{n-1}}{\K_n}\\[1ex]
    1 & 0
  \end{pmatrix}.
\end{equation}
Since the composition of two M\"obius transformations is again a M\"obius transformation, with matrix representative given by the product of the corresponding matrices, the localization problem can be reformulated as the study of a product of random \(2\times2\) matrices acting on the Riccati variable.

Since
\begin{equation}
  x_{N+1}=x_1\prod_{n=2}^{N+1} z_n=\prod_{n=2}^{N+1} z_n,
\end{equation}
one finds for real $\lambda$
\begin{equation}
  \Omega(\lambda+\I0^+)
  =
  \lim_{N\to\infty}\frac{1}{N}\sum_{n=2}^{N+1}
  \left[
    \ln|z_n|-\I\pi\,\theta(-z_n)
  \right].
\end{equation}
Assuming the distribution of $z_n$ converges to a stationary law $f(z)$, this gives the two basic formulas
\begin{equation}
  \label{eq:GammaIdosRiccati}
  \gamma(\lambda)=\int_{\mathbb R}\D z\,f(z)\ln|z|,
  \qquad
  \IDoS(\lambda)=\int_{-\infty}^0 \D z\,f(z).
\end{equation}
Thus, whenever $z_n<0$, the ratio $x_n/x_{n-1}$ is negative, the solution changes sign, and one more eigenvalue has been crossed. In other words, node counting becomes a statement about the fraction of time the Riccati variable spends on the negative half-line.

The remaining task is therefore to determine the stationary distribution of the Riccati process. For a general M\"obius transformation
\begin{equation}
  \mathcal{M}(z)=\frac{az+b}{cz+d},
\end{equation}
the density $f_n(z)$ evolves according to
\begin{equation}
  f_{n+1}(z)
  =
  \left\langle
    \frac{\D \mathcal{M}^{-1}(z)}{\D z}\,
    f_n\!\big(\mathcal{M}^{-1}(z)\big)
  \right\rangle_{\mathcal{M}}.
\end{equation}
In the stationary limit, this becomes the integral equation \cite{Tex20}
\begin{equation}
  \label{eq:DysonSchmidtInvariant}
  f(z)
  =
  \left\langle
    \frac{\D \mathcal{M}^{-1}(z)}{\D z}\,
    f\!\big(\mathcal{M}^{-1}(z)\big)
  \right\rangle_{\mathcal{M}}.
\end{equation}
This is the Dyson-Schmidt, or Furstenberg, equation for the invariant density. Once \(f(z)\) is known, the Lyapunov exponent and the integrated density of states follow from \eqref{eq:GammaIdosRiccati}. While this equation is exact and general under appropriate assumptions, it is usually difficult to solve except in a few specific cases. We now present an alternative method that allows for a much more systematic analysis.

\section{A Combinatorial Approach}
\label{sec:Combinatorics}

The idea of our method is to work directly with the polynomial structure of the spectral determinant, rather than with the invariant law of a Riccati process.  Starting from the recurrence \eqref{eq:TheRecurrence}, the solution of the initial-value problem can be expanded as
\begin{equation}
  \label{eq:MainRes}
  x_{n+1}(\lambda)=\sum_{k=0}^{n} a_{n+1}^{(k)}\,(-\lambda)^k.
\end{equation}
The coefficients admit an exact combinatorial representation; see Appendix~\ref{app:anderson-coefficients} for the derivation:
\begin{equation}
  \label{eq:MainResIntro2}
  a_{n+1}^{(k)}
  =
  \sum_{1\le j_1\le i_1<j_2\le i_2<\cdots<j_k\le i_k\le n}
  \prod_{m=1}^{k}\frac{m_{j_m}}{\K_{i_m}}.
\end{equation}
Note that this formula exhibits the symmetry between the masses and the inverse spring constants.

The only approximation of the method is introduced at the next step.
First, note that disorder averages are invariant under permutations of i.i.d.\ variables. For any function \(F\) of the masses, one can write
\begin{equation}
   \mean{ F( m_1,\cdots,m_n  ) }   
   = \mean{ F( m_{\pi(1)},\cdots,m_{\pi(n)}  )   }
   = \mean{ \ \symmetriz{ F( m_{\pi(1)},\cdots,m_{\pi(n)}  ) }_{\pi\in\mathcal{S}_n} }
\end{equation}
where \(\pi\) is a permutation and \(\symmetriz{ \cdots }_{\pi\in\mathcal{S}_n}\) denotes full symmetrization, i.e. averaging over the symmetric group \(\mathcal{S}_n\) with uniform weight. Hence we may replace any quantity by its full symmetrization over the two families of disorder variables without changing its disorder average.
Applying this idea to the Lyapunov exponent, and denoting by \(\pi\) and \(\sigma\) the permutations acting on the masses and spring constants respectively, we obtain
\begin{equation}
\gamma(\lambda)
=
\lim_{n\to\infty}\frac{1}{n}\,
\ssymmetriz{\ln x_{n+1}(\lambda)}_{\pi,\sigma\in\mathcal{S}_n}
\;\approx\;
\lim_{n\to\infty}\frac{1}{n}\,
\ln \ssymmetriz{x_{n+1}(\lambda)}_{\pi,\sigma\in\mathcal{S}_n}.
\end{equation}
In the second expression, the permutation average has been moved inside the logarithm. This amounts to assuming that the symmetrized quantity
\begin{equation}
\ssymmetriz{x_{n+1}(\lambda)}_{\pi,\sigma\in\mathcal{S}_n}
\end{equation}
captures the typical growth of \(x_{n+1}(\lambda)\). This assumption will be justified in several well-controlled examples.

This allows for significant simplifications. Indeed, noticing that 
\begin{equation}
  \symmetriz{ \prod_{m=1}^k m_{\pi(j_m)}  }_{\pi\in\mathcal{S}_n} 
  = \symmetriz{ \prod_{m=1}^k m_{\pi(m)}  }_{\pi\in\mathcal{S}_n} ,
\end{equation}
one obtains
\begin{equation}
  \ssymmetriz{a_{n+1}^{(k)}}_{\pi,\sigma}
  =
  S_k(n)\,
  \symmetriz{\prod_{j=1}^{k} m_{\pi(j)}}_{\pi\in\mathcal S_n}
  \symmetriz{\prod_{i=1}^{k} \K_{\sigma(i)}^{-1}}_{\sigma\in\mathcal S_n},
\end{equation}
where the purely combinatorial factor is $S_k(n) = \sum_{1\le j_1\le i_1<j_2\le i_2<\cdots<j_k\le i_k\le n} 1=\binom{n+k}{2k}$.

The remaining problem is therefore reduced to the asymptotics of symmetrized products of $k$ random variables chosen among $n$. Let $X_1,\dots,X_n$ be i.i.d. positive random variables. We introduce
\begin{equation}
  Z(k,n)
  =
  \sum_{1\le i_1<\cdots<i_k\le n}\prod_{m=1}^{k}X_{i_m},
\end{equation}
so that
\begin{equation}
  \symmetriz{\prod_{j=1}^{k}X_{\pi(j)}}_{\pi\in\mathcal S_n}
  =
  \frac{k!(n-k)!}{n!}\,Z(k,n).
\end{equation}
This is naturally interpreted as the canonical partition function of $k$ non-interacting fermions occupying $n$ random levels, upon writing $X_i=\EXP{-\beta \varepsilon_i}$. The corresponding grand partition function is
\begin{equation}
  \Xi(\varphi,n)
  =
  \sum_{k=0}^{n}\varphi^k Z(k,n)
  =
  \prod_{i=1}^{n}(1+\varphi X_i),
\end{equation}
where $\varphi$ is the fugacity. We can go back to the canonical partition function thanks to 
\begin{equation}
  \label{eq:FromXitoZ}
  Z(k,n) 
  = \oint \frac{\D\varphi}{2\I\pi} \, \frac{\Xi(\varphi,n) }{\varphi^{k+1}}
  \:,
\end{equation}
where the contour encircles the origin once in the complex plane of $\varphi$.
In the limit $k\gg1$, we expect that the integral \eqref{eq:FromXitoZ} is dominated by a saddle point $\varphi_*$ solving
\begin{equation}
  \label{eq:SaddleForZ}
  \frac{\partial}{\partial \varphi}
    \left[
      \ln \Xi(\varphi,n) - k \,\ln\varphi
    \right]
  \Big|_{\varphi=\varphi_*}
  =0
  \:.
\end{equation}

Introducing the fraction of occupied levels $ \theta=\frac{k}{n}$ leads to
\begin{equation}
\label{eq:SaddleForTheta}
  \theta
  =
  \frac{1}{n}\sum_{i=1}^{n}\frac{\varphi_* X_i}{1+\varphi_* X_i}
  \to
  \varphi_*(\theta)\,\mean{\frac{X}{1+\varphi_*(\theta)X}}
  =
  \theta.
\end{equation}
as \(n\to+\infty\).

Given the fugacity, we deduce the free energy per level
\begin{equation}
  \label{eq:FreeEnergy}
  \frac{\ln Z(k,n)}{n} \simeq \mean{ \ln(1 + \varphi_* \,X_i) } - \theta\,\ln\varphi_*
  \:.
\end{equation}
which can be rewritten in the form; see Appendix~\ref{app:anderson-free-energy}
\begin{equation}
  \label{eq:FreeEnergyIntegral}
  \frac{\ln Z(k,n)}{n}
  \simeq
  -\int_0^{\theta}\D t\,\ln \varphi_*(t).
\end{equation}

We now apply this generic construction twice: once with $X \to 1/\K$, and once with $X \to m$. This yields two fugacities, defined implicitly by
\begin{equation}
  \label{eq:Transcendental-K}
  \varphi_*^{(1/\K)}(\theta)\,
  \mean{
    \frac{\K^{-1}}{1+\varphi_*^{(1/\K)}(\theta)\,\K^{-1}}
  }
  =
  \theta,
\end{equation}
and
\begin{equation}
  \label{eq:Transcendental-m}
  \varphi_*^{(m)}(\theta)\,
  \mean{
    \frac{m}{1+\varphi_*^{(m)}(\theta)\,m}
  }
  =
  \theta.
\end{equation}

These two functions encode the statistical properties of the two sets of random variables. We are now able to evaluate \(a_{n+1}^{(k)}\).

In order to evaluate the Lyapunov exponent, we return to the polynomial expansion \eqref{eq:MainRes}. For fixed \(\lambda<0\), the sum over \(k\) contains only positive terms, so its exponential growth is controlled by the largest contribution. If \(k_*=k_*(n,\lambda)\) denotes the value of \(k\) for which
\begin{equation}
a_{n+1}^{(k)}(-\lambda)^k
\end{equation}
is maximal, then
\begin{equation}
a_{n+1}^{(k_*)}(-\lambda)^{k_*}
\le x_{n+1}(\lambda)
\le (n+1)\,a_{n+1}^{(k_*)}(-\lambda)^{k_*}.
\end{equation}
After taking the logarithm and dividing by \(n\), the prefactor \((n+1)\) is negligible. Hence the complex Lyapunov exponent may be obtained from the exponential growth rate of the dominant coefficient. This is the key simplification behind the method:
\begin{equation}
  \Omega(\lambda)
  =
  \lim_{n\to\infty}
  \frac{1}{n}
  \ln\!\big(a_{n+1}^{(k_*)}(-\lambda)^{k_*}\big).
\end{equation}

We begin by considering the well-defined case $\lambda<0$, and then extend the result by analytic continuation. At this point we replace the exact coefficient by its symmetrized approximation. Using the factorized form of the symmetrized coefficient, together with the canonical partition functions for $X \to 1/\K$ and $X \to m$, we obtain
\begin{align}
  \frac{1}{n}
  \ln\!\big(\ssymmetriz{a_{n+1}^{(k)}}_{\pi,\sigma}(-\lambda)^k\big)
  =\;&
  \frac{\ln S_k(n)}{n}
  +2\,\frac{\ln\!\big[k!(n-k)!/n!\big]}{n}
  +\frac{k}{n}\ln(-\lambda)
  \nonumber\\
  &+\frac{\ln Z^{(1/\K)}(k,n)}{n}
  +\frac{\ln Z^{(m)}(k,n)}{n}.
  \label{eq:AsymptoticOfAnp1k-chap}
\end{align}
The first two terms are purely combinatorial. Using Stirling's formula and \(k=\theta n\), one finds
\begin{equation}
\frac{\ln S_k(n)}{n}
+2\,\frac{\ln\!\big[k!(n-k)!/n!\big]}{n}
\simeq
(1+\theta)\ln(1+\theta)+(1-\theta)\ln(1-\theta)-2\theta\ln 2.
\end{equation}
The last two terms are the free-energy densities of the fermionic partition functions and are given by the integral representation derived above. Combining these ingredients yields
\begin{align}
  \label{eq:AsymptoticSymmetrizedCoeff-chap}
  \frac{\ln\!\big(\ssymmetriz{a_{n+1}^{(k)}}_{\pi,\sigma}(-\lambda)^k\big)}{n}
  \simeq\;&
  (1+\theta)\ln(1+\theta)
  +(1-\theta)\ln(1-\theta)
  +\theta\ln\!\frac{-\lambda}{4}
  \nonumber\\
  &-\int_0^\theta \D t\,\ln \varphi_*^{(1/\K)}(t)
  -\int_0^\theta \D t\,\ln \varphi_*^{(m)}(t).
\end{align}
In the large-$n$ limit, the dominant contribution to the polynomial \eqref{eq:MainRes} therefore comes from the value $\theta=\theta_*(\lambda)$ that maximizes \eqref{eq:AsymptoticSymmetrizedCoeff-chap}. Equivalently, one may subtract the value at $\theta=0$ and rewrite the result as an integral over $t$, which leads to the following central formula:
\begin{equation} \label{eq:resOmega}
  \boxed{
  \Omega(\lambda)
  =
  -\min_{\theta\in[0,1]}
  \int_0^\theta \D t\,
  \ln\!\left[
    \frac{4}{-\lambda}\,
    \frac{1-t}{1+t}\,
    \varphi_*^{(1/\K)}(t)\,
    \varphi_*^{(m)}(t)
  \right]}
\end{equation}
The minimizer $\theta_*(\lambda)$ is obtained by differentiating the exponent with respect to $\theta$. Since the derivative of the integral is simply the integrand evaluated at the upper bound, it yields
\begin{equation}
  \label{eq:ThetaStar}
  \varphi_*^{(1/\K)}(\theta)\,
  \varphi_*^{(m)}(\theta)
  =
  \frac{-\lambda}{4}\,\frac{1+\theta}{1-\theta}.
\end{equation}

The formula \eqref{eq:resOmega} is completely general: it provides an estimate for any disorder distribution and any value of \(\lambda\). Once the two fugacities are known, the behavior of the complex Lyapunov exponent follows. It converts a disordered spectral problem into the analysis of two scalar transcendental equations, one for the masses and one for the inverse spring constants. It is particularly practical when one wants to extract the small-\(\lambda\) behavior, as we detail below.

\begin{figure}[!h]

\centering
\includegraphics[width=0.48\textwidth]{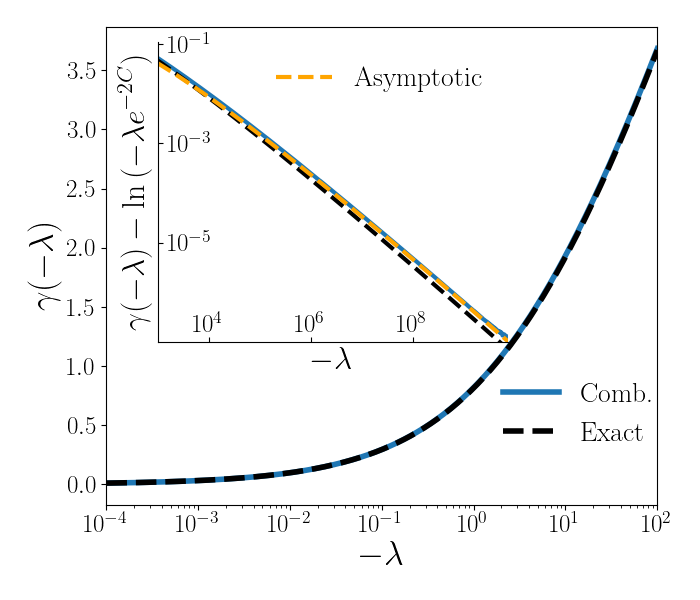}
\includegraphics[width=0.48\textwidth]{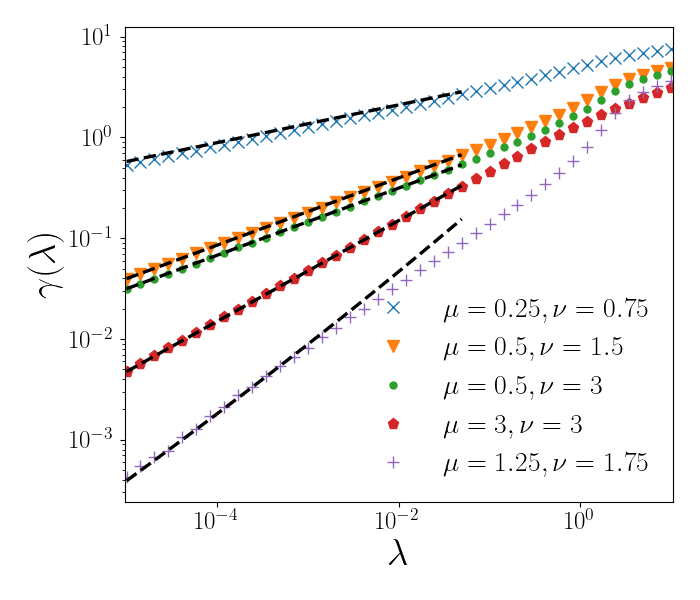}
\caption{Numerical checks of the combinatorial prediction for the Lyapunov exponent. Left: benchmark for exponentially distributed masses and inverse spring constants. The dashed curve is the exact result \eqref{eq:ComplexLyapForExpDistributions-chap}, and the solid curve is the approximation obtained from the combinatorial approach by numerically solving the fugacity and saddle-point equations. The inset shows the same comparison at large negative \(\lambda\) after subtraction of the leading logarithmic term, which the method predicts exactly. Right: numerical Lyapunov exponent for a chain with power-law disorder, compared with the asymptotic prediction \(\gamma(\lambda)\sim\lambda^\zeta\); the dashed lines are the analytical laws with no adjustable parameter.}
\label{fig:gamma-exp-benchmark-chap}
\end{figure}

Before analyzing the asymptotic regimes, let us note that Eq.~\eqref{eq:resOmega} is not exact, as it relies on replacing the true coefficients by their symmetrized versions. It is therefore instructive to benchmark our general formula against an exactly solvable example. The only known case corresponds to exponentially distributed masses, $q(m)=b\,\EXP{-bm}$, and exponentially distributed inverse spring constants, $p(\K)=\frac{a}{\K^2}\,\EXP{-a/\K}$. In this case, the complex Lyapunov exponent is known exactly in terms of MacDonald functions (cf. Eq. (4.17) of \cite{Nie84} or \cite{comtet2019representation}),
\begin{equation}
\Omega(\lambda)=\epsilon\,\frac{K_0(2/\epsilon)}{K_1(2/\epsilon)},
\qquad
\epsilon=\sqrt{\frac{-\lambda}{ab}},
\label{eq:ComplexLyapForExpDistributions-chap}
\end{equation}
and the comparison is provided in Fig.~\ref{fig:gamma-exp-benchmark-chap}.

A limitation of the combinatorial approach is that it fails to capture Lifshitz tails in the high-frequency regime \cite{LucNie88}. These tails originate from atypical disorder configurations that give rise to high-frequency modes, whereas the symmetrization of the coefficients suppresses the contribution of precisely such rare configurations.

\section{Low-Frequency Analysis}
\label{sec:anderson-lowfreq}

We now apply the general formula to the low-frequency regime, where one usually expects some universality. As we will show below, this regime depends crucially on the asymptotic behavior of the distributions of \(1/\K\) and \(m\) at large arguments. To distinguish the various possible regimes, we consider the power-law distributions
\begin{equation}
  \label{eq:PowerLawK-overview}
  p(\K)=\mu\,\K^{-1+\mu},
  \qquad
  0<\K<1,
\end{equation}
and
\begin{equation}
  \label{eq:PowerLawMass-overview}
  q(m)=\nu\,m^{-1-\nu},
  \qquad
  m>1,
\end{equation}
with $\mu,\nu>0$. The key point is that the low-frequency asymptotics is controlled by the small-$\theta$ behavior of the two fugacities appearing in \eqref{eq:Transcendental-K} and \eqref{eq:Transcendental-m}. The thresholds are set by the existence or not of the first and second moments of $1/\K$ and $m$.

The logic of the analysis is as follows and does not depend on the nature of the disorder distribution. In the low-frequency regime, one has $\lambda=-\omega^2\to0^-$, and the saddle point selected by \eqref{eq:ThetaStar} moves towards $\theta=0$. One must therefore first determine the small-$\theta$ asymptotics of the two fugacities $\varphi_*^{(1/\K)}(\theta)$ and $\varphi_*^{(m)}(\theta)$ from the transcendental equations \eqref{eq:Transcendental-K} and \eqref{eq:Transcendental-m}. These asymptotics are then inserted into the saddle equation \eqref{eq:ThetaStar}, which yields the optimal filling fraction $\theta_*(\lambda)$. Finally, substituting $\theta_*(\lambda)$ back into the variational formula \eqref{eq:MainResultForOmega} gives the small-$\lambda$ behavior of the complex Lyapunov exponent. The analytic continuation of the latter to positive $\lambda$ then allows one to retrieve the Lyapunov exponent and the IDoS $\Omega(\lambda+\I0^+)=\gamma(\lambda)-\I\pi\,\IDoS(\lambda)$.

The whole classification of low-frequency regimes therefore reduces to the small-$\theta$ behavior of the fugacities, which is controlled by the tails of the mass and spring distributions.

\subsection{Detailed example: the regime $\mean{\K^{-2}},\mean{m^2}<\infty$}

We now discuss in detail one scenario (see \cite{bernard_texier_rsccomb2025} for a detailed analysis of the other cases), in which the first two moments of the inverse spring constants and the masses are finite:
\begin{equation}
  \mean{\K^{-2}}<\infty,
  \qquad
  \mean{m^2}<\infty.
\end{equation}
For power-law distributions, this corresponds to $\mu>2$ and $\nu>2$. In this regime, the expansions of the two fugacities at small-$\theta$ up to second order are given by
\begin{equation}
  \varphi_*^{(1/\K)}(\theta)
  \simeq
  \mean{\K^{-1}}\,\theta
  \left(
    1+\frac{\mean{\K^{-2}}}{\mean{\K^{-1}}^2}\,\theta
  \right),
  \qquad
  \varphi_*^{(m)}(\theta)
  \simeq
  \frac{\theta}{\mean{m}}
  \left(
    1+\frac{\mean{m^2}}{\mean{m}^2}\,\theta
  \right).
\end{equation}
One immediately notices that this expansion must change if the first or second moments do not exist. Substituting the two fugacities into the saddle equation \eqref{eq:ThetaStar} gives
\begin{equation}
  \frac{\theta^2}{\mean{\K^{-1}}\mean{m}}
  \left(
    1+
    \left[
      \frac{\mean{\K^{-2}}}{\mean{\K^{-1}}^2}
      +
      \frac{\mean{m^2}}{\mean{m}^2}
    \right]\theta
  \right)
  \simeq
  \frac{-\lambda}{4}\,\frac{1+\theta}{1-\theta}.
\end{equation}
It is convenient to introduce
\begin{equation}
  \epsilon \eqdef \sqrt{-\mean{m}\, \mean{\K^{-1}} \, \lambda},
  \qquad
  B \eqdef
  \frac{\mean{\K^{-2}}}{\mean{\K^{-1}}^2}
  +
  \frac{\mean{m^2}}{\mean{m}^2}
  -2.
\end{equation}
The saddle equation can then be solved perturbatively in $\epsilon$, which yields
\begin{equation}
  \label{eq:ThetaStarNormalCaseSmallLambda-chap}
  \theta_*(\lambda)
  \simeq
  \frac{\epsilon}{2}
  -\frac{B}{8}\,\epsilon^2
  +\mathcal O(\epsilon^3).
\end{equation}
As expected, $\theta_*$ is small when $\lambda\to0^-$.%
We now insert this result into \eqref{eq:MainResultForOmega}. Expanding the logarithm consistently to order $\theta_*^2$, one finds
\begin{equation}
  \Omega(\lambda)
  \simeq
  2\theta_*
  \ln\!\left(
    \frac{\EXP{}\,\epsilon}{2\theta_*}
  \right)
  -\frac{B}{2}\,\theta_*^2
  +\mathcal O(\theta_*^3),
\end{equation}
and therefore
\begin{equation}
  \label{eq:Omega-SmallLambda-chap}
  \Omega(\lambda)
  \simeq
  \sqrt{-\mean{m}\, \mean{\K^{-1}} \,\lambda}
  +
  \frac{B}{8}\,\mean{m}\, \mean{\K^{-1}} \,\lambda,
  \qquad
  \lambda\to0^-.
\end{equation}
This expression contains the two low-frequency observables. The first term is purely of order $\sqrt{-\lambda}$, and after analytic continuation $\lambda\to\lambda+\I0^+$ it becomes purely imaginary:
\begin{equation}
  \Omega(\lambda+\I0^+)
  \simeq
  -\I\sqrt{\frac{\mean{m}\,\lambda}{\mean{\K^{-1}}}}
  +
  \frac{B}{8}\,\mean{m}\, \mean{\K^{-1}} \,\lambda,
  \qquad
  \lambda\to0^+.
\end{equation}
The imaginary part yields
\begin{equation}
  \IDoS(\lambda)
  \simeq
  \frac{1}{\pi}\sqrt{\mean{m}\,\mean{\K^{-1}}\,\lambda},
  \qquad
  \lambda\to0^+,
\end{equation}
hence
\begin{equation}
  \label{eq:Dos-AlphaNegBetaNeg}
  \rho(\lambda)
  \simeq
  \frac{1}{2\pi}\sqrt{\frac{\mean{m} \mean{\K^{-1}}}{\lambda}},
  \qquad
  \lambda\to0^+.
\end{equation}
Equivalently, the density of frequencies tends to the finite limit
\begin{equation}
  \varrho(\omega)=2\omega\,\rho(\omega^2)
  \longrightarrow
  \frac{1}{\pi}\sqrt{\mean{m} \, \mean{\K^{-1}}},
  \qquad
  \omega\to0.
\end{equation}
Here, the disorder is not strong enough to affect the low-frequency behavior of the density of states, which remains the same as in the homogeneous case with effective masses \(\mean{m}\) and inverse spring constants \(\mean{\K^{-1}}\).

The second term in \eqref{eq:Omega-SmallLambda-chap}, by contrast, is analytic in $\lambda$ and therefore survives as a real contribution after continuation. It directly gives the Lyapunov exponent:
\begin{equation}
  \label{eq:LyapPertub-chap}
  \gamma(\lambda)
  \simeq
  \frac{\mean{m}\mean{\K^{-1}}}{8}
  \left(
    \frac{\mean{\K^{-2}}}{\mean{\K^{-1}}^2}
    +
    \frac{\mean{m^2}}{\mean{m}^2}
    -2
  \right)\lambda,
  \qquad
  \lambda\to0^+.
\end{equation}

We retrieve the standard weak-disorder scaling \(\gamma(\lambda)\propto \lambda\). It is symmetric under the exchange \(m_n\leftrightarrow \K_n^{-1}\) and vanishes in the homogeneous limit. This shows that any small amount of disorder is enough to localize the eigenvectors. For fixed spring constants, it reduces to a result first obtained by Matsuda and Ishii \cite{MatIsh70}.

\subsection{Detailed example: the regime $\mean{\K^{-1}}=\infty$ and $\mean{m}=\infty$}

The previous regime corresponds to the standard weak-disorder case, which was already known. To illustrate the advantage of our method, let us now turn to a strong disorder regime, where the low-frequency behavior is no longer captured by standard perturbative arguments. Consider, for instance, the strong-disorder regime
\begin{equation}
  \mean{\K^{-1}}=\infty,
  \qquad
  \mean{m}=\infty,
\end{equation}
which for the power-law distributions \eqref{eq:PowerLawK-overview} and \eqref{eq:PowerLawMass-overview} corresponds to
\begin{equation}
  0<\mu<1,
  \qquad
  0<\nu<1.
\end{equation}

The logic is the same as before, but the small-\(\theta\) behavior of the fugacities is now controlled by the broad tails of \(1/\K\) and \(m\). Using the asymptotics of the transcendental equations \eqref{eq:Transcendental-K} and \eqref{eq:Transcendental-m}, one finds
\begin{equation}
  \varphi_*^{(1/\K)}(\theta)
  \simeq
  \left(
    \frac{\sin\pi\mu}{\pi\mu}\,\theta
  \right)^{1/\mu},
  \qquad
  \varphi_*^{(m)}(\theta)
  \simeq
  \left(
    \frac{\sin\pi\nu}{\pi\nu}\,\theta
  \right)^{1/\nu}.
\end{equation}
Substituting these expressions into the saddle-point equation \eqref{eq:ThetaStar} gives
\begin{equation}
  \theta_*(\lambda)\simeq \left[A\,(-\lambda)\right]^\eta,
  \label{eq:ThetaStarStrongDisorder-chap}
\end{equation}
with
\begin{equation}
  \eta=\left(\frac{1}{\mu}+\frac{1}{\nu}\right)^{-1},
  \qquad
  A=
  \frac{1}{4}
  \left(\frac{\pi\mu}{\sin\pi\mu}\right)^{1/\mu}
  \left(\frac{\pi\nu}{\sin\pi\nu}\right)^{1/\nu}.
\end{equation}
Unlike the previous case, the exponent \(\eta\) is now smaller than \(1/2\), reflecting the fact that the low-frequency sector is dominated by rare very weak springs and very large masses.
At leading order, the variational formula \eqref{eq:MainResultForOmega} simplifies. Using the small-\(\theta\) form of the fugacities together with \eqref{eq:ThetaStarStrongDisorder-chap}, the integrand becomes
\begin{equation}
  \ln\!\left[
    \frac{4}{-\lambda}\,
    \frac{1-t}{1+t}\,
    \varphi_*^{(1/\K)}(t)\,
    \varphi_*^{(m)}(t)
  \right]
  \simeq
  \frac{1}{\eta}\ln\!\left(\frac{t}{\theta_*}\right),
\end{equation}
so that
\begin{equation}
  \Omega(\lambda)\simeq \frac{\theta_*}{\eta}.
\end{equation}
One therefore obtains
\begin{equation}
  \Omega(\lambda)\simeq C_{\mu,\nu}\,(-\lambda)^\eta,
  \qquad
  \lambda\to0^-,
  \label{eq:OmegaStrongDisorder-chap}
\end{equation}
with
\begin{equation}
  C_{\mu,\nu}
  =
  \frac{4^{-\eta}}{\eta}
  \left(\frac{\pi\mu}{\sin\pi\mu}\right)^{\nu/(\mu+\nu)}
  \left(\frac{\pi\nu}{\sin\pi\nu}\right)^{\mu/(\mu+\nu)}.
\end{equation}
After analytic continuation to \(\lambda>0\), the same power law controls both observables:
\begin{align}
  \gamma(\lambda)
  &\simeq
  \cos(\pi\eta)\,C_{\mu,\nu}\,\lambda^\eta,
  \qquad
  \lambda\to0^+,
  \\
  \rho(\lambda)
  &\simeq
  \frac{\eta\sin(\pi\eta)}{\pi}\,C_{\mu,\nu}\,\lambda^{\eta-1},
  \qquad
  \lambda\to0^+.
\end{align}

\subsection{Phase diagram and extensions}

The same reasoning applies in all the other regions of the \((\mu,\nu)\) plane. One only needs to insert the appropriate small-\(\theta\) behavior of the two fugacities into Eq.~\eqref{eq:MainResultForOmega}, and then repeat the saddle-point analysis. In this way, one recovers the low-frequency laws announced in Section~\ref{sec:anderson-main-results}, including the logarithmic corrections on the marginal lines. The detailed derivations for all regimes are given in \cite{bernard_texier_rsccomb2025}. The low-frequency asymptotics of the Lyapunov exponent are in agreement with the numerical simulations, as shown in Fig.~\ref{fig:gamma-exp-benchmark-chap} (Right).

\emph{Extension to the Anderson model with random hoppings}

Before ending this chapter, let us emphasize that the combinatorial approach developed in this chapter is not restricted to the random mass-spring chain. It can be transposed, for instance, to the one-dimensional Anderson model with random hoppings (treated in detail in Section VII of \cite{bernard_texier_rsccomb2025}),
\begin{equation}
-t_n \psi_{n+1}-t_{n-1}\psi_{n-1}=\omega \psi_n ,
\label{eq:anderson-random-hoppings-conclusion}
\end{equation}
with i.i.d.\ couplings $t_n>0$. In that case, the solution of the initial-value problem can again be expanded as a polynomial in $\omega$, and the coefficients admit a combinatorial interpretation in terms of reflected random walks built from the variables $w_n=\ln t_n$. In the standard case $\mean{(\ln t)^2}<\infty$, the method recovers the well-known Dyson singularity:
\begin{equation}
\gamma(\omega)\sim \frac{1}{\ln(1/\omega)},
\qquad
\mathcal N(\omega)\sim \frac{1}{\ln^2(1/\omega)},
\qquad
\rho(\omega)\sim \frac{1}{\omega\,\ln^3(1/\omega)},
\qquad \omega\to0^+,
\end{equation}
in agreement with the classical results of Dyson and subsequent works \cite{Dys53,TheCoh76,EggRie78,Dha80,Zim82}. As noted by Dyson, there exists a mapping between spring chains and this model, although the relation between the disorder parameters is, in general, rather intricate. Remarkably, the analysis extends straightforwardly to broader hopping distributions. When \(\mean{(\ln t)^2}=\infty\) but \(\mean{\ln t}<\infty\), one obtains generalized logarithmic singularities,
\begin{equation}
\gamma(\omega)\sim \frac{1}{\ln^\mu(1/\omega)},
\qquad
\mathcal N(\omega)\sim \frac{1}{\ln^{2\mu}(1/\omega)}
\qquad (\omega\to0^+,\;1<\mu<2),
\end{equation}
while in the more extreme regime $\mean{\ln t}=\infty$ the wavefunction becomes superlocalized \cite{Bienaime_2008}, with
\begin{equation}
\ln |\psi_n| \sim n^{1/\mu},
\qquad 0<\mu<1.
\end{equation}
These results emphasize that the combinatorial method is not specific to the case treated here, but can be applied more generally to one-dimensional spectral problems.

\clearpage

 \part{Population Dynamics}
 \chapter{Generalities on Random Growth models and Motivation}
\label{ch:population}

The first part of this thesis dealt with fluctuating interfaces and anomalous roughening. The second part turns to models of random multiplicative growth, with applications to population dynamics, wealth inequality, and ecology. Although these problems may at first seem quite distinct, they are in fact closely related, as we now discuss.

\section{The KPZ equation}

A natural starting point is the Kardar-Parisi-Zhang (KPZ) equation \cite{kardar1986dynamic}, one of the central stochastic growth equations of nonequilibrium statistical physics. For a height field \(h(x,t)\), it reads, in \(d=1\),
\begin{equation}
\partial_t h(x,t)
=
\nu\,\partial_x^2 h(x,t)
+\frac{\lambda}{2}\bigl(\partial_x h(x,t)\bigr)^2
+\eta(x,t),
\label{eq:kpz-pop}
\end{equation}
where \(\nu\) and \(\lambda\) are parameters and \(\eta(x,t)\) is a Gaussian white noise. Higher-dimensional versions are obtained by replacing \(\partial_x \to \nabla\). Compared with the Edwards-Wilkinson equation, the KPZ equation \eqref{eq:kpz-pop} contains the additional nonlinear term \(\frac{\lambda}{2}(\partial_x h)^2\).

To justify this additional term, Kardar, Parisi, and Zhang argued that if growth occurs locally along the normal direction to the interface, rather than vertically, then the vertical velocity depends on the local slope. If the interface grows with normal velocity \(v\), one has
\begin{equation}
\partial_t h = v \sqrt{1+(\partial_x h)^2}.
\end{equation}
For small slopes,
\begin{equation}
\partial_t h \simeq v\left(1+\frac12(\partial_x h)^2\right).
\end{equation}
The constant part corresponds to a uniform drift of the interface, which can be absorbed, while the first nontrivial correction generates the quadratic slope term appearing in \eqref{eq:kpz-pop}. The significance of the KPZ equation lies in its universality: this scaling behavior has been observed or discussed in physically diverse growing systems such as turbulent liquid-crystal interfaces, slow-combustion fronts in paper, bacterial colony expansions, and thin-film or crystal-growth problems.
In low dimension, \(d\leq 2\), the interface is rough and $\alpha>0$. In particular, in \(d=1\), the stationary measure is Brownian, which gives the exact roughness exponent \(\alpha=1/2\). By contrast, for \(d>2\) two phases exist: a weak-noise phase, in which the effect of the noise is small and the interface is effectively flat ($\alpha=0$), and a rough phase, in which the noise remains relevant and \(\alpha>0\).
For broad reviews of the KPZ universality class and its developments, see in particular \cite{halpin1995kinetic,HalpinHealyTakeuchi2015,doussal2025}.

\section{Cole-Hopf transform, the stochastic heat equation, and directed polymers}

For the purposes of this chapter, the key point is that the KPZ equation can be mapped to a linear growth equation for a positive field. Consider the Cole-Hopf transform
\begin{equation}
Z(x,t)=\exp\!\left[\frac{\lambda}{2\nu}\,h(x,t)\right],
\label{eq:cole-hopf-pop}
\end{equation}
where \(Z(x,t)\) is a positive field. Substituting \eqref{eq:cole-hopf-pop} into \eqref{eq:kpz-pop}, one finds that \(Z\) satisfies the stochastic heat equation
\begin{equation}
\partial_t Z(x,t)
=
\nu\,\partial_x^2 Z(x,t)
+
\frac{\lambda}{2\nu}\,\eta(x,t)\,Z(x,t).
\label{eq:she-pop}
\end{equation}
Thus a nonlinear stochastic equation for a fluctuating height is transformed into a linear equation for a positive field with multiplicative white noise.

Before turning to the corresponding random-growth problem, it is useful to recall the interpretation of \eqref{eq:she-pop} in terms of directed polymers \cite{KZDirectP}. Using the Feynman-Kac formula, the solution with initial condition \(Z(x,0)=\delta(x-y)\), denoted \(Z(x,t|y,0)\), can be written as
\begin{equation}
Z(x,t|y,0)
=
\int_{x(0)=y}^{x(t)=x} \mathcal D x(\tau)\,
\exp\!\left[
-\frac{1}{T}\int_0^t d\tau
\left(
\frac12 \left|\frac{dx(\tau)}{d\tau}\right|^2
+
V(x(\tau),\tau)
\right)
\right],
\label{eq:feynman-kac-polymer}
\end{equation}
where \(T=2\nu\) plays the role of an effective temperature and \(V(x,\tau)=-\lambda \eta(x,\tau)\) is the random potential generated by the noise. Equation~\eqref{eq:feynman-kac-polymer} is therefore the partition function of a continuum directed polymer of length \(t\), fixed at \((y,0)\) and \((x,t)\), in a random medium \cite{KZDirectP}. The first term in the exponential is an elastic energy, which penalizes deviations from a vertical path, while the second favors trajectories visiting the most favorable regions of the random environment. For a given realization of the noise, the height field is proportional to (minus) the polymer free energy,
\begin{equation}
h(x,t)=T\log Z(x,t).
\end{equation}
The KPZ equation in dimension \(d\) is thus related to a directed-polymer problem in dimension \(d+1\). The rough phase of KPZ corresponds to a phase in which the directed polymer is pinned by the disorder: at positive temperature, typical trajectories remain localized around the favorable zero-temperature configurations selected by a competition between the elastic energy and the random medium. By contrast, the weak-noise phase of KPZ in \(d>2\) corresponds to a diffusive, non-localized polymer phase \cite{halpin1995kinetic}. For rigorous results on the existence of the transition, see, for instance, \cite{junk2025strongdisorder}.

\section{The Bouchaud-M\'ezard model of wealth distribution}

The stochastic heat equation \eqref{eq:she-pop} also arises in a different context, namely as a particular case of the Bouchaud-M\'ezard model of economy \cite{bouchaud_mezard_2000_wealth}. Bouchaud and M\'ezard introduced this model as a minimal stochastic description of wealth dynamics, combining random speculative returns with exchanges between agents. Their starting point is that the dynamics should remain invariant under a global change of monetary units: since the unit of money is arbitrary, multiplying all wealths by the same constant should not change the form of the evolution equation. This scale invariance naturally leads to multiplicative growth terms together with exchange terms proportional to wealth. The resulting dynamics read
\begin{equation}
\frac{{\rm d}x_i}{{\rm d}t}
=
\sigma \eta_i(t)\,x_i
+
\sum_{j\neq i}J_{ij}x_j
-
\sum_{j\neq i}J_{ji}x_i,
\label{eq:bm-intro}
\end{equation}
where \(\eta_i(t)x_i\) describes random individual returns, while the remaining two terms account for exchanges between agents: \(\sum_{j\neq i}J_{ij}x_j\) is the wealth received by agent \(i\), whereas \(\sum_{j\neq i}J_{ji}x_i\) is the wealth transferred from \(i\) to the rest of the network. 

A natural question is then how the total wealth is distributed among the agents. To quantify this, it is convenient to introduce the wealth share
\begin{equation}
p_i(t)=\frac{x_i(t)}{\sum_{j=1}^N x_j(t)},
\qquad
\sum_{i=1}^N p_i(t)=1,
\end{equation}
as well as the maximal share
\begin{equation}
p_{\max}(t)=\max_i p_i(t).
\end{equation}

When the coupling term \(J_{ij}\) is chosen to be a lattice Laplacian, Eq.~\eqref{eq:bm-intro} is precisely the spatial discretization, in dimension \(d\), of the stochastic heat equation \eqref{eq:she-pop}. In that correspondence, \(x_i(t)\) plays the role of the field \(Z(x,t)\), with the site index \(i\) replacing the spatial coordinate \(x\). From the discussion above, one thus expects a delocalized phase in which the field remains spread over many sites, and a localized phase in which it concentrates on a small number of favorable ones.
By contrast, in the fully connected mean-field limit, \(J_{ij}=\varphi/N\), the Bouchaud-M\'ezard model remains delocalized, although broad stationary wealth distributions emerge \cite{bouchaud_mezard_2000_wealth}:
\begin{equation}
P(x_i)\sim x_i^{-1-\mu},
\end{equation}
with an exponent \(\mu\) that depends on the ratio between redistribution and noise strength
\begin{equation}
\mu=1+\frac{2\varphi}{\sigma^2}.
\end{equation}
This motivates the study of more complex graph structures where phase transitions can occur \cite{Zapperi}.

More generally, random multiplicative growth naturally generates broad, often power-law, distributions. In economics, it manifests in the power-law tails of wealth distributions \cite{Pareto1897,YakovenkoRosser2009,Gabaix2009} or in the distributions of city and firm sizes \cite{Gabaix1999,Newman2005,Axtell2001}.

The same class of models also admits an ecological interpretation. In that setting, \(x_i(t)\) denotes the abundance of species \(i\), \(\eta_i(t)\) represents temporal fluctuations of fitness, for instance due to environmental variability, and the coefficients \(J_{ij}\) encode either interactions between species or, more generally, spatial couplings and migration. Ecological models usually include an additional saturation term, such as \(-x_i(t)^2\), to prevent unbounded growth. 
A central issue in ecology is the maintenance of biodiversity. Classical deterministic approaches often predict that sufficiently large and diverse communities should be unstable, in contradiction with the coexistence observed in nature. To address this problem, recent works have incorporated temporal fitness fluctuations, denoted \emph{seascape noise}. In generalized Lotka-Volterra models, these studies have examined the combined effects of environmental noise and spatial structure, and found that neither ingredient alone is generally sufficient to stabilize highly diverse communities, whereas their interplay can sustain coexistence even in the presence of strong interaction disorder \cite{MallminTraulsenDeMonte2026,Kardar2020,SwartzOttinoLofflerKardar2022,AlHiyasatSwartzGoreKardar2026}.

\section{Quenched Heterogeneity and Temporal Fluctuations}
\label{sec:population-quenched-stochastic}

In the work of Bouchaud-Mézard, the individual growth rates were taken to be Gaussian white noises $\langle \eta_i \rangle=m$, independent across sites and time. Their mean, denoted \(m\), was set to zero since a nonzero constant can be absorbed by rescaling \(x_i(t)\to x_i(t)e^{-mt}\). This description captures temporal fluctuations, but it neglects the fact that in many applications some sites or agents may have a persistent advantage. For instance, a city may benefit from a favorable location or better infrastructure. Such differences remain over long times and can be, as a first step, modeled as quenched heterogeneities.

To include this effect, we decompose the individual growth rate into two parts,
\begin{equation} \label{eq:decompgrowth}
m_i+\eta_i(t),
\end{equation}
where \(m_i\) is quenched and site-dependent, while \(\eta_i(t)\) is a zero-mean Gaussian white noise. In the directed-polymer interpretation, \(m_i\) corresponds to a columnar disorder, a contribution constant along the time direction, added on top of the point disorder \(\eta_i(t)\).

One possible interpretation is of this decomposition is at the level of individuals: the quenched variable \(m_i\) represents persistent traits such as skill or talent, whereas the temporal noise \(\eta_i(t)\) represents luck. This is the viewpoint adopted in the talent-versus-luck literature \cite{PluchinoBiondoRapisarda2018}. In particular, when past performance is used to evaluate individuals, one may ask how much of observed success reflects ability and how much is due to luck. This distinction matters both from a pragmatic point of view, lucky outcomes are not necessarily reproducible and rewarding them can distort incentives, misdirect resources, as well as from a moral one, a merit-based system should reward skill rather than chance.

In the following, we will focus on the mean-field model, where all interactions $J_{ij}=\varphi/N$ are equal:
\begin{equation}
\frac{dx_i(t)}{dt}
=
(m_i-\varphi)x_i(t)+\varphi \overline{x}(t)
+\sigma x_i(t)\,\eta_i(t),
\qquad
\overline{x}(t)=\frac{1}{N}\sum_{j=1}^N x_j(t),
\label{eq:pop-noise}
\end{equation}
where the \(m_i\) are quenched random variables, the \(\eta_i(t)\) are independent Gaussian white noises with zero mean and unit variance, and \(\varphi\) sets the strength of redistribution. Below, this equation will be interpreted in the Ito sense.
In the following, \(i\) labels the sites, \(x_i(t)\) denotes the population at site \(i\), and \(p_i(t)\) the corresponding fraction of the total population.

\section{Distribution of the mean individual growth rate $m_i$}
\label{sec:mdist}

Since the \(m_i\) are heterogeneous, one should specify a suitable distribution. While wealth might have broad distributions, this is not the case for individual growth rates, which are typically finite. An especially important feature is the behavior of the distribution near the upper edge. One natural choice is to consider distributions with compact support, in particular
\begin{equation} \label{eq:eqmcompact}
\rho(m)=\frac{\psi}{ m_>} \left(1 - \frac{m}{m_>}\right)^{\psi -1} 
\qquad\text{for } 0<m<m_>.
\end{equation}
and \(0\) elsewhere \footnote{More precisely, the compact-support assumption is only required on the positive side of the distribution, that is, near its upper edge. No restriction is necessary on the negative tail \(m<0\).}.

Another possible class of distributions is fast-decaying distributions, in particular Gaussian. However, for a system with \(N\) sites, one notices that the maximum (as well as the spacing with the second maximum) of \(N\) iid Gaussian variables with finite variance scales as \(\sim \sqrt{\log{N}}\). It is thus not bounded, and this choice will thus inevitably lead to localization in the \(N\to \infty\) limit. One must therefore scale the variance appropriately. 
For the generalized Gaussian family, this leads to
\begin{equation}
p_N(m)=\frac{1}{Z_{b,N}}
\exp\!\left[-\left(\frac{m^2}{2\Sigma_N^2}\right)^b\right],
\qquad b>0,
\label{eq:generalized-gaussian}
\end{equation}
with
\begin{equation}
\Sigma_N=\frac{\Sigma_0}{\sqrt{2(\log N)^{1/b}}},
\label{eq:SigmaN}
\end{equation}
so that the maximum obeys \(\max_i m_i\to \Sigma_0\) as \(N\to\infty\). 
The case \(b=1\) is Gaussian, while \(b=1/2\) corresponds to a double-exponential tail.

Although both classes produce \(O(1)\) maximal growth rates, the distribution of the \(m_i\)'s differs strongly. In the compactly supported case, the \(m_i\) remain spread over the support, whereas in the generalized Gaussian case they concentrate near \(0\), since the variance scales as \((\ln N)^{-1/b}\) and tends to zero as \(N\to\infty\), with only a few extreme values remaining close to \(\Sigma_0\).

In the following, it will be useful to order
\begin{equation}
m_1>m_2>\cdots>m_N.
\end{equation}

\section{Observables and Outline of Part II}
\label{sec:population-phenomenology}

First, notice that since the interactions are mean-field, all the sites have the same growth rate $x_i \sim e^{\gamma t} $ at large time. The asymptotic growth rate $\gamma$ is thus defined as 
\begin{equation}
\label{eq:gamma-def}
\gamma=\lim_{t\to\infty}\frac{1}{t}\log \overline{x}(t) \quad , \quad \overline{x}(t)=\frac{1}{N}\sum_{i=1}^N x_i(t)
\end{equation}
The first goal will be to estimate the growth rate $\gamma$, and how it depends on the parameters, especially the redistribution rate $\varphi$.

The second will be to measure concentration, for which it is useful to introduce the fractions
\begin{equation}
p_i(t)=\frac{x_i(t)}{\sum_j x_j(t)},
\qquad
\sum_{i=1}^N p_i(t)=1.
\end{equation}
The distribution of the \(p_i\) describes how the total population is shared among the sites or agents. In particular, we will be interested in the maximal fraction \(p_{\max}(t)=\max_i p_i(t)\), which detects condensation onto the best sites.
We will be interested in the asymptotic properties $t \to +\infty$ at large $N$.

The second part of this thesis is organized as follows.

In Chapter~\ref{sec:population-mf-growth}, we first study the model \eqref{eq:pop-noise} in the absence of temporal noise (\(\sigma=0\)). Even in this deterministic setting, a nontrivial competition emerges between quenched heterogeneity and redistribution. Depending on the parameters, the system may be either localized or delocalized, with a phase transition separating the two regimes.

We then turn, in Chapter~\ref{sec:population-mf-growth-noise}, to the case of multiplicative temporal noise, \(\sigma>0\). In this setting, one again finds localized and delocalized phases, but the phase diagram is enriched by the appearance of a third regime: a \emph{partially localized} phase, in which population condenses on several favorable sites that change over time.

Finally, Chapter~\ref{sec:population-resetting} considers a different kind of temporal fluctuation, namely stochastic resetting \cite{SatEvansReset}. In that setting, each site is reset to zero at random times, with a fixed rate. It suppresses the fully localized phase, but allows for a partially localized regime, suggesting that this phase is a robust feature of systems combining quenched heterogeneity with temporal fluctuations.

\chapter{Mean-Field Growth with Quenched Heterogeneity}
\label{sec:population-mf-growth}

We first study the deterministic heterogeneous mean-field growth model, hence without temporal noise. In this setting, heterogeneity favors concentration on the sites with the largest intrinsic rates \(m_i\), while redistribution tends to spread the population over the whole system.

More precisely, we consider \(N\) sites with population \(x_i(t)\ge 0\), evolving according to
\begin{equation}
\frac{{\rm d}x_i}{{\rm d}t}
=
m_i x_i+\varphi\bigl(\overline{x}(t)-x_i\bigr),
\qquad
\overline{x}(t):=\frac{1}{N}\sum_{j=1}^N x_j(t).
\label{eq:eqmo}
\end{equation}
Here \(m_i\) is the intrinsic growth rate of site \(i\), while \(\varphi\ge 0\) is the redistribution rate. 
The redistribution term conserves the total population \(N\overline{x}(t)\) and transfers it uniformly between sites.
We consider the two distributions of $m_i$ introduced in Section~\ref{sec:mdist}: compact-support laws, defined in Eq.~\eqref{eq:eqmcompact}, and the generalized-Gaussian family, defined in Eq.~\eqref{eq:generalized-gaussian} with the scaling \eqref{eq:SigmaN}.

We order the sites $m_1>\dots>m_N$. We are interested in the asymptotic growth rate \(\gamma\), defined earlier in Eq.~\eqref{eq:gamma-def}, and the stationary fractions \(p_i\).

\section{Summary of results}

The deterministic model already exhibits a transition between delocalized and localized growth. More precisely, the phase diagram is controlled by the behavior of the distribution of the \(m_i\) near the upper edge, that is, by the statistics of the most favorable sites.

For {\bf compact-support} laws, defined in Eq.~\eqref{eq:eqmcompact}, two regimes must be distinguished. If \(\psi \le 1\), near-optimal sites remain sufficiently numerous to sustain a delocalized phase for every \(\varphi>0\). In that case, each fraction is \(p_i=O(1/N)\), although the favorable sites carry a larger fraction. The growth rate is then given by \(\gamma=\gamma_{\rm deloc}\), where \(\gamma_{\rm deloc}\) is a decreasing function of \(\varphi\) defined implicitly by
\begin{equation} \label{eq:hypergeo}
\frac{\varphi}{\gamma_{\rm deloc}+\varphi}\,
{}_2F_1\!\left(
1,1;\psi+1;
\frac{m_>}{\gamma_{\rm deloc}+\varphi}
\right)
=1.
\end{equation}

If instead \(\psi>1\), a localization transition occurs at redistribution \(\varphi_c\). The growth rate and the fraction of the best site then behave as
\begin{equation} \label{eq:rescomptsig0}
\gamma=
\begin{cases}
\gamma_{\rm deloc}(\varphi), & \varphi>\varphi_c,\\[1mm]
m_>-\varphi, & \varphi<\varphi_c,
\end{cases}
\qquad
p_1= \max_i p_i =
\begin{cases}
O\!\left(\frac{1}{N}\right), & \varphi>\varphi_c,\\[1mm]
1-\frac{\varphi}{\varphi_c}, & \varphi<\varphi_c.
\end{cases}
\end{equation}
Here \(\gamma_{\rm deloc}\) is again defined by Eq.~\eqref{eq:hypergeo}, and the critical point \(\varphi_c\) is determined by the matching condition $m_>-\varphi_c=\gamma_{\rm deloc}(\varphi_c)$.
For \(\varphi<\varphi_c\), the system is localized, in the sense that a finite fraction of the total population condenses onto the best site $p_1=\max_i p_i=O(1)$, while \(p_i=O(1/N)\) for all \(i>1\).

For the {\bf generalized-Gaussian} family, defined in Eq.~\eqref{eq:generalized-gaussian}, a transition always occurs, with critical point
\begin{equation}
\varphi_c=\Sigma_0.
\end{equation}
In this case, one obtains
\begin{equation}
\gamma=
\begin{cases}
0, & \varphi>\Sigma_0,\\[1mm]
\Sigma_0-\varphi, & \varphi<\Sigma_0,
\end{cases}
\qquad
p_1=
\begin{cases}
O\!\left(\frac{1}{N}\right), & \varphi>\Sigma_0,\\[1mm]
1-\frac{\varphi}{\Sigma_0}, & \varphi<\Sigma_0.
\end{cases}
\end{equation}

Although the overall picture is similar to that of compact-support laws \eqref{eq:eqmcompact}, the delocalized phase now has a constant growth rate, \(\gamma=\overline{m}=0\). In this case, most sites have \(m_i\simeq 0\), so that the contribution of these typical sites fixes the delocalized growth rate.

More generally, redistribution always lowers the growth rate, since it transfers population away from the most favorable sites toward less favorable ones.

In the following sections, we derive these results. We begin by analyzing the generic spectral properties, and then examine in more detail the compact-support and generalized-Gaussian distributions defined in Section~\ref{sec:mdist}.

\section{Spectral properties}

Equation~\eqref{eq:eqmo} can be written in matrix form as
\begin{equation}
\frac{{\rm d}{\bf x}}{{\rm d}t}
=
\mathbb{M}\,{\bf x},
\qquad
\mathbb{M}_{ij}
=
(m_i-\varphi)\delta_{ij}
+\frac{\varphi}{N},
\label{eq:matrix_dyn}
\end{equation}
where \({\bf x}=(x_1,\dots,x_N)^T\). 
Thus \(\mathbb M\) is a diagonal matrix plus a rank-one perturbation.

Using the matrix determinant lemma, one finds
\begin{equation}
\det(z\mathbb I-\mathbb M)
=
\prod_{i=1}^N (z-m_i+\varphi)
\left(
1-\frac{\varphi}{N}\sum_{i=1}^N \frac{1}{z-m_i+\varphi}
\right).
\label{eq:determinant}
\end{equation}
Hence, for \(\varphi>0\), the eigenvalues \(\gamma_\alpha\) of \(\mathbb M\) are the roots of
\begin{equation}
\frac{1}{N}\sum_{i=1}^N
\frac{\varphi}{\gamma_\alpha+\varphi-m_i}
=1.
\label{eq:eq_gamma}
\end{equation}

If the growth rates are ordered as
\begin{equation}
m_1>m_2>\cdots>m_N,
\end{equation}
then the left-hand side of \eqref{eq:eq_gamma} is strictly decreasing between its poles, so the eigenvalues interlace with the shifted rates:
\begin{equation}
\gamma_1>m_1-\varphi,
\qquad
m_j-\varphi<\gamma_j<m_{j-1}-\varphi,
\qquad
j=2,\dots,N.
\label{eq:interlacing}
\end{equation}
The largest eigenvalue \(\gamma_1\), denoted by \(\gamma\), determines the asymptotic growth rate \eqref{eq:gamma-def}.

One can recover the same equation by assuming that the mean population grows asymptotically as
\begin{equation}
\overline{x}(t) \sim \mathrm{e}^{\gamma t}.
\end{equation}
Integrating explicitly \eqref{eq:matrix_dyn} yields at large time
\begin{equation}
x_i(t) \simeq \frac{\varphi}{\gamma+\varphi-m_i} \overline{x}(t)
\end{equation}
hence
\begin{equation}
p_i = \lim_{t \to +\infty} p_i(t)
=
\frac{1}{N}\frac{\varphi}{\gamma+\varphi-m_i}.
\label{eq:pi_asymptotic}
\end{equation}
Equation~\eqref{eq:eq_gamma} is therefore simply the normalization condition
\begin{equation}
\sum_{i=1}^N p_i=1.
\end{equation}
We thus retrieve the earlier equation as well as the stationary fractions.

It is convenient to rewrite \eqref{eq:eq_gamma} in terms of the empirical distribution of growth rates,
\begin{equation}
\rho_N(m)=\frac{1}{N}\sum_{i=1}^N \delta(m-m_i).
\label{eq:rhoN}
\end{equation}
Then
\begin{equation}
\int dm\,\rho_N(m)\,\frac{\varphi}{\gamma+\varphi-m}=1.
\label{eq:eq_gamma_rhoN}
\end{equation}
Introducing the Stieltjes transform
\begin{equation}
G_N(z)=\int dm\,\frac{\rho_N(m)}{z-m},
\label{eq:GNz}
\end{equation}
the eigenvalue equation becomes
\begin{equation}
G_N(\gamma+\varphi)=\frac{1}{\varphi}.
\label{eq:eq_gamma_GNz}
\end{equation}

This formula is the starting point of the large-\(N\) analysis. 
Its interpretation is simple: the asymptotic growth rate is determined by a balance between redistribution, encoded by \(\varphi\), and the distribution of local growth rates, encoded by \(G_N\). 
The behavior of this balance near the largest values of \(m_i\) determines whether the system remains delocalized or localizes on the best site.

\section{Compact-support laws}

We first assume that \(\rho_N\to\rho\) as \(N\to\infty\), where \(\rho\) has compact support with upper edge \(m_>\), \emph{i.e.}
\begin{equation}
\rho(m) \sim (m_> - m)^{\psi -1} 
\qquad\text{for }m<m_>.
\end{equation}

The Stieltjes transform
\begin{equation}
G(\gamma+\varphi)=\int dm\,\frac{\rho(m)}{\gamma+\varphi-m}
\end{equation}
is then well defined for \(\gamma>m_>-\varphi\), a region we now study. This corresponds to the delocalized phase, as \(p_i=\frac{1}{N}\frac{\varphi}{\gamma+\varphi-m_i}=O(1/N)\) for all \(i\) in this case. 
Taking the large-\(N\) limit in \eqref{eq:eq_gamma_GNz}, namely
\begin{equation}
G(\gamma+\varphi)=\frac{1}{\varphi},
\qquad
\gamma+\varphi>m_>.
\label{eq:R_transform}
\end{equation}
Equivalently, defining the \(R\)-transform by
\begin{equation}
R(z)=G^{-1}(z)-\frac{1}{z},
\end{equation}
one may write
\begin{equation} \label{eq:Rtrans2}
\gamma = R\!\left(\frac{1}{\varphi}\right).
\end{equation}

Because \(G(z)\) is strictly decreasing for \(z>m_>\), Eq.~\eqref{eq:R_transform} admits a solution if and only if $\frac{1}{\varphi}\le G(m_>)$. This defines the critical redistribution rate
\begin{equation}
\varphi_c=\frac{1}{G(m_>)}.
\label{eq:phic}
\end{equation}

The existence of a transition is controlled by the behavior of \(\rho\) near the upper edge. 

\begin{equation}
\rho(m)\sim (m_>-m)^{\psi-1},
\qquad m\uparrow m_>,
\label{eq:defpsi}
\end{equation}
then \(G(m_>)<\infty\) if and only if \(\psi>1\). 
Hence:
\begin{itemize}
\item if \(\psi\le1\), the delocalized solution exists for every \(\varphi>0\), and no transition occurs,
\item if \(\psi>1\), there is a transition at \(\varphi=\varphi_c\).
\end{itemize}

For \(\varphi>\varphi_c\), the system is delocalized and the asymptotic growth rate is the unique solution of \eqref{eq:R_transform}. 
For \(\varphi<\varphi_c\), however, Eq.~\eqref{eq:R_transform} ceases to have an admissible solution with \(\gamma+\varphi>m_>\). 
The reason is that the delocalized ansatz tries to place \(\gamma+\varphi\) below the edge of the support, which is impossible: in the large-\(N\) limit, the dominant mode is then forced to pin at the edge,
\begin{equation}
\gamma=m_>-\varphi.
\label{eq:gamma-localized}
\end{equation}
This is the localized phase.

Interestingly, an equation similar to Eq.~\eqref{eq:Rtrans2} appeared recently in a different context \cite{Gueneau2025}, namely for a one-dimensional Brownian particle whose diffusion coefficient \(D\) is reset at rate \(r=\varphi\) to a random value drawn from a distribution \(\rho(D)\). In that setting, the analogue of our condensation transition (dominance of $D_{\rm max}$), in the large $N$ limit, was observed by the authors of \cite{Gueneau2025} in the large-time limit.

\section{Localization as condensation onto the best site}

We now explain more precisely why the regime \(\varphi<\varphi_c\) is localized. 
At fixed \(N\), the asymptotic occupation weights are given by \eqref{eq:pi_asymptotic}. 
Suppose that \(0<\varphi<\varphi_c\), and write
\begin{equation}
\gamma = m_1-\varphi(1-\epsilon_N),
\qquad
\epsilon_N>0,
\label{eq:gamma-epsilon}
\end{equation}
with \(\epsilon_N\) to be determined. 
Then
\begin{equation}
p_1
=
\frac{1}{N\epsilon_N},
\label{eq:p1-epsilon}
\end{equation}
while the contribution of all remaining sites is
\begin{equation}
X_N
:=
\frac{1}{N}\sum_{j=2}^N
\frac{\varphi}{\varphi\epsilon_N+m_1-m_j}.
\label{eq:sum}
\end{equation}
Since \(\sum_i p_i=1\), we have
\begin{equation}
p_1=1-X_N.
\label{eq:p1-1minusX}
\end{equation}

If \(\epsilon_N\to 0\) sufficiently fast, then in the large-\(N\) limit the sum \(X_N\) converges to 
\begin{equation}
X_N
\to
\varphi\int_0^{\infty}dy\,\frac{\rho(m_>-y)}{y}
=
\varphi\,G(m_>)
=
\frac{\varphi}{\varphi_c}.
\end{equation}
Therefore
\begin{equation}
p_1
=
1-\frac{\varphi}{\varphi_c}.
\label{eq:p1phi}
\end{equation}

For \(\varphi<\varphi_c\), the best site \(i=1\) carries a finite fraction \(p_1\) of the total population: this phase is localized.
The remaining fraction \(1-p_1\) is spread over the delocalized sites. 

By contrast, in the delocalized phase \(p_1\to 0\), and no individual site carries a finite fraction of the population. 
Thus the transition at \(\varphi=\varphi_c\) separates a regime in which the asymptotic population is distributed over many sites from a regime in which a finite fraction of the population is concentrated on the best one.

The asymptotic growth rate is therefore
\begin{equation}
\gamma(\varphi)
=
\begin{cases}
m_>-\varphi, & \varphi<\varphi_c,\\[2mm]
\gamma_{\rm deloc}(\varphi), & \varphi\ge \varphi_c,
\end{cases}
\label{eq:gamma-piecewise-compact}
\end{equation}
where \(\gamma_{\rm deloc}\) is the solution of \eqref{eq:R_transform}.

\begin{figure}
    \centering
    \includegraphics[width=0.48\linewidth]{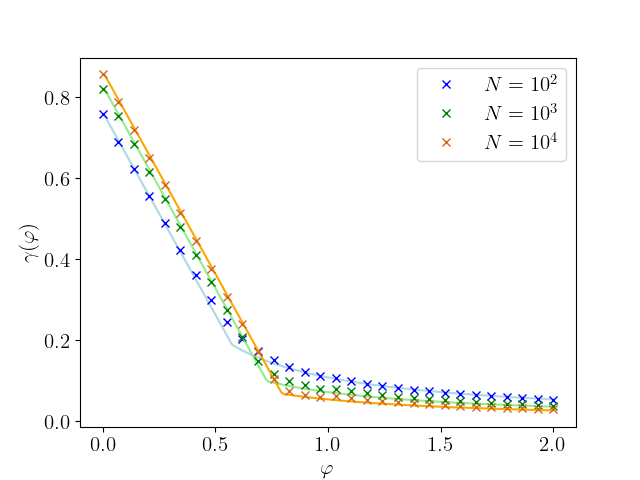}
    \includegraphics[width=0.48\linewidth]{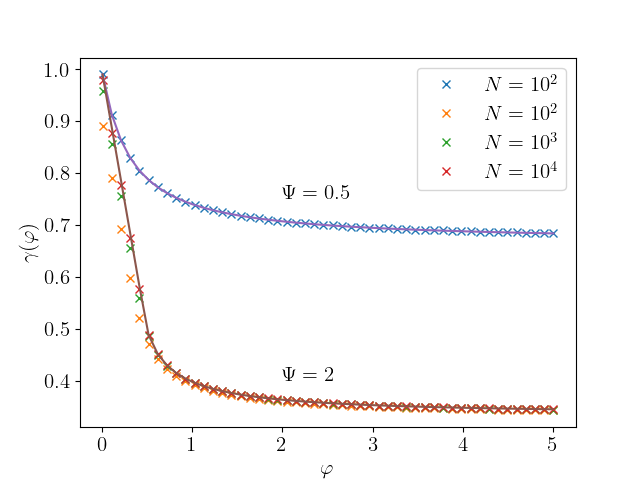}
    \caption{Deterministic case \(\sigma=0\). Numerical values of \(\gamma(\varphi)\) (crosses) compared with the analytical predictions \eqref{eq:gamma-piecewise-compact} (continuous lines). Left: Gaussian distribution for the \(m_i\) with \(\Sigma_0=1\). The prediction is \(m_1-\varphi\) in the localized phase, and \(\Sigma_0^2/2/\log{N}\) in the delocalized one. Right: \(m_i\) distributed according to \(\rho(m)=\psi(1-m)^{\psi-1}\theta(0<m<1)\), for \(\psi=\frac12\) (no localized phase) and \(\psi=2\).}
    \label{fig:gamma-deterministic-comparison}
\end{figure}

In the compact case, the $R$ transform can be written explicitly as a hypergeometric function and $\gamma_{\rm deloc}$ is determined by 
\begin{equation}
\frac{\varphi}{\gamma_{\rm deloc}+\varphi}\;
{}_2F_1\!\left(1,1;\psi+1;\frac{1}{\gamma_{\rm deloc}+\varphi}\right)
=1.
\label{eq:predDelocalpsi}
\end{equation}
These predictions are numerically checked in Fig.~\ref{fig:gamma-deterministic-comparison} (right).

\section{Discussion}

The deterministic mean-field model makes explicit the tradeoff between growth and redistribution. To maximize the total growth, one would like the population to concentrate on the sites with the largest intrinsic rates \(m_i\). Redistribution acts in the opposite direction: it transfers population away from the most favorable sites toward less favorable ones. One should therefore expect the asymptotic growth rate \(\gamma(\varphi)\) to decrease when \(\varphi\) increases.

This can be seen directly from the asymptotic fractions
\begin{equation}
p_i
=
\frac{1}{N}\frac{\varphi}{\gamma+\varphi-m_i}.
\label{eq:pi_discussion}
\end{equation}
Summing the equation $\gamma p_i=(m_i-\varphi)p_i+\frac{\varphi}{N}$
over \(i\), one obtains
\begin{equation}
\gamma=\sum_{i=1}^N p_i m_i.
\label{eq:gamma_weighted_mean}
\end{equation}
Thus the overall growth rate is the \(p_i\)-weighted average of the intrinsic growth rates. In the localized phase this average is dominated by the best site. In the delocalized phase all \(p_i\) are of order \(1/N\), but the largest rates still contribute more because \eqref{eq:pi_discussion} assigns them larger fractions. Increasing \(\varphi\) flattens these fractions and lowers the bias toward the largest growth rates.

The monotonicity can be shown explicitly. Differentiating \eqref{eq:eq_gamma} with respect to \(\varphi\) gives
\begin{equation}
0=
\frac{1}{N}\sum_{i=1}^N
\left[
\frac{1}{\gamma+\varphi-m_i}
-\frac{\varphi(1+\frac{{\rm d}\gamma}{{\rm d}\varphi})}{(\gamma+\varphi-m_i)^2}
\right].
\label{eq:dgammadphi-intermediate}
\end{equation}
Using again \eqref{eq:eq_gamma}, this becomes
\begin{equation}
\frac{{\rm d}\gamma}{{\rm d}\varphi}
=
-1+\frac{1}{\varphi^2\,\frac{1}{N}\sum_{i=1}^N(\gamma+\varphi-m_i)^{-2}} = -1+\frac{1}{N\sum_{i=1}^N p_i^2}.
\label{eq:dgammadphi}
\end{equation}

Since \(N\sum_i p_i^2\ge 1\), one gets
\begin{equation}
\frac{{\rm d}\gamma}{{\rm d}\varphi}\le 0,
\end{equation}
with equality only when all \(p_i=1/N\). The latter is achieved in the large redistribution limit as $p_i\to 1/N$ when $\varphi \to +\infty$, and from \eqref{eq:gamma_weighted_mean} one gets $\gamma(\varphi)=\bar m$.

\section{Generalized-Gaussian family}

We now study the generalized-Gaussian family defined in Eq.~\eqref{eq:generalized-gaussian}.
The localized regime can again be analyzed by writing \(\gamma=m_1-\varphi(1-\epsilon_N)\). 
As in the compact-support case \eqref{eq:eqmcompact}, one isolates the best site and writes
\begin{equation}
p_1=\frac{1}{N\epsilon_N},
\end{equation}
while the contribution of the remaining sites is
\begin{equation}
X_N
:=
\frac{1}{N}\sum_{j=2}^N
\frac{\varphi}{\varphi\epsilon_N+m_1-m_j},
\qquad
p_1=1-X_N.
\label{eq:XN-gumbel}
\end{equation}
The logic is now different from the compact-support case. For the Gaussian case \(b=1\) of the generalized-Gaussian scaling \eqref{eq:SigmaN},
\begin{equation}
\Sigma_N=\frac{\Sigma_0}{\sqrt{2\log N}},
\end{equation}
the maximum remains of order one, \(m_1\to \Sigma_0\), but the width of the bulk shrinks to zero. Hence most of the sites have vanishing rates \(m_i\to0\), with subdominant outliers with a non-vanishing \(m_i\). The sum \(X_N\) is therefore dominated by the bulk, and may be replaced by the integral over the limiting bulk distribution concentrated at \(m=0\). Equivalently, one may write
\begin{equation}
X_N \simeq \frac{1}{N}\sum_{j=2}^N \frac{\varphi}{\varphi\epsilon_N+m_1}
\simeq \frac{\varphi}{\varphi\epsilon_N+m_1}
\to \frac{\varphi}{\Sigma_0}.
\end{equation}
A more rigorous argument is given in Appendix Sec.~II of Ref.~\Pfour. One thus finds
\begin{equation}
X_N\to \frac{\varphi}{\Sigma_0},
\end{equation}
and therefore
\begin{equation}
p_1=1-\frac{\varphi}{\Sigma_0},
\qquad
\varphi<\Sigma_0.
\label{eq:p1-gumbel}
\end{equation}
Thus the critical redistribution rate is
\begin{equation}
\varphi_c=\Sigma_0.
\label{eq:phic-gumbel}
\end{equation}

The asymptotic growth rate is then
\begin{equation}
\gamma(\varphi)
=
\begin{cases}
\Sigma_0-\varphi, & \varphi<\Sigma_0,\\[2mm]
0, & \varphi>\Sigma_0.
\end{cases}
\label{eq:gammaF}
\end{equation}
Hence the generalized-Gaussian family \eqref{eq:generalized-gaussian} always exhibits a localization transition. The main difference with compact-support laws \eqref{eq:eqmcompact} is that, in the delocalized phase, the growth rate is directly controlled by the bulk. In the compact case, by contrast, the bulk remains spread over a finite interval, and the fractions shift continuously as \(\varphi\) increases, which leads to a nontrivial delocalized region.

The first finite-size correction in the delocalized phase follows from the general expansion
\begin{equation}
\gamma=\overline m+\frac{1}{N\varphi}\sum_{i=1}^N (m_i-\overline m)^2+O(\varphi^{-2}),
\end{equation}
and from the fact that, in the generalized-Gaussian scaling \eqref{eq:SigmaN}, the empirical variance is of order \(\Sigma_N^2\). For the Gaussian case \(b=1\), one has
\begin{equation}
\overline m=0,
\qquad
\frac{1}{N}\sum_{i=1}^N (m_i-\overline m)^2 \simeq \Sigma_N^2
=
\frac{\Sigma_0^2}{2\log N},
\end{equation}
so that
\begin{equation}
\gamma
\simeq
\frac{\Sigma_0^2}{2\varphi\log N},
\qquad
\varphi>\Sigma_0.
\label{eq:gamma-gaussian-deloc}
\end{equation}
Thus the approach to the delocalized limit is only logarithmically slow. These predictions have been verified numerically, as shown in Fig.~\ref{fig:gamma-deterministic-comparison} (Left).

\chapter{Mean-Field Growth with Quenched Heterogeneity and Multiplicative Noise}
\label{sec:population-mf-growth-noise}

We now extend the heterogeneous mean-field growth model by including multiplicative white noise. In the deterministic case of Section~\ref{sec:population-mf-growth}, we found a transition between a delocalized phase and a localized phase, in which a finite fraction of the population condenses onto the best site. Multiplicative noise enriches this scenario by allowing sites to become temporarily ``lucky''. %
We recall
\begin{equation} \label{eq:pop-noise2}
\frac{dx_i(t)}{dt}
=
(m_i-\varphi)x_i(t)+\varphi \overline{x}(t)
+\sigma x_i(t)\,\eta_i(t),
\qquad
\overline{x}(t)=\frac{1}{N}\sum_{j=1}^N x_j(t),
\end{equation}
where the noise is interpreted in the It\^o sense.

The chapter is organized as follows. We first recall the model and the main results. We then derive the phase diagram and the growth exponents under a self-averaging assumption. While this assumption is not valid in all regions of the phase diagram, it nevertheless provides a qualitatively correct description, and we analyze it using two complementary methods. We then study the localized phase in more detail, which enables us to recover the complete phase diagram.

\section{General phenomenology and main results}
\label{sec:population-noise-pi-approach}

We recall \(p_i(t)=\frac{x_i(t)}{\sum_j x_j(t)}\), so that \(\sum_i p_i(t) =1\). In this section, we will study the distribution of the \(p_i(t)\) in the stationary state. This will allow us to characterize the phases and estimate their associated growth rates. In the presence of noise, let us first review the different localization scenarios one can have.

{\bf General phenomenology}

Because each site has its own growth rate \(m_i\), it is natural to associate to each site its own stationary distribution \(P_i(p)\) with respect to the noise $\sigma$, for some fixed $m_i$. 
For a typically delocalized site, one expects
\begin{equation}
p_i^{\mathrm{typ}}\sim N^{-1}.
\end{equation}
However, since the growth is multiplicative, power laws naturally emerge and the tail of \(P_i(p)\) may be broad enough that the first moment is dominated by rare events. %
It is therefore natural to introduce, as in the Bouchaud-M\'ezard model, a tail exponent \(\mu_i\) which is now site-dependent defined through
\begin{equation}
P_i(p)\sim N^{-\mu_i} p^{-1-\mu_i}, \qquad \mu_i>0, \qquad N^{-1}\ll p\ll 1
\label{eq:tail_pi_mu}
\end{equation}
where the prefactor \(N^{-\mu_i}\) is required for normalization. The values \(p_i\sim 1/N\) are typical as they control the normalization, whereas values \(p_i=O(1)\), near the upper cutoff, correspond to rare events.
In the limit \(\mu_i\to 0\), the normalization is no longer controlled by typical values \(p_i\sim 1/N\), but by the region \(p_i=O(1)\), which corresponds to a localized site. Then there are three possibilities:
\begin{itemize}
\item if \(\mu_i>1\), the first moment is controlled by typical values and $\langle p_i\rangle \sim N^{-1}$;
\item if \(0<\mu_i<1\), the first moment is controlled by rare events and $
\langle p_i\rangle \sim N^{-\mu_i}$;
\item if the site is localized, one instead has $\langle p_i\rangle = O(1)$.
\end{itemize}
In summary,
\begin{equation}
\langle p_i\rangle =
\begin{cases}
N^{-1}, & \mu_i>1,\\[3pt]
N^{-\mu_i}, & 0<\mu_i<1,\\[3pt]
O(1), & \text{localized}.
\end{cases}
\label{eq:mean_pi_scaling_regimes}
\end{equation}

To enforce the normalization condition \(\sum_{i=1}^N \langle p_i\rangle =1\), we decompose the sum over all sites into three classes: sites such that \(\mu_i>1\), sites such that \(0<\mu_i<1\), and fully localized sites. In summary,
\begin{equation}
1
=
\sum_{\mu_i>1}\langle p_i\rangle
+
\sum_{0<\mu_i<1}\langle p_i\rangle
+
\sum_{\mathrm{loc.}}\langle p_i\rangle
\sim
\sum_{\mu_i>1}N^{-1}
+
\sum_{0<\mu_i<1}N^{-\mu_i}
+
\sum_{\mathrm{loc.}}O(1).
\label{eq:decomp_phases}
\end{equation}
In this notation, \(\sum_{\mu_i>1}\) denotes the sum over all sites \(i\) satisfying \(\mu_i>1\), and similarly for the other two contributions.
This decomposition naturally leads to three phases, depending on which contribution(s) to
$\sum_{i=1}^N \langle p_i\rangle=1$ remain of order unity as $N\to\infty$:

\begin{enumerate}
\item \textbf{Delocalized.}
Only the sum over the sites with $\mu_i>1$ contributes at leading order, the other two sums are negligible. 

\item \textbf{Partially localized.}
The contribution from the sum over heavy-tailed sites $0<\mu_i<1$ is relevant, i.e.\ it yields an $O(1)$ fraction of the
normalization. Individual sites are still typically delocalized ($p_i^{\mathrm{typ}}\sim N^{-1}$), but the number of sites
with broad fluctuations is large enough that, with non-vanishing probability, there exists at least one site $i$ such that $p_i =O(1)$ . In this case,
a finite fraction of the normalization $\sum_i \langle p_i \rangle=1$ is carried by a subextensive number of sites.

\item \textbf{Localized.}
The localized sector contributes to \(O(1)\), so that a finite fraction of the total population is concentrated on a finite number of sites.
\end{enumerate}

{\bf Summary of results}
For the {\bf generalized-Gaussian} family \eqref{eq:generalized-gaussian}, the phase diagram depends on the value of \(b\).
When \(b\le \frac12\), one has, as in the absence of noise $\sigma=0$, a delocalized (denoted Phase I) and a localized phase (denoted Phase II), and the growth rate is given by
\begin{equation}
\gamma=
\begin{cases}
0, & \text{delocalized},\\[1mm]
\Sigma_0-\varphi-\dfrac{\sigma^2}{2}, & \text{localized}.
\end{cases}
\end{equation}
The transition line is therefore simply
\begin{equation}
\varphi_c=\Sigma_0-\frac{\sigma^2}{2}.
\end{equation}
The nature of the localized phase changes slightly. While, at any given time, a single site carries a finite fraction of the total population \(p_{\max}=O(1)\), the identity of this dominant site may switch, at rate $1/\log{N}$, among the finitely many best sites.

When \(b>\frac12\), a third partially localized phase (denoted Phase III):
\begin{equation}
\gamma=
\begin{cases}
0, & \text{delocalized},\\[1mm]
\gamma_{\rm p.l.}, & \text{partially localized},\\[1mm]
\Sigma_0-\varphi-\dfrac{\sigma^2}{2}, & \text{localized},
\end{cases}
\end{equation}
where \(\gamma_{\rm p.l.}\) remains to be determined exactly.

The transition between the localized and delocalized phase is 
\begin{equation}
\varphi_c^{I-II}=\Sigma_0-\frac{\sigma^2}{2}.
\end{equation}
The transition to the partially localized and delocalized phase is given by
\begin{equation}
\varphi_c^{II-III}
=
\frac{2b-1}{2}\sigma^2
\left(\frac{\Sigma_0}{b\sigma^2}\right)^{\frac{2b}{2b-1}}.
\end{equation}
while the localized-partially localized boundary is given 
\begin{equation}
\frac{\Sigma_0}{2 b}\,
\frac{1-\langle p_{\max}\rangle}
{\Sigma_0(1-\langle p_{\max}\rangle)-\varphi_c^{II-III}}
=1,
\end{equation}
with
\begin{equation}
\langle p_{\max}\rangle
=
1-\mu_m e^{A_m}(A_m)^{\mu_m}\Gamma(-\mu_m,A_m),
\qquad
A_m=\frac{2\varphi}{\sigma^2},
\qquad
\mu_m=\frac{2}{\sigma^2}
\left(\Sigma_0-\varphi_c^{II-III}-\frac{\sigma^2}{2}\right).
\end{equation}
where $\Gamma(-\mu,A)$ is the incomplete Gamma function.
For the Gaussian case, \(b=1\), the phase diagram is shown in the right panel of Fig.~\ref{fig:newphaserank_mean_pi_k30} (below). 

For the {\bf compact-support} class \eqref{eq:eqmcompact}, we find, remarkably, that the growth rate is the same as in the $\sigma=0$ case. While the delocalized phase remains unchanged, we no longer have a localized phase. We now have a marginally partially localized phase, where the fraction is concentrated among a slightly sub-extensive %
set of near-optimal sites. These sites have, up to finite-size corrections, a tail exponent \(\mu_i \simeq 1\) and a growth rate \(m_i \simeq m_>\). In both phases, the noise is effectively averaged out in the calculation of the growth rate, hence \(\gamma\) is the same as in the \(\sigma=0\) case. %

\section{Equation for \texorpdfstring{$p_i$}{pi} and useful identities}

We first derive the equation satisfied by $p_i$. Starting from Eq.~\eqref{eq:pop-noise2} and using It\^o's formula, one finds:
\begin{equation}
\frac{d p_i}{dt}
=
p_i(m_i-c_m)
-\varphi\left(p_i-\frac{1}{N}\right)
+\sigma^2 p_i(Y_2-p_i)
+
\sigma p_i\left(\eta_i(t)-\sum_k p_k\,\eta_k(t)\right).
\label{eq:eqpi}
\end{equation}
where
\begin{equation}
c_m:=\sum_j m_j p_j,
\qquad
Y_2:=\sum_j p_j^2.
\end{equation}
\(Y_2\) is the inverse participation ratio. Note that the effective noise \(\sum_k p_k\,\eta_k(t)\) is Gaussian with covariance \(Y_2\,\delta(t-t')\).

We will repeatedly use the following two exact identities, which relate the stationary fractions \(p_i=x_i/\sum_j x_j\) to the asymptotic growth rate \(\gamma\):
\begin{align}
\gamma
&=
m_i-\varphi-\frac{\sigma^2}{2}
+\varphi \left\langle \frac{1}{N p_i} \right\rangle,
\qquad \text{for every } i,
\label{eq:gammaexact1}
\\
\gamma
&=
\sum_j m_j \langle p_j \rangle
-\frac{\sigma^2}{2}
\left\langle \sum_j p_j^2 \right\rangle.
\label{eq:gammaexact2}
\end{align}
valid in the stationary state and where \(\langle \dots \rangle\) is the average with respect only to the noise \(\eta_i(t)\) (and not with respect to the \(m_i\)). 
Let us briefly indicate how they are obtained. Applying It\^o's formula to \(\log x_i\) and using Eq.~\eqref{eq:pop-noise2}, one gets
\begin{equation}
\frac{d}{dt}\log x_i
=
m_i-\varphi+\frac{\varphi}{N}\frac{\sum_j x_j}{x_i}
-\frac{\sigma^2}{2}
+\sigma\,\eta_i(t)
=
m_i-\varphi-\frac{\sigma^2}{2}
+\frac{\varphi}{N p_i}
+\sigma\,\eta_i(t).
\end{equation}
Since $\lim_{t\to\infty}\frac{1}{t}\log x_i(t)=\gamma$, averaging the previous equation over noise gives \eqref{eq:gammaexact1}.
For the second identity, let $X(t)=\sum_j x_j(t)$. Summing Eq.~\eqref{eq:pop-noise2} over \(j\), the redistribution terms cancel exactly and one finds
\begin{equation}
\frac{dX}{dt}=\sum_j m_j x_j+\sigma\sum_j x_j\,\eta_j(t).
\end{equation}
Applying It\^o's formula to \(\log X\) yields
\begin{equation}
\frac{d}{dt}\log X
=
\frac{\sum_j m_j x_j}{X}
-\frac{\sigma^2}{2}\frac{\sum_j x_j^2}{X^2}
+\sigma\sum_j \frac{x_j}{X}\,\eta_j(t).
\end{equation}
Since \(\gamma=\lim_{t\to\infty} \frac{\log X(t)}{t}\), averaging gives \eqref{eq:gammaexact2}.

\section{Self-averaging and Delocalized phase}
\label{sec:population-noise-deloc}

The exact equation \eqref{eq:pop-noise2} couples each fraction \(p_i\) to the whole system through the two collective observables
\begin{equation}
c_m=\sum_j m_j p_j,
\qquad
Y_2=\sum_j p_j^2.
\end{equation}
In this section, we study the case where these quantities become self-averaging 
\begin{equation}
c_m=\sum_{i=1}^N m_i p_i
\;\xrightarrow[N\to\infty]{}\;
\sum_{i=1}^N m_i\langle p_i\rangle \qquad Y_2=\sum_j p_j^2 \;\xrightarrow[N\to\infty]{}\; \langle Y_2 \rangle
\end{equation}
A detailed discussion of the region of validity of these assumptions is given at the end of the section. In particular, we will see that it describes a delocalized phase, for which one also has $\langle Y_2 \rangle=0$. At leading order, one may thus replace
\begin{equation}
Y_2 \to 0, \qquad
c_m \to \langle c_m \rangle = \gamma.
\end{equation}
where we used $\gamma= \langle c_m \rangle-\frac{\sigma^2}{2} \langle Y_2 \rangle$.
The self-averaging picture, in particular $Y_2=o(1)$, is only possible when all $p_i$ are $o(1)$ and thus corresponds to a delocalized phase. We therefore neglect the higher-order terms $+\sigma^2 p_i(Y_2-p_i)$ and $\sigma p_i\sum_k p_k\,\eta_k(t)$ in \eqref{eq:eqpi}, since the effective noise \(\sum_k p_k\,\eta_k(t)\) has covariance \(Y_2\,\delta(t-t')\), leading to
\begin{equation}
\label{eq:SDE_deloc}
\frac{d p_i}{dt}
=
p_i(m_i-\gamma)-\varphi\Big(p_i-\frac{1}{N}\Big)
+\sigma p_i\,\eta_i(t).
\end{equation}
This is valid for all sites $i$, and we have also replaced $c_m\to \gamma$, since its fluctuations are of higher order.
The stationary density \(P_i(p)\) of \(p_i\) is then given by the inverse-gamma law
\begin{equation}
\label{eq:pdf_inverse_gamma}
P_i(p)
=
\frac{A^{\mu_i}}{N^{\mu_i}\Gamma(\mu_i)}
\,p^{-1-\mu_i}
\exp\!\left(-\frac{A}{N p}\right),
\qquad
A:=\frac{2\varphi}{\sigma^2},
\qquad
\mu_i:=1+\frac{2(\gamma+\varphi-m_i)}{\sigma^2}.
\end{equation}

Let us now state more precisely the domain of validity of the self-averaging assumption. We required in particular 
\begin{equation}
\langle Y_2\rangle = \sum_{i=1}^N \langle p_i^2 \rangle \to 0.
\end{equation}
There can be three cases. (i) A sufficient condition for this hypothesis to be valid is that all the \(\mu_i>1\), or equivalently \(\mu_1>1\). Indeed, sites with \(\mu_i>2\) have, from \eqref{eq:pdf_inverse_gamma}, \(\langle p_i^2\rangle \sim N^{-2}\) and sites with \(1<\mu_i<2\) have \(\langle p_i^2\rangle\sim N^{-\mu_i}\). Both are \(o(N^{-1})\), and therefore cannot give a finite contribution to \(Y_2\). 
(ii) On the other hand, this condition is no longer satisfied when $\mu_1 \to 0$, corresponding to localization. When it exists, $\mu_1 = 0$, i.e. $m_1-\varphi-\sigma^2/2=0$, thus predicts the transition line between the localized and delocalized phase.
(iii) The third case occurs when several sites have $0<\mu_i<1$. In that case, since for $0<\mu_i<1$, \eqref{eq:pdf_inverse_gamma} leads to $\langle p_i^2\rangle \sim \langle p_i\rangle \sim N^{-\mu_i}$, the self-averaging assumption remains valid as long as 
\begin{equation} 
\sum_{0<\mu_i<1}\langle p_i^2\rangle \sim \sum_{0<\mu_i<1} \langle p_i\rangle
\sim
\sum_{0<\mu_i<1}N^{-\mu_i}
=o(1).
\end{equation}
so that the contribution of these sites is negligible not only for \(Y_2\), but also for \(c_m=\sum_i m_i p_i\) and the normalization condition. In that case, \(c_m\) may then be replaced by its typical value \(\langle c_m\rangle\). Conversely, since the sites with \(1<\mu_i<2\) always give irrelevant contributions to \(\langle Y_2 \rangle\), the self-averaging of \(\langle c_m \rangle\) implies \(\langle Y_2\rangle \to 0\). Thus sites with \(0<\mu_i<1\) may still be present, as long as their total contribution is negligible
\begin{equation}
\sum_{0<\mu_i<1}N^{-\mu_i}=o(1).
\end{equation}
Thus, self-averaging implies \(Y_2\to 0\), and therefore corresponds to a delocalized phase. In particular, the presence of some sites with \(0<\mu_i<1\) is not inconsistent with delocalization as their total contribution is negligible. For sites with \(\mu_i>1\), the first moment exists \footnote{Near \(\mu_i=1\), the \(O(1/N)\) contribution written here must be combined with the contribution from the cutoff, proportional to \(N^{-\mu_i}/(1-\mu_i)\). The apparent divergence as \(\mu_i\to1^+\) then cancels against the corresponding divergence from the tail term, leaving only a logarithmic contribution at the transition.} and, from \eqref{eq:pdf_inverse_gamma}, is given by
\begin{equation}
\label{eq:mean_pi}
\langle p_i\rangle
=
\frac{\varphi}{N(\gamma+\varphi-m_i)}.
\end{equation}
Using \(\sum_i\langle p_i\rangle=1\) and neglecting sites with \(\mu_i<1\), one obtains the equation that determines \(\gamma\) in the self-averaging phase:
\begin{equation}
\label{eq:normalization_mu}
1
\simeq
\frac{1}{N}\sum_{\mu_i>1}\frac{\varphi}{\gamma+\varphi-m_i}.
\end{equation}
In particular, if all sites satisfy \(\mu_i>1\), then the growth rate is determined by exactly the same equation as in the noiseless case \(\sigma=0\).

\section{Estimation of the full phase diagram under the self-averaging assumption}
\label{sec:phasdiag-Selfavg}

We now go beyond the delocalized phase and derive the full phase diagram. We use the self-averaging solution as a starting point, by assuming that the tail exponent \(\mu_i\) from the inverse-gamma distribution remains valid even in regimes where the full self-averaging assumption is no longer controlled. This provides a first prediction for the localized, partially localized, and delocalized phases, as well as for the transition lines between them. This leads to the phase diagram shown in the left panel of Fig.~\ref{fig:figplext}. We then go beyond this approximation by analyzing the localized phase more precisely. This refined description corrects the phase diagram obtained from the self-averaging argument and yields the correct one, shown in the right panel of Fig.~\ref{fig:newphaserank_mean_pi_k30}.

\subsection{Self-averaging}

The inverse-gamma form \eqref{eq:pdf_inverse_gamma} was obtained under the self-averaging assumption appropriate to the delocalized phase. However, it is instructive to assume that, in the other phases, the distribution of the \(p_i\) is still given by a power-law distribution with tail exponent
\begin{equation}
\mu_i=1+\frac{2(\gamma+\varphi-m_i)}{\sigma^2}
\label{eq:mu_i_general}
\end{equation}
It yields a qualitatively good estimation of the growth rate $\gamma$ and of the different phases. In particular, it predicts correctly the transition lines between the self-averaging/delocalized phase and the other phases.

\paragraph{Localized phase.}
A site becomes localized when its fraction is of order one. For the best site \(i=1\), this occurs when the tail exponent reaches
\begin{equation}
\mu_1=0,
\end{equation}
which gives
\begin{equation}
\gamma_{\rm loc}=m_1-\varphi-\frac{\sigma^2}{2} \to \Sigma_0-\varphi-\frac{\sigma^2}{2}.
\label{eq:gamma_loc_phase}
\end{equation}
since $m_1(N)\underset{N\to+\infty}{\to}\Sigma_0$.

\paragraph{Partially localized phase.}

As seen from \eqref{eq:decomp_phases}, the partially localized phase occurs when enough sites satisfy
\begin{equation}
0<\mu_i<1.
\end{equation}

To simplify the discussion, here and in the rest of the thesis we replace the random variables \(m_i\) by their deterministic quantiles
\begin{equation}
\frac{i}{N} = \int_{m_i}^{+ \infty} \rho(m) dm.
\end{equation}
Note that the $m_i$ are ordered \(m_1\ge m_2\ge\cdots\ge m_N\).
This avoids having to keep track of the fluctuations of the \(m_i\), while leading to the same large-\(N\) results.

For the generalized-Gaussian family \eqref{eq:generalized-gaussian}, one has at large $x$
\begin{equation}
\int_{x}^{+ \infty} \rho(m) dm
\sim
\exp\!\left[-\left(\frac{x^2}{2\Sigma_N^2}\right)^b\right].
\end{equation}
For $i \ll N$, imposing \(\int_{m_i}^{+ \infty} \rho(m) dm = i/N\) therefore gives $-\left(\frac{m_i^2}{2\Sigma_N^2}\right)^b
\simeq
\ln\!\left(\frac{i}{N}\right)$.
Using the scaling \eqref{eq:SigmaN}, one obtains, for \(i\ll N\),
\begin{equation}
m_i
\simeq
\Sigma_0
\left(1-\frac{\ln i}{\ln N}\right)^{\frac{1}{2b}}.
\label{eq:mi-gauss-scaling}
\end{equation}

To study the partially localized phase, we first identify the sites which have $\mu_i<1$. Let \(j\sim N^\alpha\) be such that \(\mu_j=1\). Since \(\mu_j=1\) is equivalent to \(m_j=\gamma+\varphi\), one finds
\begin{equation}
\Sigma_0(1-\alpha)^{\frac{1}{2b}}=\gamma+\varphi.
\label{eq:alpha_cutoff}
\end{equation}
Since $\gamma \ge 0$, if \(\varphi>\Sigma_0\), there is no such \(j\): all \(\mu_i>1\), and the system remains delocalized. The nontrivial region is therefore \(\varphi \le \Sigma_0\).

We now estimate the contribution of the sites with \(0<\mu_i<1\), namely \(i\lesssim N^\alpha\), to the normalization condition. This contribution is the quantity that decides whether these sites are relevant: if it is \(o(1)\), they do not contribute to the normalization, whereas in the partially localized regime it must be of order one. We therefore define
\begin{equation}
S_N
:=
\sum_{i=1}^{N^\alpha}\langle p_i\rangle
\sim
\sum_{i=1}^{N^\alpha}
N^{-\mu_i}.
\label{eq:partiallocsum}
\end{equation}
Using Eq.~\eqref{eq:mi-gauss-scaling}, we write, for \(i=N^u\),
\begin{equation}
\mu_i\simeq \mu(u)
:=
1+\frac{2}{\sigma^2}
\left[
\gamma+\varphi-\Sigma_0(1-u)^{\frac{1}{2b}}
\right].
\end{equation}
Approximating this sum by an integral and setting \(i=N^u\), \(di=(\ln N)N^u\,du\), gives
\begin{equation}
S_N
\simeq
\ln N
\int_0^\alpha
\exp\!\left[
\ln N\, g(u)
\right]du,
\qquad
g(u)=u-\mu(u).
\label{eq:Su}
\end{equation}
The asymptotics of \(S_N\) are controlled by the maximum of \(g(u)\) on \([0,\alpha]\). The contribution to the normalization is relevant, i.e. $S_N=O(1)$, only if this maximum vanishes. It cannot be positive, since \(S_N\le 1\) by normalization, and if it is negative then the contribution is negligible, \(S_N=o(1)\).
There are four possibilities:
\begin{itemize}
\item if \(b\le \frac12\), the function \(g\) has no maximum strictly inside \(]0,\alpha]\). The maximum is at the lower bound \(u=0\). It is of order one only if \(g(0)=0\), or equivalently \(\mu_1 \simeq \mu(0)=0\). One recovers the localized phase \eqref{eq:gamma_loc_phase};

\item if \(b>\frac12\) but \(\sigma^2\le \Sigma_0/b\), the maximum is still at \(u=0\). The contribution is again of order one only if \(\mu(0)=0\), which gives the localized phase \eqref{eq:gamma_loc_phase};

\item if \(b>\frac12\) and \(\sigma^2>\Sigma_0/b\), the maximum is reached at the saddle point $g'(u^\star)=0$,
\begin{equation} \label{eq:ustar}
u^\star
=
1-\left(\frac{\Sigma_0}{b\sigma^2}\right)^{\frac{2b}{2b-1}},
\end{equation}
and
\begin{equation}
S_N\sim \sqrt{\ln N}\,N^{g(u^\star)}.
\end{equation}
For the contribution $S_N$ in Eq.~\eqref{eq:partiallocsum} to be of order one, one must impose
\begin{equation}
g(u^\star)=0,
\end{equation}
which yields
\begin{equation}
\gamma_{\rm p.l.} =\frac{2b-1}{2}\sigma^2 \left(\frac{\Sigma_0}{b\sigma^2}\right)^{\frac{2b}{2b-1}} -\varphi.
\end{equation}

\item the region where the integral is dominated by the upper cutoff cannot lead to an \(S_N=O(1)\) contribution and is therefore discarded.
\end{itemize}

Let us now discuss in more detail the regime selected by the saddle point \eqref{eq:ustar}, which corresponds to the partially localized phase (Phase III). In that regime, the growth rate is
\begin{equation}
\gamma_{\rm p.l.}= \frac{2b-1}{2}\sigma^2 \left(\frac{\Sigma_0}{b\sigma^2}\right)^{\frac{2b}{2b-1}} -\varphi,
\label{eq:gammapl}
\end{equation}
and this phase exists only for \(b>\frac12\) and \(\sigma^2>\Sigma_0/b\).

In this phase, the contribution \(S_N\) to the normalization is dominated by the saddle point \(u^\star\). The main contribution therefore comes from \(\sim N^{u^\star}\) favorable sites, with
\begin{equation}
m_i \simeq \Sigma_0 (1-u^\star)^{\frac{1}{2b}}.
\end{equation}
Thus, each site selected by the saddle point has
\begin{equation}
\mu_i \simeq \mu(u^\star)=u^\star<1.
\end{equation}
For the sites $i$ in the saddle-point, the probability of being localized, that is, of having \(p_i=O(1)\), is obtained by integrating the tail its distribution $P(p_i) \sim N^{-\mu_i} p^{-1-\mu_i}$ over the \(O(1)\) region, between a fixed threshold \(v=O(1)\) and the \(O(1)\) upper cutoff \(p_c>v\):
\begin{equation}
\int_v^{p_c} dp\, N^{-\mu(u^\star)} p^{-1-\mu(u^\star)}
\sim N^{-\mu(u^\star)}
= N^{-u^\star}.
\end{equation}
Thus each site $i$ selected by the saddle-point is individually localized only with small probability \(N^{-u^\star}\), but there are \(N^{u^\star}\) such candidate sites. The expected number of sites carrying an \(O(1)\) fraction of the total weight at a given time is therefore $N^{u^\star}\, N^{-u^\star}=O(1)$. This corresponds to the partially localized phase, where many sites can localize, leading to high turnover of the sites on which one is localized. 

For \(b\le \frac12\), or for \(b>\frac12\) with \(\sigma^2\le \Sigma_0/b\), the partially localized phase is no longer possible, and one can have instead the localized one.

\begin{figure}
    \centering
    \includegraphics[width=0.48\linewidth]{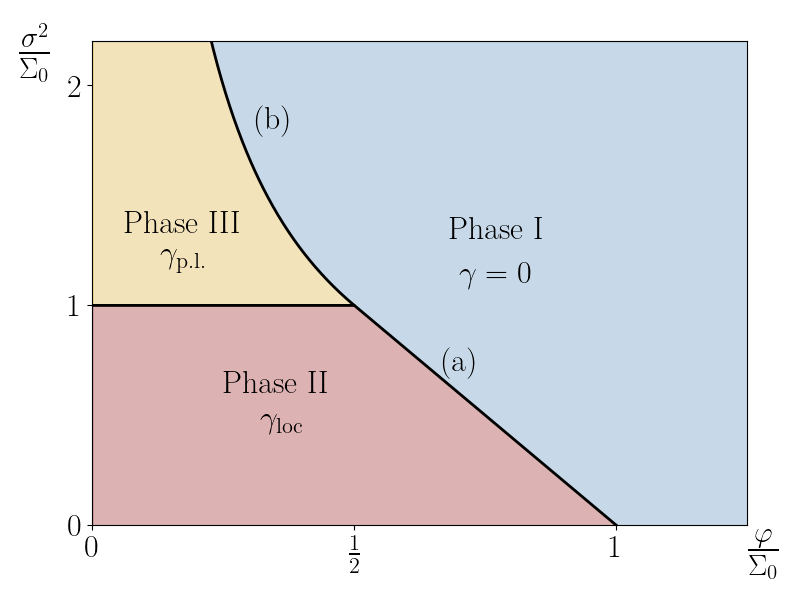}
    \includegraphics[width=0.48\linewidth]{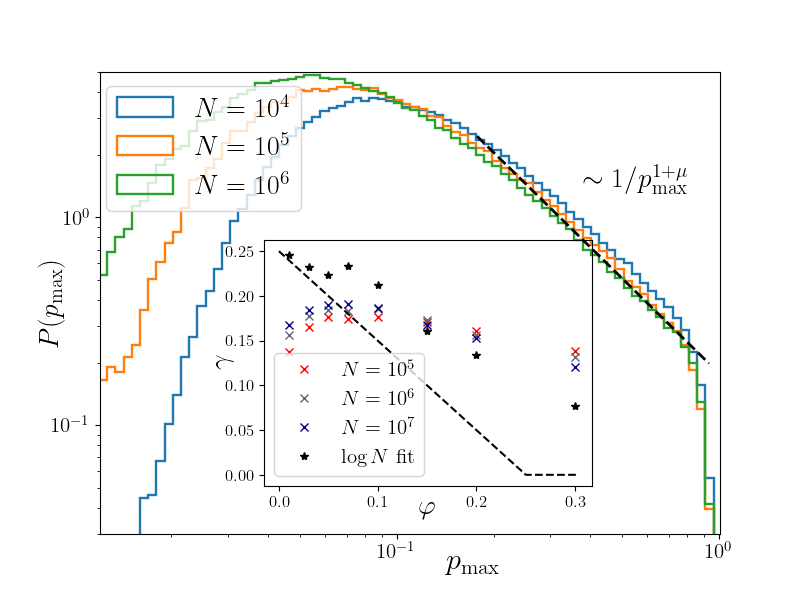}
    \includegraphics[width=0.58\linewidth]{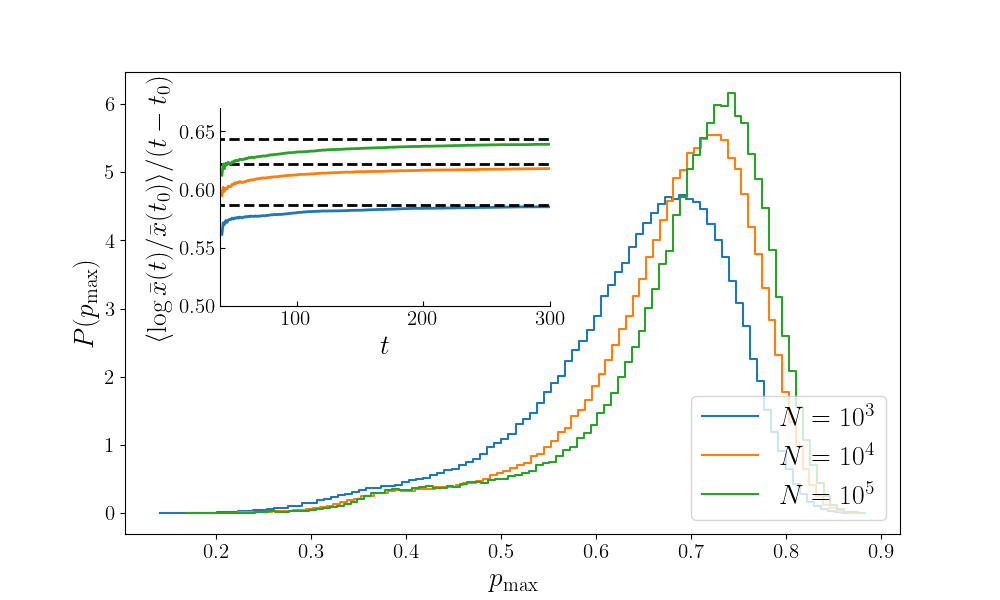}
    \caption{Gaussian case under the self-averaging description. Left: phase diagram in the plane \((\varphi/\Sigma_0,\sigma^2/\Sigma_0)\), with phase I delocalized (\(\gamma=0\)), phase II localized, and phase III partially localized. Right: distribution of \(p_{\max}=\max_i p_i\) in phase III for \(\varphi=0.15\), \(\Sigma_0=1\), and \(\sigma^2=1.44\). The broad tail is consistent with the self-averaging exponent \(\mu=1-\Sigma_0^2/\sigma^2\). Inset: \(\gamma(\varphi)\) for \(\sigma^2=2\) and several $N$, compared with the self-averaging prediction \(\gamma_{\rm p.l.}=\Sigma_0^2/(2\sigma^2)-\varphi\); black stars denote extrapolations of the form \(\gamma(N)=\gamma_\infty+c/\log N\). Bottom: distribution of \(p_{\max}\) in phase II for \(\varphi=0.2\) and \(\sigma=0.3\), showing concentration around an \(O(1)\) value. Inset: corresponding finite-size estimates of the growth rate, compared with the localized prediction \(\gamma=m_1-\varphi-\sigma^2/2\). (Here, the $m_i$ are taken to be the quantiles of the distribution)}
    \label{fig:figplext}
\end{figure}

When \(\gamma_{\rm loc}\) or \(\gamma_{\rm p.l.}\) reach \(0\), the solution \(\gamma_{\rm deloc.}=0\) becomes relevant and one transitions to the delocalized phase.
The corresponding self-averaging phase diagram is shown in the left panel of Fig.~\ref{fig:figplext}. In the Gaussian case, the distributions of \(p_{\max}\) in phases III and II are shown in the right and bottom panels, respectively.

\paragraph{Compact-support law.}

For the compact-support law \eqref{eq:eqmcompact}, in the delocalized phase, all sites satisfy \(\mu_i>1\), so that \(\gamma\) is the same as in the case without noise \(\sigma=0\), Eq.~\eqref{eq:hypergeo}. The transition occurs when \(\mu_1 \to 1\), at which point one enters the marginally partially localized phase and \(\mu_1\) freezes
\begin{equation}
\mu_1=1.
\end{equation}
This equality is equivalent to setting $\gamma=m_>-\varphi$, which, in the model without noise $\sigma=0$, characterizes the localized phase \eqref{eq:rescomptsig0}.

We again consider the case where the $m_i$ are fixed, given by the quantiles of \eqref{eq:eqmcompact}: 
\begin{equation}
m\!\left(\frac{i}{N}\right)
=
m_>\left[1-\left(\frac{i}{N}\right)^{1/\psi}\right].
\label{eq:mi-compact-quantile}
\end{equation}
 To see what happens at that threshold, suppose that \(\mu_1<1\). For \(i=O(1)\), one has
\begin{equation}
m_i\simeq m_>\left[1-\left(\frac{i}{N}\right)^{1/\psi}\right].
\end{equation}
Denoting $y_i$ the distance to the upper edge,
\begin{equation}
y_i:=m_>-m_i\simeq m_>\left(\frac{i}{N}\right)^{1/\psi}.
\end{equation}
Then the sites \(\mu_i<1\) are those such that $0<y_i<m_>-\gamma-\varphi$. For such sites,
\begin{equation}
\mu_i
=
1+\frac{2(\gamma+\varphi-m_i)}{\sigma^2}
=
1-\frac{2\big[m_>-\gamma-\varphi-y_i\big]}{\sigma^2}.
\end{equation}
Their contribution to the normalization is
\begin{equation}
S_N
:=
\sum_{0<\mu_i<1}\langle p_i\rangle
\sim
\sum_{0<\mu_i<1}N^{-\mu_i}.
\end{equation}
Using the edge behavior $\rho(m_>-y)\simeq \frac{\psi}{m_>^\psi}y^{\psi-1}$, this becomes
\begin{equation}
S_N
\sim
\frac{\psi}{m_>^\psi}
\int_{0}^{m_>-\gamma-\varphi}
dy\,
y^{\psi-1}
\exp\!\left[\frac{2\ln N}{\sigma^2}\big(m_>-\gamma-\varphi-y\big)\right].
\label{eq:SNcompactint-clean}
\end{equation}
This integral is dominated by the lower edge \(y=0\). Therefore, in order for \(S_N\) to remain of order one, one must impose
\begin{equation}
m_>-\gamma-\varphi=0,
\end{equation}
which is equivalent to
\begin{equation}
\mu_1
=
1+\frac{2(\gamma+\varphi-m_>)}{\sigma^2}
=1.
\end{equation}
and the upper-bound of \eqref{eq:SNcompactint-clean} goes to $0$, consistent with the fact that there no sites with $\mu_i<1$. Hence the system cannot reach into a genuine regime with \(\mu_1<1\): as soon as the best site reaches \(\mu_1=1\), it gets pinned there.
This corresponds to a \emph{marginally partially localized} phase, dominated by the sites close to the upper cutoff \(m_>\). Multiplicative noise prevents condensation onto the single best site. The growth rate is again the same as the model without noise $\sigma=0$
\begin{equation}
\gamma_{\rm p.l.}=m_>-\varphi.
\end{equation}
The phase diagram is thus the same as in the case \(\sigma=0\), although the localized phase is replaced by a marginally partially localized.

\subsection{Alternative REM approach}
\label{sec:pop-noise-rem}

The previous two subsections derived the phase diagram by first obtaining the stationary distribution of the \(p_i\), and then using the normalization condition to determine \(\gamma\). 
An alternative route consists in reversing these two steps: summing first and then estimating the integral. 
This second route allows us to make a nice connection with the random-energy model, which we detail below. 

We focus here on $m_i$ given by a Gaussian distribution.
For convenience, we introduce the Brownian motions
\begin{equation}
B_i(t):=\int_0^t du\,\eta_i(u),
\end{equation}
so that \(B_i(t)-B_i(t')=\int_{t'}^t du\,\eta_i(u)\).
The exact It\^o solution of Eq.~\eqref{eq:pop-noise} reads
\begin{equation}
\label{eq:ODE_growth2}
x_i(t)
=
x_i(0)e^{(m_i-\varphi-\sigma^2/2)t+\sigma B_i(t)}
+
\varphi\int_0^t {\rm d}t' \,
\overline{x}(t')\,
e^{(m_i-\varphi-\sigma^2/2)(t-t')+\sigma(B_i(t)-B_i(t'))}.
\end{equation}
We assume the asymptotic growth law
\begin{equation}
\overline{x}(t)\sim \Gamma(t) e^{\gamma t}.
\end{equation}
Here \(\Gamma(t)\) is the term that couples \(x_i\) to the rest of the system, playing the same role as \(c_m\) and \(Y_2\) in the equation for \(p_i\). To obtain a self-consistent equation for \(\gamma\), we approximate \(\Gamma(t)\) by a constant, thereby neglecting its temporal fluctuations induced by the noise. This is the analogue, in the \(x_i\) formulation, of the self-averaging assumption made for \(c_m\) and \(Y_2\) in the previous subsection. We therefore expect this approach to reproduce the same phase diagram.

Summing over \(i\) and neglecting the first initial condition term, one obtains the self-consistent condition
\begin{equation}
\label{eq:pop-rem-Z}
N \overline{x}(t)
\simeq
\varphi \Gamma e^{\gamma t}
\int_0^t {\rm d}u \,
e^{-(\varphi+\gamma+\sigma^2/2)u}
\sum_{i=1}^N e^{m_i u+\sigma(B_i(t)-B_i(t-u))},
\end{equation}
where \(u=t-t'\).

For fixed \(u\), the random variable
\begin{equation}
m_i u+\sigma(B_i(t)-B_i(t-u))
\end{equation}
is Gaussian with variance
\begin{equation}
S^2(u)=\Sigma_N^2u^2+\sigma^2 u.
\end{equation}
The sum over \(i\) in Eq.~\eqref{eq:pop-rem-Z} is therefore a sum such as considered in the Random Energy Model (REM) \cite{derrida1980random}. 
We recall the standard REM behavior at inverse temperature $S$
\begin{equation}
\sum_{i=1}^N e^{S \Omega_i}
\approx
\begin{cases}
N e^{S^2/2}, & S<\sqrt{2\log N},
\\[2mm]
e^{S\sqrt{2\log N}}\,\chi^{(\mu)}, & S>\sqrt{2\log N},
\end{cases}
\label{eq:REM}
\end{equation}
where the \(\Omega_i\) are independent Gaussian random variables. The first line corresponds to the high-temperature REM phase: many terms contribute to the sum, and the result is self-averaging. The second line corresponds to the frozen phase: the sum is dominated by large values of \(\Omega_i\), so sample-to-sample fluctuations remain important even as \(N\to\infty\). In the frozen phase, \(\chi^{(\mu)}\) is a one-sided L\'evy random variable of index \(\mu=S_c/S<1\), where \(S_c=\sqrt{2\log N}\), with tail \(P(\chi)\sim \chi^{-1-\mu}\).

In our case, the REM temperature depends on the parameter $u$ which is integrated in \eqref{eq:ODE_growth2}. The integral should thus be cut in two parts, where, in the first contribution, the sum is controlled by the high-temperature REM while, in the second, it is controlled by the frozen phase of the REM. Introducing \(\ell=\log N\) and the scaling \(u=\ell\tau\), one finds the crossover time which delimits the two contributions
\begin{equation}
u_c=\ell\tau_c,
\qquad
\tau_c=\frac{\sqrt{\sigma^4+4\Sigma_0^2}-\sigma^2}{\Sigma_0^2}.
\end{equation}
The integral now reads
\begin{align}
1
&=
\varphi \Bigg[
\int_0^{u_c} {\rm d}u \,
e^{-(\varphi + \gamma +\frac{\sigma^2}{2})u + \frac{S^2(u)}{2}}
\nonumber\\
&\hspace{2em}
+ \frac{1}{N} \int_{u_c}^t {\rm d}u \, \chi_{\mu(u)} \,
e^{-(\varphi + \gamma +\frac{\sigma^2}{2})u
 + \sqrt{2 \ell}\sqrt{\Sigma_0^2 u^2/(2 \ell) + \sigma^2 u}
}
\Bigg]
\label{eqI}
\end{align}
Analyzing the behavior of this integral, one retrieves the three phases found previously, as detailed in \Pfour. In the delocalized phase, the first integral, corresponding to high-temperature, self-averaging REM, dominates and yields $\gamma_{\rm deloc}=0$. The second integral is either (i) dominated by a saddle-point with a REM parameter
\begin{equation}
\mu
=
1-\frac{\Sigma_0^2}{\sigma^4},
\label{eq:mufrozen}
\end{equation}
which corresponds to the saddle-point $u^\star$ found in the previous section. It corresponds to the partially localized phase (phase III) and gives us the tail of the term that dominates the sum. It yields as expected the same $\gamma_{\rm p.l.}$. In this interpretation, the REM variable \(\chi\) describes the fluctuations of the dominant contribution to the sum. Its tail, $P(\chi)\sim \chi^{-1-\mu}$, matches the tail exponent of the sites that dominate the normalization in the previous subsection.
(ii) Or it is dominated by the limit $u\to +\infty$, which corresponds to the $\mu=0$ REM and to the localized phase. It yields $\gamma_{\rm loc} = m_1-\varphi-\frac{\sigma^2}{2}$. %

\section{Beyond self-averaging: effective description of the localized region}
\label{sec:population-noise-effective-localized}

The self-averaging approach of Sections~\ref{sec:phasdiag-Selfavg} is exact in the delocalized phase, where \(p_i=O(N^{-1})\) for all sites and the collective observables entering Eq.~\eqref{eq:eqpi} become self-averaging at large \(N\). It therefore gives the correct growth rate in that phase. More generally, it also correctly predicts the transition lines at which the delocalized solution disappears, since these lines are reached from the delocalized side, where self-averaging still holds.

Beyond that region, the approximation which was used in the previous section captures qualitatively the phase diagram, but it misses part of the fluctuations in the localized and partially localized phases, which thus require a more detailed treatment. For a site $i$, there are two sources of noise in \eqref{eq:pop-noise2}: the internal one, coming from the local multiplicative term \(\sigma x_i\eta_i\), and the external one, coming from the exchange term \(\overline{x}(t)\). In the localized and partially localized phases, the second source is no longer self-averaging, and treating it by its mean underestimates the effect of fluctuations.

In the compact-support case \eqref{eq:eqmcompact}, the noise is effectively averaged out, hence it does not affect the phase diagram. In particular, this does not affect the marginally partially localized phase, where the growth rate
\begin{equation}
\gamma_{\rm p.l.}=m_>-\varphi,
\end{equation}
is independent of \(\sigma\).%

In the generalized-Gaussian family \eqref{eq:generalized-gaussian}, the growth rate in the localized phase satisfies the exact relation
\begin{equation}
\gamma_{\rm loc}=m_1-\varphi-\frac{\sigma^2}{2} +\varphi \langle \frac{1}{N p_1} \rangle \to \Sigma_0-\varphi-\frac{\sigma^2}{2},
\end{equation}
since \(\langle \frac{1}{N p_1} \rangle=o(1)\) in this phase. This estimate remains correct within the localized phase. This is not the case for the partially localized phase. As discussed in Section~\ref{sec:phasdiag-Selfavg}, it is determined by a saddle point, and the position of this saddle point depends on the tail exponents \(\mu_i\) of the one-site distributions. These exponents depend explicitly on \(\sigma\), which can be corrected by additional fluctuations.%
Hence the self-averaging estimate of \(\gamma_{\rm p.l.}\) is not expected to be exact, and the transition line between the localized and partially localized phases is expected to shift.

The aim of this subsection is to develop an effective description of the localized regime. We proceed in two steps. We first study a simplified two-block model, which already contains the essential mechanism. We then use this picture to revisit the localized phase in the generalized-Gaussian family \eqref{eq:generalized-gaussian}. This allows us to estimate when this phase can exist and obtain an adjusted estimate of the transition line between the localized and partially localized phases.

\subsection{Description of the localized phase in terms of a two-block model}

As explained in the self-averaging analysis of Section~\ref{sec:phasdiag-Selfavg}, the normalization condition \eqref{eq:decomp_phases} contains two relevant contributions: one, which is always relevant, comes from the delocalized sites, while the second comes from the best, localized sites with \(\langle p_i \rangle=O(1)\). Moreover, the delocalized contribution is dominated by the sites with \(m_i\) close to \(0\), since most sites are close to \(0\) because the variance is \(\sim 1/\log{N}\), whereas the localized contribution comes from the best sites, \(i=O(1)\). For these sites, expanding the quantiles at first order in \(1/\log N\), one finds 
\begin{equation} \label{eq:eqmiiO1}
m_i \simeq
\Sigma_0
\left(1-\frac{\ln i}{\ln N}\right)^{\frac{1}{2b}}
\simeq
\Sigma_0-\frac{\Sigma_0}{2 b}\frac{\log i}{\log N}. 
\end{equation}
Hence the relevant sites separate into two groups: the sites from the bulk of the distribution, with \(m_i=o(1)\), and the best sites, with \(m_i=\Sigma_0+o(1)\). This suggests introducing a simplified two-block model:
\begin{itemize}
\item a first block of \(k=O(1)\) ``good'' sites with \(m_i=\Sigma_0+o(1)\);
\item a second block of \(N-k\) ``bad'' sites with \(m_i=o(1)\).
\end{itemize}
We expect this setting to reproduce the localized phase behavior of the full generalized-Gaussian model \eqref{eq:generalized-gaussian} with appropriate $m_i$, while being simpler to analyze. It will allow for a better theoretical understanding and extensive numerical checks. We focus on the localized phase, since the delocalized one is already captured by the self-averaging analysis.

\subsection*{The case \(k=1\)}
\begin{figure}
    \centering
	\includegraphics[width=0.48\linewidth]{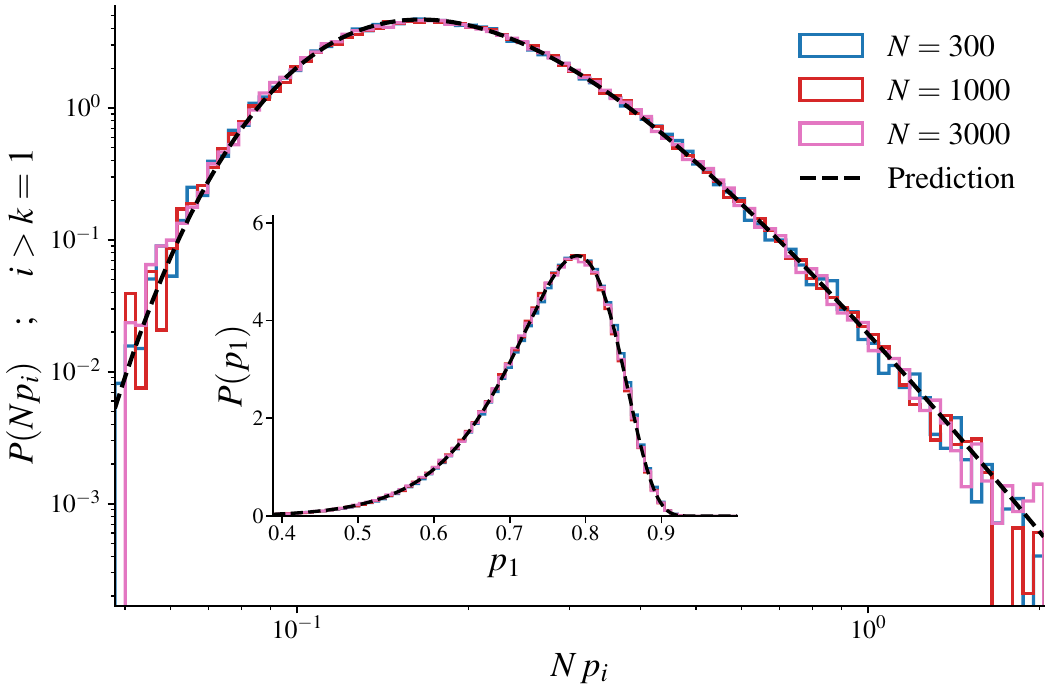}
	\includegraphics[width=0.48\linewidth]{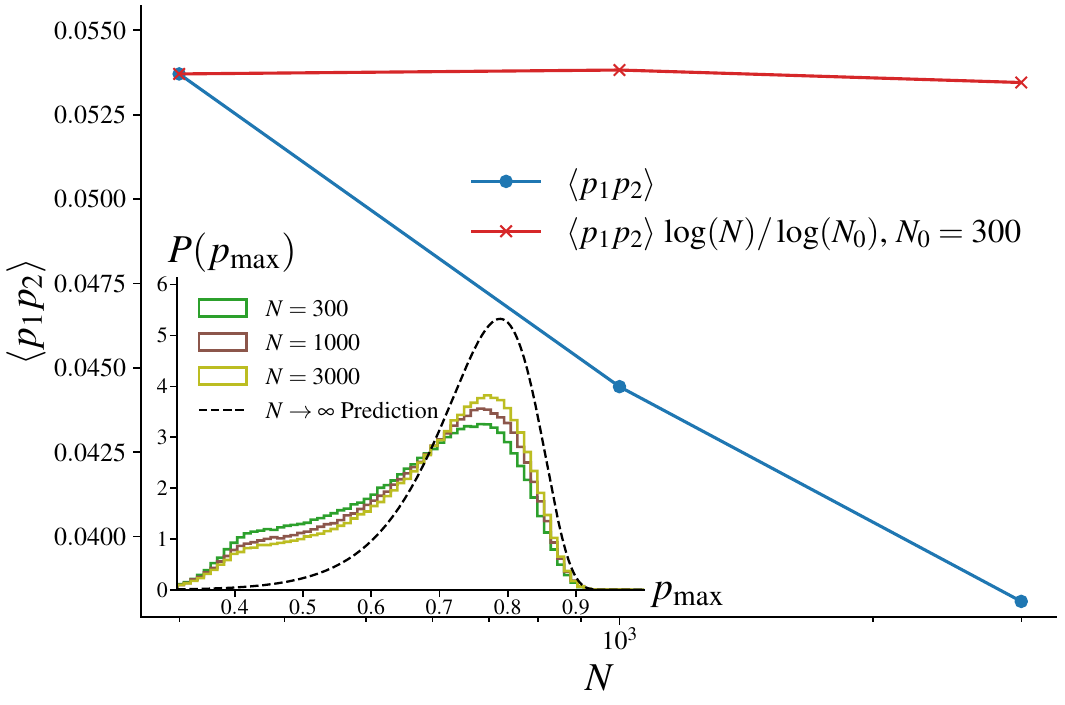}
    \caption{Two-block model with \(m_i=1\) for \(i\le k\) and \(m_i=0\) otherwise. Left (\(k=1\)): distribution of \(N p_i\) for the background sites \(i>1\), together with the theoretical prediction; inset: distribution of the dominant fraction \(p_1\). Right (\(k=2\)): \(\langle p_1 p_2\rangle\) as a function of \(N\), showing the predicted \(1/\log N\) decay associated with rare switching events; inset: distribution of \(p_{\max}\) compared with the large-\(N\) prediction. The convergence is slow due to the logarithmic corrections.}
    \label{fig:k1k2}
\end{figure}

We begin with the simplest case: a single ``good'' site with \(m_1=\Sigma_0\). In this setting, the system is localized on \(p_1=O(1)\), and the ``bad'' sites, \(j>1\), are delocalized, \(p_j \sim 1/N\). Writing Eq.~\eqref{eq:eqpi} for the site \(i=1\) and neglecting the negligible \(p_j\), \(j>1\), terms yields a closed equation for \(p_1\)
\begin{equation}
\frac{d p_1}{dt}
\simeq
\Sigma_0 p_1(1-p_1)
-\varphi\left(p_1-\frac{1}{N}\right)
-\sigma^2 p_1^2(1-p_1)
+
\sigma p_1(1-p_1)\eta_1(t).
\label{eq:p1loc}
\end{equation}
Since the localized site is also the one with largest fraction, we will denote its stationary distribution \(P_{\max}(p)\), which is given by
\begin{equation}
\label{eq:distp1}
P_{\max}(p)
=
\mathcal N\,
p^{-1-\mu_1}(1-p)^{\mu_1-1}
\exp\!\left(-\frac{A}{N p}-\frac{B}{1-p}\right),
\qquad p\in(0,1),
\end{equation}
where \(\mathcal N\) is a normalization constant and
\begin{equation}
A=\frac{2\varphi}{\sigma^2},
\qquad
B=\frac{2\varphi(1-\frac{1}{N})}{\sigma^2}
=
A+O(N^{-1}),
\end{equation}
and
\begin{equation}
\mu_1
=
-\frac{2}{\sigma^2}\left(
\Sigma_0-\varphi-\frac{\sigma^2}{2}
\right)
+O(N^{-1}).
\label{eq:mu_gamma_m1}
\end{equation}
We are thus able to obtain the distribution of \(p_1\) in the relevant \(p_1=O(1)\) region, which is not possible with the self-averaging approach. Note in particular that the tail exponent \(\mu_1<0\) differs strongly from the self-averaging prediction \(1+\frac{2(\gamma+\varphi-\Sigma_0)}{\sigma^2}=0\). In particular, contrary to the self-averaging case, the tail exponent does not depend on \(\gamma\), which has not been introduced to derive the distributions.
It can be easily recovered since \(p_1=O(1)\) from
\begin{equation}
\gamma
=
\Sigma_0-\varphi-\frac{\sigma^2}{2}
+\varphi\left\langle \frac{1}{N p_1}\right\rangle \to \Sigma_0-\varphi-\frac{\sigma^2}{2},
\label{eq:gamma_identity}
\end{equation}
which is correctly predicted by the self-averaging argument.
For the remaining delocalized sites \(j>1\), it is not possible to write a self-consistent equation, since they are coupled to the localized site \(p_1=O(1)\), which is not negligible. To obtain their distribution, we make the ansatz, supported numerically, that their distribution has the same form as in the delocalized phase, namely an inverse-gamma law, but with parameters corrected by the presence of \(p_{1}\) and its fluctuations:
\begin{equation} \label{eq:ansatzdeloc}
P_j(p)\sim p^{-1-\mu_j}\exp\!\left(-\frac{A_d}{N p}\right),
\qquad j>1,
\end{equation}
with parameters \(\mu_j\) and \(A_d\). These parameters \(\mu_j\) and \(A_d\) are constrained by the system
\begin{align}
\sum_{j>1} \langle p_j \rangle &= 1-\langle p_1 \rangle, \\
\varphi \left\langle \frac{1}{N p_j} \right\rangle &= \Sigma_0-m_j,
\end{align}
where the first equation comes from the normalization condition, and the second from the relation
\(\gamma= m_j-\varphi-\frac{\sigma^2}{2} + \varphi \left\langle \frac{1}{N p_j} \right\rangle\). Using the distribution \eqref{eq:ansatzdeloc} to obtain \(\langle p_j\rangle\) and \(\langle 1/(N p_j)\rangle\), one obtains
\begin{align}
\mu_j &= \frac{\Sigma_0-m_j}{\varphi}\,A_d,
\\
A_d &= \frac{1-\langle p_1\rangle}{\frac{1}{N}\sum_{j>k}\frac{1}{\mu_j-1}} \Rightarrow \quad
A_d \simeq \frac{\Sigma_0\,(1-\langle p_1\rangle)}{\varphi-\Sigma_0(1-\langle p_1\rangle)}.
\label{eq:paramdelock1}
\end{align}
where we used in the second line that $m_j=o(1)$. \(\langle p_1 \rangle\) is given in \eqref{eq:meanpmax} ($p_1 $ is denoted $p_{\max}$ there) and hence $A_d$ and $\mu_j$ are known. %
We thus obtain, up to the cutoffs of the delocalized sites, a complete description of the distributions in this regime. These predictions are in excellent agreement with numerics, as shown in the left panel of Fig.~\ref{fig:k1k2}, both for the localized site and for the delocalized ones, thus validating the inverse-gamma ansatz.

\subsection*{Several favorable sites: an effective description in terms of \texorpdfstring{$p_{\max}$}{pmax}}
\begin{figure}
    \centering
    \includegraphics[width=0.49\linewidth, height=4cm]{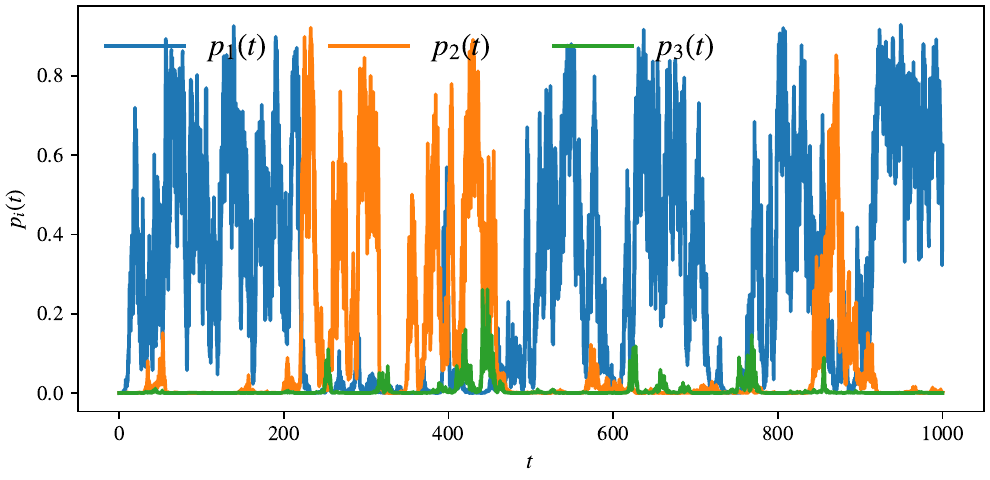}
    \includegraphics[width=0.49\linewidth, height=4cm]{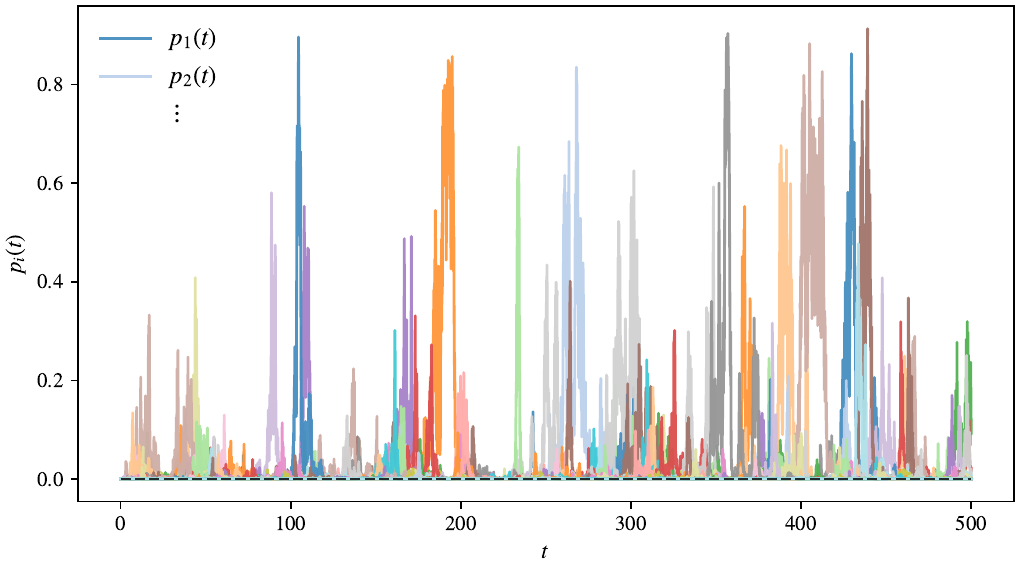}
    \caption{Time evolution of the fractions \(p_i(t)\) for Gaussian \(m_i\) with \(\Sigma_0=1\) and \(\varphi=0.2\). Left: localized regime (\(\sigma=1.2\)), where one site dominates at a given time and switches are rare. In this realization the competition is essentially between the two best sites. Right: partially localized regime (\(\sigma\simeq 0.71\)), with \(1\le i \le 30\), where several favorable sites successively acquire an \(O(1)\) fraction and the dominant site changes rapidly over time. A large number of favorable sites can localize.}
    \label{fig:pi-time-loc-deloc}
\end{figure}

We now consider a block of \(k>1\) ``good'' sites with comparable growth rates. Numerically, one finds that even in this case the system is typically localized on only one of them at any given time: one site carries an \(O(1)\) fraction of the total population, while the others remain at fractions of order \(1/N\). The identity of the dominant site changes over time, but these switching events are rare.
This is illustrated qualitatively in the left panel of Fig.~\ref{fig:pi-time-loc-deloc}: a single site dominates for long periods before a rare switch occurs.%

The maximum fraction is typically given by one of the good sites, hence
\begin{equation}
p_{\max}(t) \simeq \max_{i\le k} p_i(t).
\end{equation}
In the localized regime, the total fraction carried by the good block can then be approximated by
\begin{equation}
\sum_{i\le k} p_i(t)\simeq p_{\max}(t),
\qquad
Y_2=\sum_{i=1}^N p_i(t)^2\simeq p_{\max}(t)^2.
\end{equation}
The reason is that simultaneous events in which two favorable sites both carry \(O(1)\) fractions are rare. Indeed, if one favorable site is dominant, while another remains delocalized at \(p_j\sim N^{-1}\), then the latter needs, assuming exponential growth \(\sim e^{a t}\), a time of order
\begin{equation}
t_{\mathrm{switch}}\sim \frac{\log N}{a}
\end{equation}
to grow from \(O(N^{-1})\) to \(O(1)\). Thus the switching events are rare, and since we have $k$ sites, we expect these to happen at a rate $(k-1)/\log{N}$. In particular, the product $p_i p_j$ ($i \neq j$) is $O(1)$ only if one of them corresponds to $p_{\rm max}$ and we are near a switching event (otherwise it is $O(1/N)$). Since it happens at rate $1/\log{N}$, we thus expect in particular
\begin{equation}
\langle p_i p_j\rangle  \sim \frac{1}{\log N},
\qquad i\neq j,\quad i,j\le k,
\end{equation}
which is checked numerically in Fig.~\ref{fig:k1k2} (right) for \(k=2\).

The law of \(p_{\max}\) at large \(N\) is thus expected to be given by the distribution \(P_{\max}(p)\) given in \eqref{eq:distp1} and obtained in the \(k=1\) case. This is checked numerically for \(k=2\) in the inset of Fig.~\ref{fig:k1k2} (right). While the qualitative behavior seems coherent, \(1/\log{N}\) corrections make the quantitative validation difficult. Likewise, one expects the distribution of the ``bad'' sites to still be given by their \(k=1\) distribution \eqref{eq:ansatzdeloc}.
While we know that the distribution of a good site is given by \(P_{\max}\) whenever it is localized, we do not yet know its full stationary distribution. It is therefore natural to condition on whether or not site \(i\) is localized, and to decompose the distribution of \(p_i\) as
\begin{equation}
P_i(p)
=
P(i,\mathrm{loc})\,P_{\max}(p)
+
\bigl(1-P(i,\mathrm{loc})\bigr)\,P_i^{\mathrm{deloc}}(p),
\end{equation}
where \(P(i,\mathrm{loc})\) is the probability that site \(i\) is the localized one, \(P_{\max}\) is the distribution of the localized site, and \(P_i^{\mathrm{deloc}}\) is the conditional distribution of \(p_i\) given that site \(i\) is not localized.

In the special case where the \(k\) favorable sites have the same $m_i$, symmetry implies $P(i,\mathrm{loc})=\frac{1}{k}$.
Thus the only remaining quantity to determine is \(P_i^{\mathrm{deloc}}(p)\). Since such a site is delocalized, we again assume that its distribution has the inverse-gamma form
\begin{equation}
P_i^{\rm deloc}(p)\sim p^{-1-\mu_d}\exp\!\left(-\frac{A_d}{N p}\right),
\end{equation}
with an \(O(1)\) upper cutoff set by \(p_{\max}\). The exponent \(\mu_d\) can no longer be obtained from \eqref{eq:paramdelock1}, since that relation was derived for the ``bad'' sites. 
As argued earlier, the switching of identity of the localized site, i.e. when a ``good'' delocalized site becomes localized, occurs at a rate \(1/\log{N}\). Equivalently, for a given good delocalized site, the probability of being \(p_i=O(1)\) at a given time, that is, on the verge of becoming the localized site, should scale as \(1/\log N\). On the other hand, under the inverse-gamma ansatz, this probability is obtained by integrating the tail of $ P_i^{\rm deloc}(p)$ over the \(O(1)\) region:
\begin{equation}
\int_v^{c} P_i^{\rm deloc}(p) \sim \int_v^{p_c} dp\, p^{-1-\mu_d}\exp\!\left(-\frac{A_d}{N p}\right),
\end{equation}
where \(v=O(1)\) is fixed and \(c=O(1)\) is the upper cutoff set by \(p_{\max}\). Since the exponential factor is irrelevant in this range, one gets
\begin{equation}
\int_v^{c} P_i^{\rm deloc}(p) \sim N^{-\mu_d}.
\end{equation}
Matching this with the estimate \(1/\log N\) gives
\begin{equation}
N^{-\mu_d}\sim \frac{1}{\log N},
\end{equation}
and therefore
\begin{equation}
\mu_d \to 0,
\end{equation}
completing the characterization of the distributions.

\subsection*{Distinct good sites and rank-dependent localization probabilities}
As a last step before returning to the generalized-Gaussian family \eqref{eq:generalized-gaussian}, we now consider the two-block scenario where the good sites have different \(m_i\) and are no longer equivalent. We consider growth rates of the form
\begin{equation}
m_i=\Sigma_0-\frac{\Sigma_0}{2 b}\frac{\log i}{\log N},
\end{equation}
which corresponds to the tail of the generalized-Gaussian family \eqref{eq:generalized-gaussian}, with the Gaussian case recovered for \(b=1\).
Because the corrections of \(m_i\) are only of order \(1/\log N\), the distribution of the fraction \(p_{\max}\) is still given by \(P_{\max}\) up to logarithmic corrections. The main difference is that each favorable site now has its own tail exponent \(\mu_i\) and therefore its own probability of becoming localized.
The goal of the following argument is to determine the distribution of the delocalized ``good'' sites and to obtain how the probability for a site to be localized depends on its rank \(i\).

As earlier, we make an inverse-gamma ansatz for the delocalized good sites
\begin{equation}
P_i^{\mathrm{deloc}}(p)\sim p^{-1-\mu_i}\exp\!\left(-\frac{A_d}{Np}\right),
\label{eq:Pi_deloc_ansatz}
\end{equation}
where now \(\mu_i\) depends on the rank \(i\), and with an \(O(1)\) upper cutoff \(c\) set by \(p_{\max}\).

In order to estimate $\mu_i$, we use the exact identity \eqref{eq:gammaexact1}:
\begin{align}
\gamma+\varphi+\frac{\sigma^2}{2}
&= m_i+\varphi \left\langle \frac{1}{N p_i} \right\rangle \nonumber\\
&\simeq m_i+\varphi\,\bigl(1-P(i,\mathrm{loc})\bigr)
\left\langle \frac{1}{N p_i} \right\rangle_\mathrm{deloc},
\label{eq:gamma_rank_dep}
\end{align}
where $\langle \dots \rangle_\mathrm{deloc}$ denotes the average given that it is delocalized, i.e. with respect to the distribution $P_i^{\rm deloc}(p)$. We used in the last line that the localized contribution to \(\langle 1/(N p_i)\rangle\) is negligible: when the site \(i\) is localized, \(p_i=O(1)\), so \(1/(N p_i)=O(1/N)\).%
For sufficiently large rank \(i\), one may assume that \(1-P(i,\mathrm{loc})\simeq 1\), such that 
\begin{equation}\label{eq:eqlabmi}
m_i+\varphi \left\langle \frac{1}{N p_i} \right\rangle_\mathrm{deloc} = \text{const.}
\end{equation}
is independent of \(i\). In Appendix~\ref{app:sigma-noise-deloc-moments}, we estimate \(\left\langle \frac{1}{N p_i} \right\rangle_\mathrm{deloc}\) with \eqref{eq:Pi_deloc_ansatz}. We also show, that since the $i$ dependence must cancel out in \eqref{eq:eqlabmi}, that one obtains at large $i$
\begin{equation}
\mu_i
\simeq
\frac{A_d\Sigma_0}{2 b\,\varphi}\,
\frac{\log i}{\log N}.
\label{eq:mui_largei}
\end{equation}

\begin{figure}[t]
    \centering
    \includegraphics[width=0.49\linewidth]{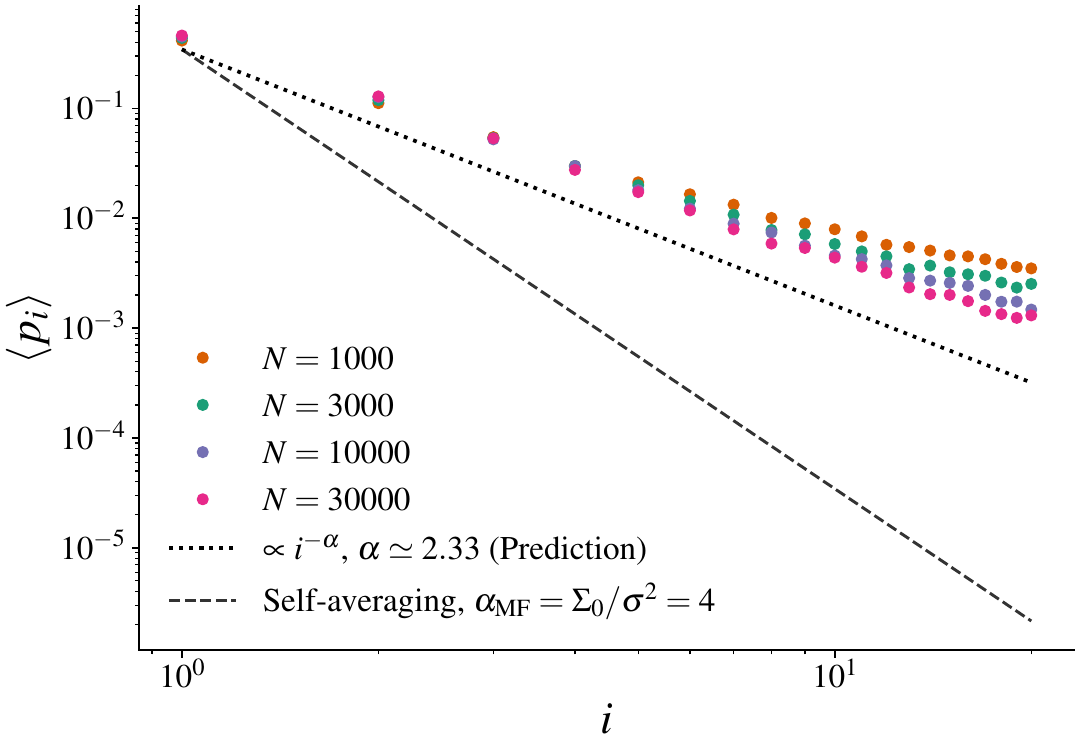}
    \includegraphics[width=0.49\linewidth]{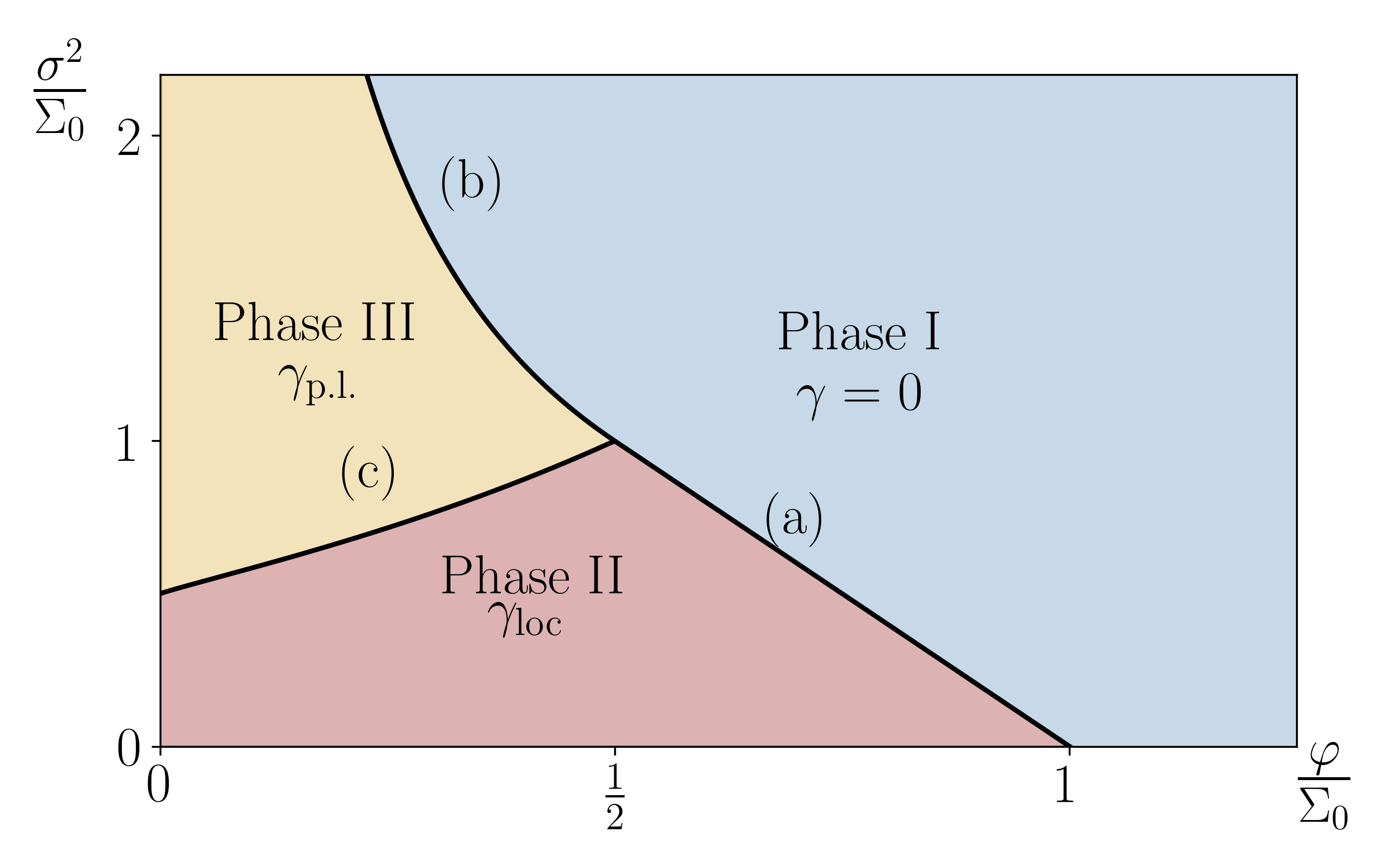}
    \caption{
    Left: mean stationary weights \(\langle p_i\rangle\) as a function of the rank \(i\) for a block of \(k=30\) favorable sites with \(m_i=\Sigma_0(1-\log i/(2\log N))\) for \(i\le k\) and \(m_i=0\) for \(i>k\), with \(\Sigma_0=1\), \(\varphi=0.2\), and \(\sigma=0.5\) for several $N$. The black dotted line is the effective localized prediction \(\langle p_i\rangle \propto i^{-\alpha}\), with \(\alpha=A_d\Sigma_0/(2b\varphi)\), while the black dashed line shows the self-averaging estimate \(\langle p_i\rangle \propto i^{-\Sigma_0/\sigma^2}\). Right: Adjusted Gaussian phase diagram. Line (a) is the localized-delocalized boundary, line (b) the partially localized-delocalized boundary, and line (c) the corrected localized-partially localized boundary obtained from the effective \(p_{\max}\) description.
    }
    \label{fig:newphaserank_mean_pi_k30}
\end{figure}

To quantify how close a site is to localization, we use as a proxy the probability that it reaches the upper edge of the delocalized distribution. Specifically, for a favorable site \(i\), whose typical fraction is \(p_i=O(1/N)\), we define
\begin{equation}
P_>(i):=\int_v^{c} P_i^{\rm deloc}(p)\,dp,
\qquad v=O(1),
\end{equation}
with \(v\) fixed of order \(O(1)\), but below the effective cutoff determined by \(p_{\max}\). The quantity \(P_>(i)\) therefore measures the probability that a delocalized site is observed with an \(O(1)\) fraction. Using \eqref{eq:Pi_deloc_ansatz},
\begin{align}
P_>(i)
&=
\frac{1}{\mathcal N_i^{\mathrm{deloc}}}
\int_v^c dp\, p^{-1-\mu_i}\exp\!\left(-\frac{A_d}{Np}\right) \simeq
\frac{1}{\mathcal N_i^{\mathrm{deloc}}}
\int_v^c dp\, p^{-1-\mu_i} \\
\end{align}
where \(c=O(1)\) is the upper cutoff set by \(p_{\max}\).
Now \(\mu_i \to 0\) as \(N \to +\infty\), hence \(\int_v^c dp\, p^{-1-\mu_i}
=
\frac{v^{-\mu_i}-c^{-\mu_i}}{\mu_i}
\sim
\log\!\left(\frac{c}{v}\right)\) is \(O(1)\) and therefore
\begin{equation}
P_>(i)
\sim
\frac{1}{\mathcal N_i^{\mathrm{deloc}}} \sim \frac{\log i}{\log N}\,
i^{-\frac{A_d\Sigma_0}{\varphi 2 b}}.
\label{eq:Pgt_i_intermediate}
\end{equation}
The details of the estimation of \(\mathcal N_i^{\mathrm{deloc}}\) are in Appendix~\ref{app:sigma-noise-deloc-moments}. We retrieve the \(1/\log{N}\) as in the identical \(m_i\) case. Equation~\eqref{eq:Pgt_i_intermediate} also shows that the probability for the \(i\)-th favorable site to lie near the localized sector decays as a power of its rank, up to a logarithmic correction. In particular, for \(i\le k\), one expects \(\langle p_i \rangle \simeq P(i, {\rm loc}) \langle p_{\rm max} \rangle\) such that it should scale as
\begin{equation}
\langle p_i \rangle \sim i^{-\frac{A_d\Sigma_0}{\varphi 2 b}}.
\end{equation}
The rank dependence is verified numerically in the left panel of Fig.~\ref{fig:newphaserank_mean_pi_k30}. Note that under the self-averaging assumption in Section~\ref{sec:phasdiag-Selfavg}, we had obtained for the localized phase and for the localized sites, \(\langle p_i \rangle \sim N^{-\mu_i}\) with \(\mu_i=1+\frac{2(\gamma+\varphi-m_i)}{\sigma^2}\) and \(\mu_1=0\). This leads to a different, incorrect, scaling \(\langle p_i \rangle \sim i^{-\frac{\Sigma_0}{\sigma^2}}\), for \(i=O(1)\). 
Thus, the effective description of the localized phase correctly captures the observed dependence on the rank \(i\), in contrast with the self-averaging prediction, which, in particular, underestimates the amplitude of the fluctuations.

It is this rank-dependent decay that controls the transition between the localized and partially localized phases, as we now show.

\subsection{Adjusted transition line for Phase II-III for the generalized-Gaussian family}
We now return to the generalized-Gaussian case \eqref{eq:generalized-gaussian}, in which all the \(m_i\) are given by the quantiles of the distribution. In the localized regime, the asymptotic growth rate remains
\begin{equation}
\gamma=\Sigma_0-\varphi-\frac{\sigma^2}{2}.
\end{equation}

The main difference with the two-block model is that the \(m_i\) are now spread over the whole interval \([m_N,m_1]\). However, in the localized phase, the normalization is still controlled by two regions: the finitely many best sites, for which
\begin{equation}
m_i \simeq \Sigma_0-\frac{\Sigma_0}{2 b}\frac{\log i}{\log N},
\end{equation}
and the extensively many sites for which \(m_i\simeq 0\). As a result, the predictions obtained in the two-block case remain valid.

Indeed, for the delocalized site, which might now have \(m_i=O(1)\), the parameter \(A_d\) is fixed by the normalization condition and is therefore unchanged, \(A_d \simeq \frac{\Sigma_0\,(1-\langle p_{\max}\rangle)}{\varphi-\Sigma_0(1-\langle p_{\max}\rangle)}\). Likewise, the relation
\begin{equation}
\mu_j = \frac{\Sigma_0-m_j}{\varphi}\,A_d
\end{equation}
holds for any delocalized site, regardless of the value of \(m_j\). The distribution of the delocalized sites is therefore still described by the formulas obtained in the two-block case. 

As for the potentially localized sites, their time evolution remains effectively decoupled from the delocalized sites, and is thus again correctly described by the two-block case.

We now wish to establish the transition line from the localized to the partially-localized phase.
A key property of the localized phase is that, at a given time, the system is, in the large \(N\) limit, localized on only one site. The identity of this site switches at a vanishing rate \(\sim 1/\log{N}\). On the other hand, if the number of sites that can localize starts diverging, the switching rate no longer vanishes and the system is no longer localized on a single site. One enters the partially localized phase, as illustrated in the right panel of Fig.~\ref{fig:pi-time-loc-deloc}. In order to obtain the transition line between the localized and partially localized phase, one should therefore estimate when the number of sites that can localize starts diverging.
Let us now consider the localized phase, close to the transition to the partially localized phase. We have established in the previous section that, for a site \(i=O(1)\), its probability to be localized scales as
\be
P(i, {\rm loc}) \sim i^{-\frac{A_d\Sigma_0}{\varphi 2 b}}.
\ee
The total number of sites that can localize thus scales as 
\be
\sum_{i=O(1)} P(i, {\rm loc}) \sim \sum_{i=O(1)} i^{-\frac{A_d\Sigma_0}{\varphi 2 b}}.
\ee
Here \(i=O(1)\) means that the sum is restricted to the best sites, say \(1\le i\le i_c\), where \(i_c\gg 1\) but remains \(i_c=O(1)\) with respect to \(N\). If \(\frac{A_d\Sigma_0}{\varphi 2 b}>1\), this sum converges, the effective number of sites that can localize is finite and we are in the localized phase. It starts diverging at 
\begin{equation}
\frac{A_d\Sigma_0}{\varphi 2 b}=1, \qquad A_d = \frac{\Sigma_0\,(1-\langle p_{\max}\rangle)}{\varphi-\Sigma_0(1-\langle p_{\max}\rangle)}
\end{equation}
which thus yields the transition line with the partially localized phase.
The transition line is thus given by $\varphi_c^{II-III}$ satisfying
\begin{equation}
\frac{\Sigma_0}{ 2 b}\,
\frac{1-\langle p_{\max}\rangle}
{\Sigma_0(1-\langle p_{\max}\rangle)-\varphi_c^{II-III}}
=1,
\label{eq:translin}
\end{equation}
where, from the distribution $P_{\max}$ \eqref{eq:distp1}
\begin{align} \label{eq:meanpmax}
\langle p_{\max}\rangle
&=
1-\mu_m e^{A_m}(A_m)^{\mu_m}\Gamma(-\mu_m,A_m), \\
A_m &\equiv \frac{2\varphi}{\sigma^2}, \nonumber\\
\mu_m &\equiv \frac{2}{\sigma^2}
\left(\Sigma_0-\varphi-\frac{\sigma^2}{2}\right). \nonumber
\end{align}
This yields the adjusted transition line between the localized and partially localized phases, shown as line (c) in the right panel of Fig.~\ref{fig:newphaserank_mean_pi_k30}.

\section{Discussion}
The main conclusion of Section~\ref{sec:population-noise-effective-localized} is that, in the presence of noise, the localized phase is effectively controlled by the maximal fraction \(p_{\max}\), but not by a fixed site as in the deterministic case \(\sigma=0\). As long as only finitely many favorable sites can compete for localization, the system is localized at any given time on a single site. The transition to the partially localized phase occurs when the set of potentially localized sites becomes too large for this effective description to remain valid.

This picture also clarifies the limitation of the self-averaging analysis of Section~\ref{sec:phasdiag-Selfavg}. That approach correctly captures the existence of a partially localized phase and the transition lines involving the delocalized phase, but it underestimates the extent of the partially localized region because it neglects the fluctuations associated with the localized site.

An important remaining question concerns the role of redistribution in the growth rate. In the case \(\sigma=0\), and also for \(\sigma>0\) in the localized and delocalized phases, redistribution is always detrimental to growth. In the localized phase, growth is essentially controlled by the most favorable site, so transferring population away from it lowers \(\gamma\). In the delocalized phase, the largest \(m_i\) still carry the largest fractions, so increasing \(\varphi\) again reduces the growth rate. In the partially localized phase, under the self-averaging approximation, \(\gamma\) is likewise a decreasing function of \(\varphi\). A remaining open question is to determine the true expression of \(\gamma\) in the partially localized phase.

Note that for two sites (\(N=2\)) with \(m_1>m_2\), Ref.~\cite{bouchaudtax} showed that \(\gamma(\varphi)\) is not monotonically decreasing in \(\varphi\), but instead reaches a maximum at an intermediate value. Excessive redistribution suppresses growth by continuously draining population from the favorable site, whereas a small amount of redistribution allows the system to benefit from occasional favorable fluctuations of the disadvantaged site. One may therefore expect a similar mechanism to persist in the partially localized phase.

In particular, at large but finite \(N\), the inset of the right panel of Fig.~\ref{fig:figplext} shows that there is still an optimal redistribution rate \(\varphi\) maximizing growth. By contrast, in the limit \(N\to\infty\), the estimated growth rate \(\gamma_\infty\), obtained from a fit of the form
\begin{equation}
\gamma(N)=\gamma_\infty+\frac{c}{\log N},
\end{equation}
appears to be a strictly decreasing function of \(\varphi\). A possible explanation is that, in the \(N\to\infty\) regime, most of the redistributed population is absorbed by the typical sites with \(m_i\simeq 0\), so that a temporarily favorable site recovers only a fraction \(1/N\) of the total redistributed population. The scaling of the optimal redistribution rate still remains to be understood in more detail.

\chapter{Mean-Field Growth with Quenched Heterogeneity and Resetting}
\label{sec:population-resetting}

\begin{figure}
    \centering
    \includegraphics[width=0.62\linewidth]{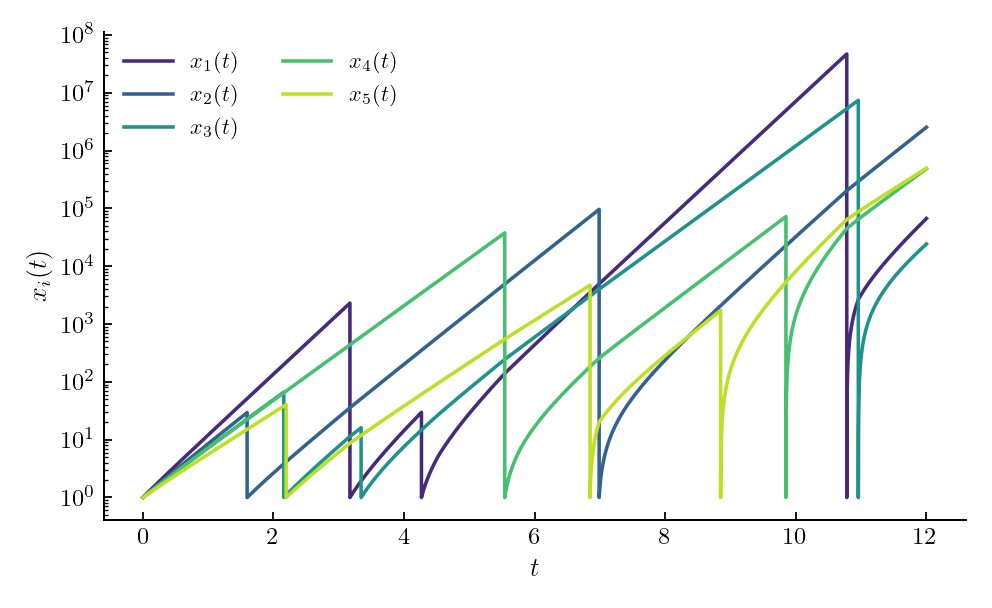}
    \caption{Example of trajectories \(x_1(t),\dots,x_5(t)\) in the resetting model, for Gaussian heterogeneous growth rates and parameters \(\varphi=0.2\), \(r=0.2\). The intermittent downward jumps correspond to resetting events, which interrupt growth episodes and prevent a single site from remaining dominant forever.}
    \label{fig:resetting-example-trajectories}
\end{figure}

We now study a different stochastic extension of the deterministic growth model, namely resetting \cite{SatEvansReset}. In a population context, resetting may be interpreted as local extinction followed by recolonization through migration. In an economic context, it may represent bankruptcy followed by restart through redistribution. Unlike multiplicative noise, which continuously modifies the growth, resetting acts intermittently by interrupting growth sequences. Its main effect is therefore to prevent persistent accumulation on a single site, while still allowing broad distributions and, under suitable conditions, a partially localized regime. An example of a realization is shown in Fig.~\ref{fig:resetting-example-trajectories}.

In this section, we first define the model and present the main results. We then derive the stationary one-site distribution in the self-averaging regime and obtain the corresponding equation for the asymptotic growth rate. We discuss why complete localization is no longer sustainable in the presence of resetting. Finally, we estimate the partially localized regime for both the generalized-Gaussian and compact-support classes defined in Section~\ref{sec:mdist}.

\section{Model}

We consider \(N\) sites with populations \(x_i(t)\), deterministic heterogeneous growth rates \(m_i\), uniform redistribution at rate \(\varphi\), and Poissonian resetting:
\begin{equation}
\frac{d x_i(t)}{dt}
=
(m_i-\varphi)x_i(t)+\varphi \overline{x}(t),
\qquad
x_i \to x_0 \ \text{at rate } r_i,
\label{eq:pop-reset}
\end{equation}
where
\begin{equation}
\overline{x}(t)=\frac{1}{N}\sum_{j=1}^N x_j(t).
\end{equation}
The resetting rates \(r_i\) may in principle depend on \(i\), although most explicit formulas below will be given for the simpler case of a uniform resetting rate \(r_i=r\).

As in the previous sections, the two main observables are the asymptotic growth rate \(\gamma\) and the stationary fractions
\begin{equation}
p_i(t)=\frac{x_i(t)}{\sum_j x_j(t)}.
\end{equation}

\section{Main results}

 Resetting suppresses the fully localized phase: a favorable site can become localized only if it avoids resetting for a time of order \(\log N\), which becomes unlikely for a finite set of sites as \(N\to\infty\). There can thus only be a delocalized phase and a partially localized one.

For the {\bf generalized-Gaussian} family \eqref{eq:generalized-gaussian}, one finds a delocalized phase and a partially localized phase. The corresponding growth rate is
\begin{equation}
\gamma=
\begin{cases}
0, & \text{delocalized},\\[1mm]
\Sigma_0(1-u^\star)^{\frac{1}{2b}}-\varphi-\dfrac{r}{u^\star}, & \text{partially localized},
\end{cases}
\end{equation}
where \(u^\star\) is determined implicitly by
\begin{equation}
r=\frac{\Sigma_0}{2b}\,(u^\star)^2(1-u^\star)^{\frac{1}{2b}-1}.
\end{equation}
At fixed \(x_0>0\), the transition to the delocalized phase is then obtained by imposing \(\gamma=0\), which gives
\begin{equation}
\varphi_c=\Sigma_0(1-u)^{\frac{1}{2b}}-\frac{r}{u},
\qquad
r=\frac{\Sigma_0}{2b}\,u^2(1-u)^{\frac{1}{2b}-1}.
\end{equation}

The complete phase diagram for the Gaussian case \(b=1\) is shown in the left panel of Fig.~\ref{fig:resetting-phase-and-growth}.

For the {\bf compact-support} class \eqref{eq:eqmcompact}, the results and their interpretation are the same as in the multiplicative-noise case, up to the shift \(m\to m-r\) induced by resetting. One finds a delocalized phase and a marginally partially localized regime. The growth rate is obtained from the deterministic compact-support formulas \eqref{eq:rescomptsig0}	after the replacement \(m_i\to m_i-r\). In particular,
\begin{equation}
\gamma=
\begin{cases}
\gamma_{\rm deloc}^{(r)}(\varphi), & \text{delocalized},\\[1mm]
m_>-\varphi-r, & \text{marginally partially localized},
\end{cases}
\end{equation}
where \(\gamma_{\rm deloc}^{(r)}\) is the delocalized result with the shift.  %

\section{Self-averaging regime}

We begin with the self-averaging regime, in which \(\overline{x}(t)\simeq \langle \overline{x}(t)\rangle\) and relative fluctuations vanish at large \(N\). In the large time limit, we write
\begin{equation}
x_i(t)=\tilde x_i(t)\,e^{\gamma t},
\end{equation}
where \(\gamma\) is chosen so that the rescaled variables \(\tilde x_i(t)\) remain of order one. Between resets, the rescaled dynamics reads
\begin{equation}
\frac{d\tilde x_i(t)}{dt}
=
(m_i-\varphi-\gamma)\tilde x_i(t)
+\frac{\varphi}{N}\sum_{j=1}^N \tilde x_j(t).
\label{eq:tilde_dynamics_selfavg_reset}
\end{equation}
In the self-averaging regime, we assume that
\begin{equation}
\frac{1}{N}\sum_{j=1}^N \tilde x_j(t)\simeq x_m,
\end{equation}
with \(x_m = \frac{1}{N}\sum_{j=1}^N \langle \tilde x_j(t) \rangle\) approximately constant at large \(N\). Equation~\eqref{eq:tilde_dynamics_selfavg_reset} then reduces to
\begin{equation}
\frac{d\tilde x_i(t)}{dt}
=
(m_i-\varphi-\gamma)\tilde x_i(t)+\varphi x_m.
\label{eq:tilde_dyn_reduced_reset}
\end{equation}

Writing $p_i(t) = \frac{\tilde x_i(t)}{N x_m}$, one obtains
\begin{equation}
\frac{d p_i}{dt}
=
(m_i-\varphi-\gamma)p_i+\frac{\varphi}{N}.
\label{eq:eqp_selfavg_reset}
\end{equation}

\paragraph{Stationary distribution under resetting.}
Equation~\eqref{eq:eqp_selfavg_reset} has the generic form
\begin{equation}
\frac{d p_i}{dt}=a_i p_i+b_i,
\qquad
a_i:=m_i-\varphi-\gamma,
\qquad
b_i:=\frac{\varphi}{N}.
\label{eq:dp_aibi_reset}
\end{equation}
With post-reset value \(p_{0,i}=p_0\), its solution between two reset events is
\begin{equation}
p_i(t)=p_i^\star +(p_0-p_i^\star)e^{a_i t},
\qquad a_i\neq 0.
\label{eq:p_sol_reset}
\end{equation}
where
\begin{equation}
p_i^\star=-\frac{b_i}{a_i}.
\end{equation}

For Poissonian resetting at rate \(r_i\), the stationary distribution is given by the renewal formula \cite{Evans_2020}
\begin{equation} 
P_i(p)=r_i\,\big|\tau'(p)\big|\,e^{-r_i\tau(p)},
\label{eq:state_resetting_pi}
\end{equation}
where \(\tau(p)\) is the time needed for the deterministic dynamics to reach the value \(p\) after a reset. Two cases must be distinguished.

If \(a_i>0\), the dynamics grows away from the reset point and the support is \(p\ge p_0\). Inverting Eq.~\eqref{eq:p_sol_reset}, one gets
\begin{equation}
\tau(p)=\frac{1}{a_i}\log\!\left(\frac{p+\frac{b_i}{a_i}}{p_0+\frac{b_i}{a_i}}\right),
\label{eq:tau_of_p_reset}
\end{equation}
and hence
\begin{equation}
\big|\tau'(p)\big|=\frac{1}{a_i p+b_i}.
\label{eq:tauprime_reset}
\end{equation}
Substituting \eqref{eq:tau_of_p_reset} and \eqref{eq:tauprime_reset} into \eqref{eq:state_resetting_pi}, one obtains
\begin{equation}
P_i(p)
=
\mu_i\left(p_0+\frac{b_i}{a_i}\right)^{\mu_i}
\left(p+\frac{b_i}{a_i}\right)^{-1-\mu_i},
\qquad
\mu_i:=\frac{r_i}{m_i-\varphi-\gamma}.
\label{eq:stat_p_growth_final_mu_reset}
\end{equation}
Thus the stationary distribution has a power-law tail,
\begin{equation}
P_i(p)\sim p^{-1-\mu_i}.
\end{equation}

If instead \(a_i<0\), the deterministic dynamics relaxes towards the stable fixed point
\begin{equation}
p_{c,i}:=\frac{b_i}{|a_i|}=\frac{\varphi}{N(\gamma+\varphi-m_i)},
\end{equation}
and the support is the compact interval \([p_0,p_{c,i}]\). In that case,
\begin{equation}
\tau(p)=\frac{1}{|a_i|}\log\!\left(\frac{p_{c,i}-p_0}{p_{c,i}-p}\right),
\end{equation}
which yields
\begin{equation}\label{eq:distimu}
P_i(p)
=
\frac{r_i}{|a_i|}(p_{c,i}-p_0)^{-|\mu_i|}(p_{c,i}-p)^{|\mu_i|-1},
\qquad
\mu_i=\frac{r_i}{m_i-\varphi-\gamma}<0.
\end{equation}
Near the upper edge \(p_{c,i}\), the stationary density behaves as
\begin{equation}
P_i(p)\sim (p_{c,i}-p)^{|\mu_i|-1},
\end{equation}
which might diverge but is always integrable.
These expressions were derived under the self-averaging assumption, which breaks when \(p_i=O(1)\), since fluctuations then become relevant. They should therefore not be expected to describe the \(p_i=O(1)\) region.

\paragraph{Mean fraction and self-consistency equation.}
The stationary mean can be computed from the distribution \eqref{eq:stat_p_growth_final_mu_reset} or \eqref{eq:distimu}, or directly from the renewal equation:
\begin{equation}
\langle p_i\rangle
=
\int_0^\infty r_i e^{-r_i t}\,p_i(t)\,dt.
\end{equation}
Using Eq.~\eqref{eq:p_sol_reset}, one finds
\begin{equation}
\langle p_i\rangle
=
\frac{\varphi/N+r_i p_0}{\gamma+\varphi+r_i-m_i}.
\label{eq:mean_pi_mu_gt1_reset}
\end{equation}
Under the self-averaging assumption, the growth rate \(\gamma\) and the mean rescaled population \(x_m\) are then determined by the normalization condition
\begin{equation}
\sum_{i=1}^N \langle p_i\rangle=1,
\end{equation}
with $p_0=\frac{x_0}{N x_m}$, that is,
\begin{equation}
\sum_{i=1}^N
\frac{\varphi/N+r_i p_0}{\gamma+\varphi+r_i-m_i}
=1.
\label{eq:eqgammaSA_reset}
\end{equation}
Equivalently, one may write the self-consistency condition directly for \(x_m\):
\begin{equation}
x_m
=
\frac{1}{N}\sum_{i=1}^N
\frac{\varphi x_m+r_i x_0}{\gamma+\varphi+r_i-m_i}.
\label{eq:xm_selfcons_reset}
\end{equation}

In the growing self-averaging phase, \(x_m\) grows exponentially while the reset value \(x_0\) remains fixed, so that \(p_0=x_0/(N x_m)\to 0\). The self-consistency equation then reduces to
\begin{equation}
1
=
\frac{1}{N}\sum_{i=1}^N
\frac{\varphi}{\gamma+\varphi+r_i-m_i}.
\label{eq:xm_selfcons_resetgrowth}
\end{equation}
In that regime, resetting simply amounts to the replacement
\begin{equation}
m_i\longrightarrow m_i-r_i
\end{equation}
in the deterministic equation for \(\gamma\), Eq.~\eqref{eq:eq_gamma}.

When \(\gamma=0\), Eq.~\eqref{eq:xm_selfcons_reset} instead yields a self-consistent equation for \(x_m\),
\begin{equation}
x_m
=
\frac{1}{N}\sum_{i=1}^N
\frac{\varphi x_m+r_i x_0}{\varphi+r_i-m_i}.
\end{equation}
In that case, the reset value \(x_0\) remains relevant and fixes the scale of the mean population.

\paragraph{Homogeneous case}
In the homogeneous case, where \(m_i=m\) and \(r_i=r\) for all \(i\), the system is growing for \(m>r\), in which case
\begin{equation}
\gamma=m-r>0,
\end{equation}
and the reset point is asymptotically irrelevant. For \(m<r\), the growth rate saturates at \(\gamma=0\), and Eq.~\eqref{eq:xm_selfcons_reset} gives
\begin{equation}
x_m=\frac{r x_0}{r-m}.
\end{equation}
The exponents are then
\begin{equation}
\mu=
\begin{cases}
\dfrac{r}{r-\varphi}, & m>r,\\[2mm]
\dfrac{r}{m-\varphi}, & m<r.
\end{cases}
\end{equation}
In both cases, one finds
\begin{equation}
\mu\in (-\infty,0)\cup[1,+\infty),
\end{equation}
where we recall that $\mu<0$ corresponds to a distribution with compact support. The regime \(0<\mu<1\), which would signal partial localization, never occurs. Thus the homogeneous system is always self-averaging, as expected.

\section{Partially localized regime}

We now turn to the heterogeneous case. A fully localized phase is no longer possible: under exponential growth, it takes a time $\sim \log{N}$ for a site to go from a fraction $p_i \sim 1/N$ to $p_i = O(1)$, while resetting interrupts this growth on \(O(1)\) timescales.
The minimum number of favorable sites required to sustain some localization must therefore grow with the system size, while remaining subextensive. An estimate is given in Appendix~\ref{app:resetting-selfavg}, and provides an alternative viewpoint on the regime discussed below. One can thus only have a partially localized phase, which we now study.

We are interested in the sites that may contribute to this phase, namely those whose tail exponent satisfies \(0<\mu_i<1\), which implies \(a_i=m_i-\varphi-\gamma>0\). Recall that, in the self-averaging regime, the stationary one-site distribution has the power-law form \eqref{eq:stat_p_growth_final_mu_reset}. Even when self-averaging breaks down, we argue that the large-\(p\) tail remains controlled by the same exponent,
\begin{equation}
P_i(p)\sim p^{-1-\mu_i},
\qquad
\mu_i\simeq \frac{r_i}{m_i-\varphi-\gamma},
\label{eq:stat_p_growth_tail_mu_reset}
\end{equation}
provided \(p\) is large enough.

The reason is that the exchange term \(\varphi \bar x(t)\), which now fluctuates, only affects the early stage of the growth, whereas the tail is generated by rare events in which the time between two resets is anomalously large. In that regime, the behavior is controlled only by the deterministic growth rate \(a_i\). This can be seen directly from the renewal representation \eqref{eq:state_resetting_pi}: the distribution is determined by the time \(\tau(p)\) needed to reach \(p\), together with its Jacobian. Using \eqref{eq:tau_of_p_reset} and \eqref{eq:tauprime_reset}, one finds at large \(p\)
\begin{equation}
\tau(p)\sim \frac{\log p}{a_i},
\qquad
|\tau'(p)|\sim \frac{1}{a_i p}.
\end{equation}
Both quantities depend only on \(a_i\), and not on the exchange term \(b_i\). Thus the fluctuations of the exchange term do not modify the large-\(p\) behavior of the distribution, and in particular do not change its tail exponent.
The normalization condition may then be decomposed into two terms:
\begin{equation}
1=\sum_i \langle p_i\rangle
=
\sum_{\mu_i>1 \,\text{or}\, \mu_i<0}\langle p_i\rangle
+
\sum_{0<\mu_i<1}\langle p_i\rangle.
\label{eq:selfconstpartloc_reset}
\end{equation}
The first sum corresponds to delocalized sites, for which \(\langle p_i\rangle\sim N^{-1}\). The second contains the partially localized contributions, for which \(\langle p_i\rangle\sim N^{-\mu_i}\). The partially localized phase is present when the second sum contributes at order one.

\subsection{Generalized-Gaussian class}

We will consider the case \(r_i=r\) to simplify the discussion, although our method is not restricted to this case. Here we consider $m_i$ from the generalized-Gaussian family defined in Eq.~\eqref{eq:generalized-gaussian}. With the scaling \eqref{eq:SigmaN}, the maximum remains of order one, \(m_1\simeq \Sigma_0\), and the ordered realization \(m_1>\cdots>m_N\) satisfies, as derived earlier in Eq.~\eqref{eq:mi-gauss-scaling},
\begin{equation}
m_i \simeq \Sigma_0\left(1-\frac{\log i}{\log N}\right)^{\frac{1}{2b}}
\label{eq:mi_quantile_stretched_reset}
\end{equation}
for \(i=o(N)\). Writing \(i\sim N^u\) with \(u\in(0,1)\), this becomes
\begin{equation}
m_i=m(u)\simeq \Sigma_0(1-u)^{\frac{1}{2b}}.
\end{equation}

We first estimate which sites satisfy $\mu_i<1$. The cutoff of the partially localized sector is defined by \(\mu_i=1\), i.e.
\begin{equation}
m_i=r+\varphi+\gamma.
\end{equation}
Equivalently, if \(i\sim N^{\alpha_c}\),
\begin{equation}
\alpha_c = 1-\Big(\frac{r+\varphi+\gamma}{\Sigma_0}\Big)^{2b}.
\label{eq:alphac_def_reset}
\end{equation}

The contribution of the sites with \(0<\mu_i<1\) is then
\begin{equation}
S_N:=\sum_{0<\mu_i<1}\langle p_i\rangle
\sim
\sum_{i=1}^{N^{\alpha_c}} N^{-\mu_i},
\qquad
\mu_i=\frac{r}{m_i-\varphi-\gamma}.
\label{eq:S_def_reset}
\end{equation}
Using \(i=N^u\), so that \(di=(\log N)N^u\,du\), one obtains
\begin{equation}
S_N \simeq \log N \int_0^{\alpha_c}
\exp\!\Big(\log N\, g(u)\Big)\,du,
\qquad
g(u):=u-\frac{r}{m(u)-\varphi-\gamma}.
\label{eq:S_integral_saddle_reset}
\end{equation}
Thus the asymptotics are controlled by the maximum of \(g(u)\) on \([0,\alpha_c]\). If the maximum occurs at \(u^\star\) in the interval, Laplace's method gives
\begin{equation}
S_N \simeq 
\sqrt{\log N}\,
\frac{\sqrt{2\pi}}{\sqrt{|g''(u^\star)|}}\,
N^{g(u^\star)}.
\label{eq:S_saddle_asympt_reset}
\end{equation}
For this contribution to remain of order one, one must impose
\begin{equation}
g(u^\star)=0,
\qquad
g'(u^\star)=0,
\qquad
g''(u^\star)<0.
\label{eq:saddle_system_reset}
\end{equation}
The first condition yields
\begin{equation}
\gamma=m(u^\star)-\varphi-\frac{r}{u^\star},
\label{eq:gamma_from_u_reset}
\end{equation}
while the second gives
\begin{equation}
r=-m'(u^\star)(u^\star)^2.
\label{eq:r_from_u_general_reset}
\end{equation}
Since
\begin{equation}
m'(u)=-(\Sigma_0/2b)(1-u)^{1/(2b)-1},
\end{equation}
this becomes a self-consistent equation for $u^\star$.
\begin{equation}
r=\frac{\Sigma_0}{2b}\,(u^\star)^2\,(1-u^\star)^{\frac{1}{2b}-1}.
\label{eq:r_of_u_reset}
\end{equation}

Equations \eqref{eq:gamma_from_u_reset} and \eqref{eq:r_of_u_reset} characterize the partially localized phase. In particular, \(u^\star\) depends only on \(r\), \(\Sigma_0\), and \(b\), so \(\gamma\) decreases linearly in \(\varphi\).

The transition to the delocalized phase is obtained by imposing \(\gamma=0\), which yields the set of equations
\begin{equation}
\left\{
\begin{aligned}
\varphi_c
&=\Sigma_0(1-u)^{\frac{1}{2b}}-\frac{r}{u},\\[2pt]
r
&=\frac{\Sigma_0}{2b}\,u^2\,(1-u)^{\frac{1}{2b}-1}.
\end{aligned}
\right.
\label{eq:gamma0_curve_reset}
\end{equation}

\begin{figure}
    \centering
    \includegraphics[width=0.68\linewidth]{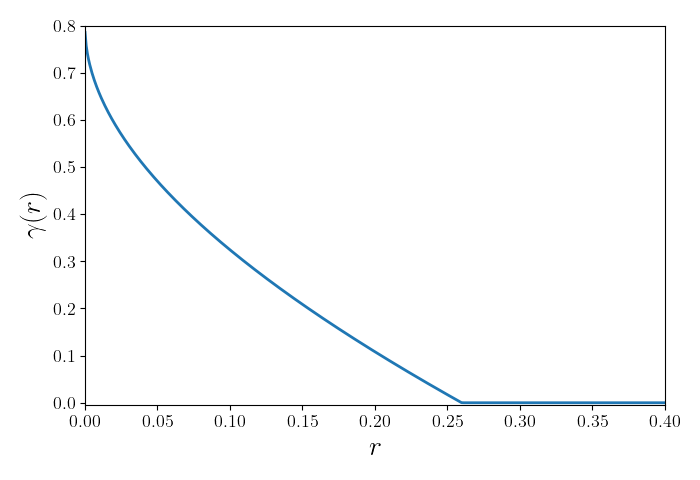}
    \caption{Predicted growth rate \(\gamma(r)\) as a function of the resetting rate for $m_i$ Gaussian and  \(\varphi=0.2\). Near \(r=0\), \(\gamma(r)\) is non-analytic and decreases abruptly at \(r \simeq 0\).}
    \label{fig:gammarphi02b1}
\end{figure}

For small \(r\), one has \(u^\star \simeq \sqrt{2 b r/\Sigma_0}\), and therefore
\begin{equation}
\gamma
\simeq
\Sigma_0-\varphi-\sqrt{\frac{2\Sigma_0 r}{b}}.
\label{eq:gamma_small_r_reset}
\end{equation}
The growth rate is in particular non-analytic \(\propto -\sqrt{r}\): the effect of resetting on growth is much stronger than multiplicative noise. An illustration of the predicted value of \(\gamma\) is shown in Fig.~\ref{fig:gammarphi02b1}.

\subsection{Compact-support class}

In the compact-support class \eqref{eq:eqmcompact}, the situation is very similar to the one with multiplicative noise \(\sigma>0\).All sites satisfy in the delocalized phase
\begin{equation}
\mu_i=\frac{r}{m_i-\varphi-\gamma}>1
\end{equation}
and the growth rate is determined by Eq.~\eqref{eq:eqgammaSA_reset}. In the growing delocalized phase, where \(p_0\to0\), this reduces to the deterministic equation \eqref{eq:eq_gamma} with the shifted rates \(m_i\to m_i-r\). The first site to become marginal is the best one, \(i=1\). The transition to the marginally partially localized phase occurs when
\begin{equation}
\mu_1=1.
\end{equation}
Beyond that point, the system remains pinned at this marginal value, \(\mu_1=1\), and therefore
\begin{equation}
\gamma = m_>-\varphi-r.
\end{equation}

\begin{figure}
	\centering
    \includegraphics[width=0.48\linewidth]{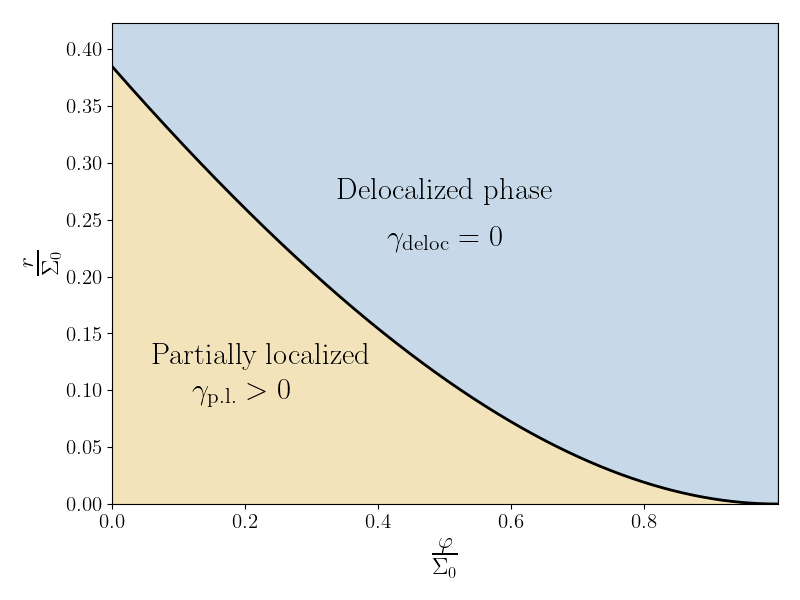}
     \includegraphics[width=0.48\linewidth]{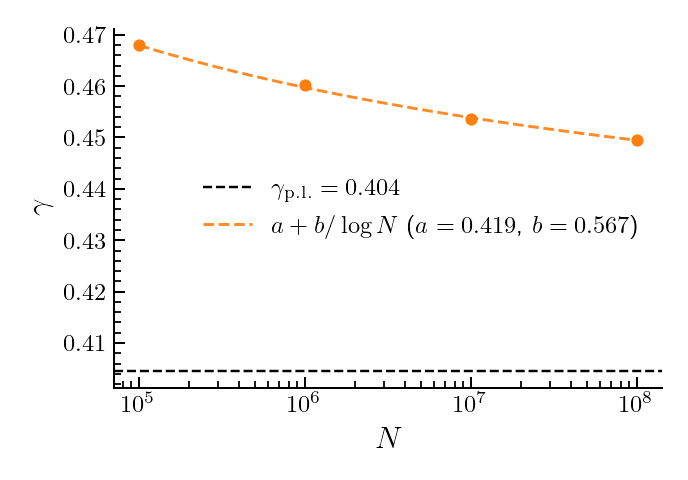}
    \caption{Heterogeneous growth with resetting. Left: Phase diagram with Gaussian $m_i$ showing the delocalized and partially localized regions. The transition line is given by Eq.\eqref{eq:gamma0_curve_reset}  Right: Numerical estimate of the growth rate (dots) for \(r=0.2\), \(\varphi=0.2\) and $\Sigma_0=\sqrt{2}$ in the partially localized regime, as well as a fit assuming logarithmic finite size corrections. The fit predicts an asymptotic value $\simeq 0.419$, close to the predicted $0.404$.}
    \label{fig:resetting-phase-and-growth}
\end{figure}

To show that this is the only possible transition, suppose, as for the model with \(\sigma>0\), that \(\mu_1<1\). Denoting again $y_i:=m_>-m_i$ the distance to the upper edge and using the compact-support quantiles \eqref{eq:mi-compact-quantile}, for \(i\ll N\) one has
\begin{equation}
y_i \simeq m_>\left(\frac{i}{N}\right)^{1/\psi}.
\end{equation}
The sites with \(\mu_i<1\) satisfy
\begin{equation}
0<y_i<m_>-(r+\varphi+\gamma),
\end{equation}
and their contribution to the normalization is
\begin{equation}
S_N
:=
\sum_{0<\mu_i<1}\langle p_i\rangle
\sim
\sum_{0<\mu_i<1}N^{-\mu_i}.
\end{equation}
Using the edge behavior \(\rho(m_>-y)\simeq \frac{\psi}{m_>^\psi}y^{\psi-1}\), this becomes
\begin{equation}
S_N
\sim
\frac{\psi}{m_>^\psi}
\int_0^{m_>-(r+\varphi+\gamma)}
dy\,
y^{\psi-1}
\exp\!\left[
\frac{(m_>-r-\varphi-\gamma-y)\ln N}{m_>-\varphi-\gamma-y}
\right].
\end{equation}
This integral is dominated by the lower edge \(y=0\). Therefore, for \(S_N\) to remain of order one, one must impose
\begin{equation}
\gamma\to m_>-\varphi-r,
\end{equation}
which is precisely the condition \(\mu_1=1\). Hence the system cannot be in a genuine \(\mu_1<1\) phase: as soon as the best site reaches the marginal value \(\mu_1=1\), it gets pinned there.
The phase diagram is therefore the same as in the deterministic compact-support case \eqref{eq:eqmcompact}, up to the shift by \(r\), with, however, the localized phase replaced by a marginally partially localized phase.
For the Gaussian case, a numerical estimate of the growth rate in the partially localized regime, together with a logarithmic fit of the finite size corrections, are shown in the right panel of Fig.~\ref{fig:resetting-phase-and-growth} and yield a good agreement.

{\bf Remarks}

The most significant effect of resetting is to suppress the fully localized phase. A site with favorable growth rate can dominate the system only if it avoids resetting for a time of order \(\log N\), and this becomes unlikely for a finite set of sites as \(N\to\infty\). It is replaced by a partially localized phase. This seems to show that this partially localized phase is quite generic when one combines quenched heterogeneities and temporal fluctuations. An interesting future subject of study is the case where resetting rates scale as $\sim 1/\log{N}$, which could allow a localized phase.

The same picture extends to more realistic variants in which the \(m_i\) are redrawn at each reset or in which the number of sites can fluctuate slightly in time (provided their mean remains fixed). Likewise, heterogeneous resetting rates \(r_i\) can be incorporated straightforwardly through the tail exponent
\begin{equation}
\mu_i=\frac{r_i}{m_i-\gamma-\varphi}.
\end{equation}

 \chapter*{Outlook}
\addcontentsline{toc}{chapter}{Outlook}
\label{ch:conclusion}

The results of this thesis open several questions.

Chapter~\ref{sec:interfaces-elastic} suggests that, for interfaces, anomalous scaling as a statistical effect may be a rather general mechanism. As discussed in that chapter, it is relevant to experiments on Hele-Shaw cells \cite{soriano2002anomalous,soriano2003anomalous,soriano2005anomalous} and to recent simulations of Ising interfaces \cite{Rodriguez-Fernandez_2025}. It would be interesting to understand more systematically in which other experimental or numerical settings the same mechanism is present. In order to identify it, one should look at the full family of local exponents \(\alpha_{\rm loc}(q)\), and in particular for the bifractal form
\begin{equation}
q \,\alpha_{\rm loc}(q)=
\begin{cases}
q \, \alpha, & q<1/\alpha,
\\[1mm]
1, & q>1/\alpha.
\end{cases}
\end{equation}
Looking only at \(q=2\) is therefore misleading: in the elastic-line model, \(\alpha_{\rm loc}(2)=1/2\) arises for purely statistical reasons, whereas in Chapter~\ref{sec:interfaces-spme-main} the same value is found for the stochastic porous medium equation, where it reflects a genuine linear local scaling.

Regarding faceted anomalous scaling, reviewed in Section~\ref{sec:intro-faceted}, it is still poorly understood beyond the simplest examples discussed there. Faceting is usually described at the level of the height field \(h(x,t)\), where it is characterized by large changes of slope. This naturally suggests looking instead at the slope field itself, \(u(x,t)=\partial_x h(x,t)\), where faceting is expected to induce large jumps in \(u(x,t)\), similar to what happens in anomalous scaling. It would be interesting to explore this direction.

Regarding the chapter on Anderson localization, a first question is whether the combinatorial method introduced here can be adapted to other one-dimensional localization problems, beyond the random mass-spring chains considered in this thesis and the tight-binding Anderson model. A second one is whether the key symmetrization approximation used in the discrete setting admits a continuum formulation and can be incorporated into more standard analytical methods. A third direction is to go beyond the \(\langle \log \cdot \rangle \simeq \log \langle \cdot \rangle\) approximation when averaging over permutations. It would be particularly interesting to know whether methods such as replicas could refine this step and recover the effects that the present approach misses, in particular Lifshitz tails.

The multiplicative-growth setting also suggests several extensions. A first direction is to consider random growth rates \(m_i(t)\) that are correlated in time. For a nontrivial large \(N\) limit, the relevant correlation time is expected to scale as \(\log N\), i.e. the timescale required for localization. A similar question arises for resetting, where one expects a rate of order \(1/\log N\) to allow the existence of a localized phase. A second direction is to understand the partially localized phase beyond the self-averaging picture. The effect of redistribution in that phase remains to be studied in more detail: finite \(N\) numerics suggest the existence of an optimal redistribution rate maximizing growth, which may vanish in the limit \(N\to+\infty\). Finally, related ecological models include saturation terms while displaying the same multiplicative-growth structure. It would therefore be interesting to determine to what extent the methods developed here can be extended to that setting.

\appendix
\chapter{Technical Details for Interface Scaling}
\label{app:interfaces-tech}

\section{Flat spectrum for randomly distributed slope flips}
\label{app:faceted-q-spectrum}

This appendix details the estimate used in Section~\ref{sec:intro-faceted}. Starting from the curvature field in Eq.~\eqref{eq:intro-qx-faceted},
\begin{equation}
  q(x)=\sum_{n=1}^N a_n\,\delta(x-X_n),
\end{equation}
its Fourier transform reads
\begin{equation}
  \hat q(k)
  =
  \frac{1}{\sqrt L}\sum_{n=1}^N a_n\,e^{-ikX_n},
  \qquad k\neq 0.
\end{equation}
Hence
\begin{equation}
  |\hat q(k)|^2
  =
  \frac{1}{L}
  \sum_{n,m=1}^N
  a_n a_m\,e^{-ik(X_n-X_m)}.
\end{equation}

If the flip positions \(X_n\) are independent and uniformly distributed on \([0,L]\), then for nonzero Fourier modes,
\begin{equation}
  \left\langle e^{-ik(X_n-X_m)} \right\rangle
  =
  \delta_{nm},
\end{equation}
since \(\langle e^{-ikX}\rangle=0\) for \(k\neq 0\). Therefore only the diagonal terms survive after averaging:
\begin{equation}
  \left\langle |\hat q(k)|^2 \right\rangle
  =
  \frac{1}{L}\sum_{n=1}^N a_n^2.
\end{equation}
If the jumps \(a_n\) are of order unity, for instance \(a_n=\pm a\), this gives
\begin{equation}
  \left\langle |\hat q(k)|^2 \right\rangle
  =
  \frac{a^2 N}{L}
  \sim \frac{N}{L},
\end{equation}
which shows that the spectrum of \(q\) is flat. Combined with Eq.~\eqref{eq:intro-Sk-from-q} of the main text, this gives the \(\propto k^{-4}\) structure factor quoted in Eq.~\eqref{eq:intro-kminus4}.

\chapter{Technical Details for Anderson Localization}
\label{app:anderson-tech}

\section{Combinatorial representation of the coefficients \texorpdfstring{$a_{n+1}^{(k)}$}{a(n+1,k)}}
\label{app:anderson-coefficients}

We derive here the exact combinatorial formula stated in Eq.~\eqref{eq:MainResIntro2} of the main text:
\begin{equation}
  a_{n+1}^{(k)}
  =
  \sum_{1\le j_1\le i_1<j_2\le i_2<\cdots<j_k\le i_k\le n}
  \prod_{m=1}^{k}\frac{m_{j_m}}{\K_{i_m}},
\end{equation}
used in Chapter~\ref{sec:Combinatorics}. We consider the initial-value problem with
\begin{equation}
  x_0=0,
  \qquad
  x_1=1,
\end{equation}
and the recurrence
\begin{equation}
  x_{n+1}(\lambda)
  =
  \left(
    1+\frac{\K_{n-1}-m_n\lambda}{\K_n}
  \right)x_n(\lambda)
  -
  \frac{\K_{n-1}}{\K_n}\,x_{n-1}(\lambda).
\end{equation}
Expanding \(x_{n+1}(\lambda)\) as a polynomial in \(\lambda\),
\begin{equation}
  x_{n+1}(\lambda)=\sum_{k=0}^{n} a_{n+1}^{(k)}\,(-\lambda)^k,
\end{equation}
and identifying powers of \(\lambda\), one finds
\begin{equation}
  a_{n+1}^{(0)}-a_n^{(0)}
  =
  \frac{\K_{n-1}}{\K_n}\bigl(a_n^{(0)}-a_{n-1}^{(0)}\bigr),
\end{equation}
while for \(k\geq1\),
\begin{equation}
  \label{app:eq:coeff-recursion}
  a_{n+1}^{(k)}-a_n^{(k)}
  =
  \frac{\K_{n-1}}{\K_n}\bigl(a_n^{(k)}-a_{n-1}^{(k)}\bigr)
  +
  \frac{m_n}{\K_n}\,a_n^{(k-1)}.
\end{equation}

For the boundary conditions used in the main text, \(\K_0=0\), so \(a_2^{(0)}=a_1^{(0)}\). Since \(a_1^{(0)}=1\), it follows that
\begin{equation}
  a_n^{(0)}=1,
  \qquad n\geq1.
\end{equation}
It is then convenient to introduce the discrete increments
\begin{equation}
  u_{n+1}^{(k)}\eqdef a_{n+1}^{(k)}-a_n^{(k)},
  \qquad
  a_{n+1}^{(k)}=\sum_{m=0}^{n}u_{m+1}^{(k)}.
\end{equation}
The case \(k=0\) corresponds to
\begin{equation}
  u_n^{(0)}=\delta_{n,1}.
\end{equation}
For \(k\geq1\), Eq.~\eqref{app:eq:coeff-recursion} becomes
\begin{equation}
  u_{n+1}^{(k)}
  =
  \frac{\K_{n-1}}{\K_n}\,u_n^{(k)}
  +
  \frac{m_n}{\K_n}\sum_{j=1}^{n}u_j^{(k-1)}.
\end{equation}
This recursion can be solved explicitly. Writing it as an arithmetic recurrence for \(\K_n u_{n+1}^{(k)}\), one gets
\begin{equation}
  \K_n u_{n+1}^{(k)}
  =
  \K_{n-1}u_n^{(k)}
  +
  m_n\sum_{j=1}^{n}u_j^{(k-1)},
\end{equation}
hence
\begin{equation}
  \label{app:eq:u-iteration-1}
  u_{n+1}^{(k)}
  =
  \frac{1}{\K_n}
  \sum_{j=1}^{n}m_j\sum_{p=1}^{j}u_p^{(k-1)}
  =
  \frac{1}{\K_n}
  \sum_{p=1}^{n}u_p^{(k-1)}\sum_{j=p}^{n}m_j.
\end{equation}
Since \(u_p^{(k-1)}=0\) for \(p<k\), the first nonzero term occurs at \(p=k\), and it is more convenient to rewrite this as
\begin{equation}
  \label{app:eq:u-iteration-2}
  u_{n+1}^{(k)}
  =
  \frac{1}{\K_n}
  \sum_{p=k-1}^{n-1}u_{p+1}^{(k-1)}
  \sum_{j=p+1}^{n}m_j.
\end{equation}

The first iterations are already suggestive. For \(k=1\), one finds
\begin{equation}
  u_{n+1}^{(1)}
  =
  \frac{1}{\K_n}\sum_{j=1}^{n}m_j.
\end{equation}
For \(k=2\),
\begin{align}
  u_{n+1}^{(2)}
  &=
  \frac{1}{\K_n}\sum_{p=1}^{n-1}
  \left(\frac{1}{\K_p}\sum_{j_1=1}^{p}m_{j_1}\right)
  \left(\sum_{j_2=p+1}^{n}m_{j_2}\right)
  \nonumber\\
  &=
  \sum_{1\le j_1\le i_1<j_2\le n}
  \frac{m_{j_1}m_{j_2}}{\K_{i_1}\K_n}.
\end{align}
The general pattern follows by induction from Eq.~\eqref{app:eq:u-iteration-2}:
\begin{equation}
  \label{app:eq:u-general}
  u_{n+1}^{(k)}
  =
  \sum_{1\le j_1\le i_1<j_2\le i_2<\cdots<j_{k-1}\le i_{k-1}<j_k\le n}
  \frac{1}{\K_n}
  \prod_{m=1}^{k-1}\frac{m_{j_m}}{\K_{i_m}}\,m_{j_k}.
\end{equation}
In other words, the last spring index is fixed to \(n\), while the earlier mass and spring indices are interlaced in increasing order.

Finally, summing over the last index \(n\) gives
\begin{equation}
  a_{n+1}^{(k)}
  =
  \sum_{p=0}^{n}u_{p+1}^{(k)}
  =
  \sum_{1\le j_1\le i_1<j_2\le i_2<\cdots<j_k\le i_k\le n}
  \prod_{m=1}^{k}\frac{m_{j_m}}{\K_{i_m}},
\end{equation}
which is precisely Eq.~\eqref{eq:MainResIntro2}. This representation makes explicit the symmetry between the masses \(m_j\) and the inverse spring constants \(\K_i^{-1}\), and it is the starting point of the subsequent symmetrization and saddle-point analysis.

\section{Integral form of the free energy}
\label{app:anderson-free-energy}

We derive here the integral representation quoted after Eq.~\eqref{eq:FreeEnergy} in the main text:
\begin{equation}
  \frac{\ln Z(k,n)}{n}
  \simeq
  -\int_0^{\theta}\D t\,\ln \varphi_*(t),
  \qquad
  \theta=\frac{k}{n},
\end{equation}
used in Chapter~\ref{sec:Combinatorics}. Starting from the saddle-point expression
\begin{equation}
  \label{app:eq:free-energy-saddle}
  \frac{\ln Z(k,n)}{n}
  \simeq
  \mean{\ln\!\bigl(1+\varphi_* X\bigr)}
  -
  \theta \ln \varphi_*,
\end{equation}
we regard \(\varphi_*\) as a function of \(\theta\), determined implicitly by the saddle-point equation
\begin{equation}
  \label{app:eq:free-energy-saddle-eq}
  \theta
  =
  \varphi_*(\theta)\,
  \mean{\frac{X}{1+\varphi_*(\theta)X}}.
\end{equation}

If we introduce the free energy per level
\begin{equation}
  f(\theta)\eqdef -\frac{\ln Z(k,n)}{n},
\end{equation}
then we have the thermodynamic identity
\begin{equation}
  \label{app:eq:free-energy-identity}
  \frac{\partial f(\theta)}{\partial \theta}=\ln\varphi_*(\theta),
\end{equation}
which is simply the canonical chemical potential. The relation is proven by differentiating Eq.~\eqref{app:eq:free-energy-saddle} with respect to \(\theta\):
\begin{align}
  \frac{\partial}{\partial \theta}
  \left[\frac{\ln Z(k,n)}{n}\right]
  &=
  \varphi_*'(\theta)\,
  \mean{\frac{X}{1+\varphi_*(\theta)X}}
  -\ln\varphi_*(\theta)
  -\theta\,\frac{\varphi_*'(\theta)}{\varphi_*(\theta)}.
\end{align}
Using Eq.~\eqref{app:eq:free-energy-saddle-eq} to replace the average, the terms proportional to \(\varphi_*'(\theta)\) cancel exactly, leaving
\begin{equation}
  \label{app:eq:free-energy-derivative}
  \frac{\partial}{\partial \theta}
  \left[\frac{\ln Z(k,n)}{n}\right]
  =
  -\ln\varphi_*(\theta).
\end{equation}
This is equivalent to Eq.~\eqref{app:eq:free-energy-identity}.

Noticing that \(Z(0,n)=1\), we have
\begin{equation}
  \left.\frac{\ln Z(k,n)}{n}\right|_{\theta=0}=0.
\end{equation}
and therefore we can integrate Eq.~\eqref{app:eq:free-energy-identity} from \(0\) to \(\theta=k/n\) to obtain
\begin{equation}
  \label{app:eq:free-energy-integral}
  \frac{\ln Z(k,n)}{n}
  \simeq
  -\int_0^{\theta}\D t\,\ln \varphi_*(t),
\end{equation}
which is the integral form stated after Eq.~\eqref{eq:FreeEnergy}.

\chapter{Technical Details for the Population Dynamics Part}
\label{app:mfg-tech}

This appendix contains a few technical derivations for Part~II.

\section{Multiplicative noise: delocalized conditional law for favorable sites}
\label{app:sigma-noise-deloc-moments}

In the effective description of the localized phase developed in Chapter~\ref{sec:population-mf-growth-noise}, good but non-localized sites are described by the law
\begin{equation}
P_i^{\mathrm{deloc}}(p)\sim p^{-1-\mu_i}\exp\!\left(-\frac{A_d}{Np}\right),
\end{equation}
with an \(O(1)\) upper cutoff \(c\); this is the conditional ansatz introduced in Eq.~\eqref{eq:ansatzdeloc} of the main text and later generalized in Eq.~\eqref{eq:Pi_deloc_ansatz}. As we will show below, the relevant scaling is
\begin{equation}
\mu_i=\frac{d_i}{\log N},
\qquad d_i=O(1).
\end{equation}
We derive here the normalization \(\mathcal N_i^{\mathrm{deloc}}\) and the asymptotic behavior of
\(\bigl\langle 1/(N p_i^{\mathrm{deloc}})\bigr\rangle\), which are used in the main text to obtain the large-rank estimate \eqref{eq:mui_largei} and the localization probability \eqref{eq:Pgt_i_intermediate}.

\paragraph{Normalization.}
One starts from
\begin{equation}
\mathcal N_i^{\mathrm{deloc}}
=
\int_0^c dp\, p^{-1-\mu_i}\exp\!\left(-\frac{A_d}{N p}\right).
\end{equation}
Using the change of variable \(u=A_d/(Np)\), so that \(p=A_d/(Nu)\) and \(dp=-(A_d/N)u^{-2}du\), this becomes
\begin{equation}
\mathcal N_i^{\mathrm{deloc}}
=
\left(\frac{N}{A_d}\right)^{\mu_i}
\int_{A_d/(Nc)}^\infty u^{\mu_i-1}e^{-u}\,du
=
\left(\frac{N}{A_d}\right)^{\mu_i}
\Gamma\!\left(\mu_i,\frac{A_d}{Nc}\right),
\end{equation}
where \(\Gamma(\mu,x)\) is the incomplete Gamma function. For \(\mu_i\to0\) and \(N\to\infty\),
\begin{equation}
\Gamma(\mu_i,x)\simeq \frac{1-x^{\mu_i}}{\mu_i},
\qquad
x=\frac{A_d}{Nc}\ll1,
\end{equation}
so that
\begin{equation}
\mathcal N_i^{\mathrm{deloc}}
\simeq
\left(\frac{N}{A_d}\right)^{\mu_i}
\frac{1-\left(\frac{A_d}{Nc}\right)^{\mu_i}}{\mu_i}.
\end{equation}
Using \(\mu_i=d_i/\log N\), one obtains
\begin{equation}
\mathcal N_i^{\mathrm{deloc}}
\simeq
e^{d_i}\frac{1-e^{-d_i}}{d_i}\,\log N
=
\frac{e^{d_i}-1}{d_i}\,\log N.
\label{app:eq:Ni_deloc}
\end{equation}

\paragraph{Inverse moment.}
Next,
\begin{equation}
\left\langle \frac{1}{N p_i^{\mathrm{deloc}}} \right\rangle
=
\frac{1}{N\,\mathcal N_i^{\mathrm{deloc}}}
\int_0^c dp\, p^{-2-\mu_i}\exp\!\left(-\frac{A_d}{Np}\right).
\end{equation}
With the same change of variable,
\begin{equation}
\left\langle \frac{1}{N p_i^{\mathrm{deloc}}} \right\rangle
=
\frac{1}{N\,\mathcal N_i^{\mathrm{deloc}}}
\left(\frac{N}{A_d}\right)^{1+\mu_i}
\int_{A_d/(Nc)}^\infty u^{\mu_i}e^{-u}\,du
\simeq
\frac{1}{A_d\,\mathcal N_i^{\mathrm{deloc}}}
\left(\frac{N}{A_d}\right)^{\mu_i}\Gamma(1+\mu_i).
\end{equation}
Since \(\Gamma(1+\mu_i)\to 1\) and \((N/A_d)^{\mu_i}\simeq e^{d_i}\), Eq.~\eqref{app:eq:Ni_deloc} gives
\begin{equation}
\left\langle \frac{1}{N p_i^{\mathrm{deloc}}} \right\rangle
\simeq
\frac{1}{A_d}\,
\frac{e^{d_i}}{(e^{d_i}-1)\log N}\,d_i
=
\frac{1}{A_d}\,
\frac{d_i}{1-e^{-d_i}}\,
\frac{1}{\log N}.
\label{app:eq:invp_mean_di}
\end{equation}

\paragraph{Large-rank asymptotics of the tail exponent.}
We now derive the large-\(i\) behavior of \(\mu_i\). In the main text, the exact identity
\begin{equation}
\gamma+\varphi+\frac{\sigma^2}{2}
\simeq
m_i+\varphi \left\langle \frac{1}{N p_i} \right\rangle_\mathrm{deloc}
\end{equation}
is used for sufficiently large rank \(i\), where the probability that site \(i\) is itself localized is negligible; this is Eq.~\eqref{eq:eqlabmi} in the main text. Since the left-hand side is independent of \(i\), the rank dependence of
\(\left\langle 1/(N p_i)\right\rangle_\mathrm{deloc}\) must compensate that of \(m_i\). For the generalized-Gaussian tail,
\begin{equation}
m_i=\Sigma_0-\frac{\Sigma_0}{2 b}\frac{\log i}{\log N}.
\end{equation}
This immediately shows that \(\left\langle 1/(N p_i)\right\rangle_\mathrm{deloc}\) must be of order \(1/\log N\), which is only possible if
\begin{equation}
\mu_i=\frac{d_i}{\log N},
\qquad d_i=O(1).
\end{equation}
Substituting Eq.~\eqref{app:eq:invp_mean_di} into the identity above gives
\begin{equation}
-\frac{\Sigma_0}{2 b}\frac{\log i}{\log N}
+
\frac{\varphi}{A_d}
\frac{d_i}{1-e^{-d_i}}
\frac{1}{\log N}
=
\mathrm{const.}
\label{app:eq:balance_di_exact}
\end{equation}
For large \(i\), one expects \(d_i\) to be large enough that \(e^{-d_i}\ll1\), so that
\(\frac{d_i}{1-e^{-d_i}}\simeq d_i\). Since the $i$ dependence must cancel out, one then finds
\begin{equation}
\mu_i = \frac{d_i}{\log{N}}
\simeq
\frac{A_d\Sigma_0}{2 b\,\varphi}\,
\frac{\log i}{\log N}
\label{app:eq:mui_largei}
\end{equation}
which is the estimate quoted in Eq.~\eqref{eq:mui_largei} of the main text.

\section{Resetting: minimum number of favorable sites needed to localize}
\label{app:resetting-selfavg}

Here, we estimate the number of favorable sites needed to obtain a localized phase. This provides the quantitative version of the argument summarized at the beginning of Section~\ref{sec:population-resetting}. Consider a simple setting constituted of \(k\) favorable sites with growth rate \(m>\varphi\) and \(N-k\) unfavorable sites with \(m=0\). For one of the favorable sites to carry an \(O(1)\) fraction of the total population, it must first grow from \(O(1)\) to \(O(N)\), which takes a time
\begin{equation}
t_N=\frac{\log N}{m-\varphi}.
\end{equation}
Since resetting is Poissonian with rate \(r\), the probability that a given site avoids reset up to that time is
\begin{equation}
P(\tau>t_N)=e^{-rt_N}=N^{-\frac{r}{m-\varphi}}.
\end{equation}
For \(k\) such sites, the probability that at least one of them survives that long is
\begin{equation}
P_{\max}(\tau>t_N)
=
1-\bigl(1-e^{-rt_N}\bigr)^k
\simeq
1-\exp\!\left[-k\,N^{-\frac{r}{m-\varphi}}\right].
\end{equation}
Thus one needs
\begin{equation}
k\gg N^{\frac{r}{m-\varphi}}
\end{equation}
for this probability to approach one. A finite number of favorable sites is therefore not enough to sustain growth and have a localized phase. It is however possible with a subextensive but diverging number of favorable sites, which precisely corresponds to the partially localized phase.

\printbibliography[heading=bibintoc,title={Bibliography}]

\end{document}